\documentclass[twocolumn,times]{aastex63}

\received{XXXX XX, 2020}

\shorttitle{The $\shb$-based \mdot and its connection with the corona for AGN}
\shortauthors{Chen, Liu \& Bian}

\usepackage{graphicx}
\usepackage{multirow}
\usepackage{color}
\usepackage[figuresright]{rotating}
\usepackage{longtable}
\usepackage{mathrsfs,amsmath}
\usepackage{epstopdf}
\newsavebox{\tablebox}
\usepackage{hyperref}
\usepackage{times}
\usepackage{longtable}
\usepackage{array}
\usepackage{booktabs}

\usepackage{lineno}

\newcommand{\lv}{\ifmmode L_{5100} \else $L_{5100}$\ \fi}
\newcommand{\kms}{\ifmmode {\rm km\ s}^{-1} \else km s$^{-1}$\ \fi}
\newcommand{\ergs}{\ifmmode {\rm erg\ s}^{-1} \else erg s$^{-1}$\ \fi}
\newcommand{\lb}{\ifmmode L_{\rm Bol} \else $L_{\rm Bol}$\ \fi}
\newcommand{\ledd}{\ifmmode L_{\rm Edd} \else $L_{\rm Edd}$\ \fi}
\newcommand{\hb}{\ifmmode H\beta \else H$\beta$\ \fi}
\newcommand{\ha}{\ifmmode H\alpha \else H$\alpha$\ \fi}

\newcommand{\oiii}{[O {\sc iii}]}
\newcommand{\heii}{He {\sc ii}}
\newcommand{\sii}{[S {\sc ii}]}
\newcommand{\nii}{[N {\sc ii}]}
\newcommand{\feii}{Fe {\sc ii}\ }
\newcommand{\mbh}{\ifmmode M_{\rm BH}  \else $M_{\rm BH}$\ \fi}
\newcommand{\msun}{M_{\odot}}
\newcommand{\rfe}{\ifmmode R_{\rm Fe} \else $R_{\rm Fe}$\ \fi}
\newcommand{\sst}{\ifmmode \sigma_{\rm \ast}\else $\sigma_{\rm \ast}$\ \fi}
\newcommand{\dhb}{\ifmmode \mathscr{D}_{\rm H\beta} \else $\mathscr{D}_{\rm H\beta}$\ \fi}
\newcommand{\leddR}{\ifmmode L_{\rm Bol}/L_{\rm Edd} \else $L_{\rm Bol}/L_{\rm Edd}$\ \fi}
\newcommand{\mdot}{\ifmmode \dot{\mathscr{M}}  \else $\dot{\mathscr{M}}$\ \fi}
\newcommand{\rhb}{\ifmmode R_{\rm BLR}({\rm H\beta})  \else $R_{\rm BLR}({\rm H\beta})$ \ \fi}
\newcommand{\shb}{\ifmmode \sigma_{\rm H\beta} \else $\sigma_{\rm \hb}$\ \fi}
\newcommand{\RL}{\ifmmode R_{\rm BLR}({\rm H\beta}) - L_{\rm 5100} \else $R_{\rm BLR}({\rm H\beta}) - L_{\rm 5100}$ \ \fi}
\newcommand{\ms}{\ifmmode M_{\rm BH}-\sigma_{\ast} \else $M_{\rm BH}-\sigma_{\ast}$\ \fi}
\newcommand{\sm}{\ifmmode \sigma_{\rm H\beta,mean} \else $\sigma_{\rm H\beta,mean}$\ \fi}
\newcommand{\sr}{\ifmmode \sigma_{\rm H\beta,rms} \else $\sigma_{\rm H\beta,rms}$\ \fi}
\newcommand{\fwm}{\ifmmode \rm FWHM_{\rm mean} \else $\rm FWHM_{\rm mean}$\ \fi}
\newcommand{\fwr}{\ifmmode \rm FWHM_{\rm rms} \else $\rm FWHM_{\rm rms}$\ \fi}
\newcommand{\vpfm}{\ifmmode \rm VP_{\rm F, mean} \else $\rm VP_{\rm F, mean}$\ \fi}
\newcommand{\vpsm}{\ifmmode \rm VP_{\rm \sigma, mean} \else $\rm VP_{\rm \sigma, mean}$\ \fi}
\newcommand{\vpfr}{\ifmmode \rm VP_{\rm F, rms} \else $\rm VP_{\rm F, rms}$\ \fi}
\newcommand{\vpsr}{\ifmmode \rm VP_{\rm \sigma, rms} \else $\rm VP_{\rm \sigma, rms}$\ \fi}
\newcommand{\ffm}{\ifmmode f_{\rm F, mean} \else $f_{\rm F, mean}$\ \fi}
\newcommand{\fsm}{\ifmmode f_{\rm \sigma, mean} \else $f_{\rm \sigma, mean}$\ \fi}
\newcommand{\ffr}{\ifmmode f_{\rm F, rms} \else $f_{\rm F, rms}$\ \fi}
\newcommand{\fsr}{\ifmmode f_{\rm \sigma, rms} \else $f_{\rm \sigma, rms}$\ \fi}
\newcommand{\fc}{\ifmmode f_{\rm c} \else $f_{\rm c}$\ \fi}
\newcommand{\dhbm}{\ifmmode D_{\rm H\beta,mean} \else $D_{\rm H\beta,mean}$\ \fi}
\newcommand{\dhbr}{\ifmmode D_{\rm H\beta,rms} \else $D_{\rm H\beta,rms}$\ \fi}

\newcommand{\lxm}{\ifmmode L_{\rm 14-195keV} \else  $L_{\rm 14-195keV}$ \fi}
\newcommand{\fx}{\ifmmode F_{\rm X} \else  $F_{\rm X}$ \fi}
\newcommand{\lx}{\ifmmode L_{\rm X} \else  $L_{\rm X}$ \fi}

\begin{document}

\title{The $\shb$-based dimensionless accretion rate and its connection with the corona for AGN}

\correspondingauthor{W. -H. Bian}
\email{whbian@njnu.edu.cn}

\author{Yu-Qin Chen}
\affiliation{School of Physics and Technology, Nanjing
Normal University, Nanjing 210046, China}

\author{Yan-Sheng Liu}
\affiliation{School of Physics and Technology, Nanjing
Normal University, Nanjing 210046, China}

\author[0000-0002-2121-8960]{Wei-Hao Bian}
\affiliation{School of Physics and Technology, Nanjing
Normal University, Nanjing 210046, China}

\begin{abstract}
With respect to the \hb full width at half-maximum ($\rm FWHM_{\hb}$), the broad \hb line dispersion (\shb) was preferred as a velocity tracer to calculate the single-epoch supermassive black hole mass (\mbh) suggested by \cite{Yu2020b}.  For a compiled sample of 311 broad-line active galactic nuclei (AGN) 
with measured hard X-ray photon index ($z<0.7$),  \shb and the optical \feii relative strength (\rfe) are measured from their optical spectra, which are used to calculate $\shb$-based virial \mbh and dimensionless accretion rate (\mdot).  
With respect to $\rm FWHM_{\hb}$, it is found that the mean value of $\shb$-based \mbh is on average larger by 0.26 dex, and the mean value of $\shb$-based \mdot is on average smaller by 0.51 dex.
It is found that there exists a non-linear relationship between the Eddington ratio  ($\leddR$) and \mdot, i.e., $\leddR \propto \mdot^{0.56\pm 0.01}$. This non-linear relationship comes from the accretion efficiency $\eta$, which is smaller for AGN with higher \mdot.
We find a strong bivariate correlation of the fraction of energy released in the corona \fx with \mdot and \mbh, $\fx\propto \mdot^{-0.57\pm 0.05} M_{\rm BH}^{-0.54\pm 0.06}$. The flat slope of $-0.57\pm 0.05$ favours the shear stress tensor of the accretion disk being proportional to the geometric mean of gas pressure and total pressure. 
We find a strong bivariate relation of $\Gamma$ with \mdot and \fx, $\Gamma \propto \mdot^{-0.21\pm 0.02}\fx^{0.02\pm 0.04}$. The hard X-ray spectrum becomes softer with increasing of \fx, although the scatter is large.


\end{abstract}




\section{Introduction} \label{sec:intro}
Supermassive black holes (SMBHs) are believed to exist at the centers of nearly all galaxies \citep[e.g.,][]{KH13}. There are mainly two key parameters for SMBHs, i.e., mass (\mbh) and spin. Understanding the properties of SMBHs will clarify  the physics of active galactic nuclei (AGN)  \citep[e.g.,][]{Bian2002, KH13, N2013}. Based on the accretion disk theory, the accretion rates can be calculated from the emitted spectra of AGN \citep[e.g.,][]{N2013}.  

A model with a hot corona surrounding a cold accretion disk in AGN describes the X-ray emission through Compton up-scattering of disk UV photons by the relativistic electrons in the hot corona \citep{Liang1979, Haardt1991, N2013}. There are various models of the coronal geometry,  such as  hot parallel planes covering the cold accretion disk, a hot sphere around the central SMBH, and an inner hot sphere plus an inner warm disk \citep[e.g.,][]{Haardt1991, Haardt1994, N2013, Kubota2018}. 
For the standard accretion disk model, \cite{SS73}  assumed that angular momentum transport was carried out by turbulence and that the stress tensor scaled with the disk pressure, $t_{r\phi} = -\alpha P$ , with $\alpha$ being the viscosity. Other expressions of the viscous stress $t_{r\phi}$  have been discussed, with varying dependence on the gas pressure $P_{\rm gas}$, radiation pressure $P_{\rm rad}$ or both  \citep[e.g.,][]{Wang2004, Yang2007}.
It is believed that the strong buoyancy and magnetic-field reconnection inevitably lead to the formation of a hot corona \citep{Stella1984}. A fraction of total dissipated energy is transferred vertically outside the disk, and released in the hot, magnetically dominated corona \citep[e.g.][]{Haardt1991, Svensson1994}. 
The fraction of the energy transported by magnetic buoyancy is $\fx=P_{\rm mag}v_{p}/Q$, where $P_{\rm mag} = B^{2}/8\pi$ is the magnetic pressure, $v_{p}$ is the transport velocity, and the dissipated energy  flux density in the disk $Q=-(3/2)c_{\rm s}t_{\rm r\phi}$ \citep{Merloni2002}. 
The viscous stress is assumed to be $t_{r\phi}=-k_0 P_{\rm mag}$, and 
\begin{equation}
\fx=\frac {2v_{p}}{3k_{0}c_{s}} =
\frac{2^{\frac{3}{2}}b}{3k_{0}}\sqrt{\frac{P_{\rm mag}}{P_{\rm tot}}} =
\frac{2^{\frac{3}{2}}b}{3k_{0}^{\frac{3}{2}}}\sqrt{\frac{(-t_{r\phi})}{P_{\rm tot}}}
= C\sqrt{\frac{(-t_{r\phi})}{P_{\rm tot}}}
\end{equation}
where $b = v_{p}/v_{A}$, the $ Alfv\acute{e}n$ velocity $v_{A} = B/\sqrt{4\pi\rho}$, $P_{\rm tot} = P_{\rm rad} + P_{\rm gas}$, $c_{s}^{2}=P_{\rm tot}/\rho$ and $C = 2^{3/2}b/(3k_{0}^{3/2})\sim 1$. 
Therefore, if the shear stress is assumed, the fraction \fx can be theoretically obtained for different accretion rates \citep[e.g.][]{Svensson1994, Merloni2002, Wang2004, Yang2007}. Through X-ray and optical observations, $\fx \equiv \lx/\lb$ (\lx and \lb are the X-ray luminosity and the bolometric luminosity, respectively) can be estimated. The dependence from Equation 1 gives us a possible opportunity to determine the working magnetic stress from hard X-ray observations.

The relation between \fx and the Eddington ratio ($\leddR$, where \ledd is the Eddington luminosity) was investigated by some authors with different AGN samples, suggesting a relation between \fx\ and  $\leddR$, $\fx\propto (\leddR)^{\alpha}$, where $\alpha$ is between $-0.64$ and $ -0.74$  \citep[e.g.][]{Merloni2002, Wang2004, Yang2007, Wang2019}. These results supported the magnetic stress tensor being of the form $t_{\rm r\phi}\propto P_{\rm gas}$. 
The relation between the optical-to-X-ray power-law slope parameter $\alpha_{\rm OX}$ \footnote{$\alpha_{\rm OX}$  is defined as $\alpha_{\rm OX} = 0.3838 \log(L_{\rm 2keV}/ L_{\rm 2500})$, where $L_{\rm 2keV}$ and $L_{\rm 2500}$ are the monochromatic luminosities at rest-frame 2 keV and 2500 \AA, respectively.} and $L_{\rm 2500}$ shows that the accretion disk and X-ray corona are connected \citep[e.g., ][]{Avni1982, Jin2012, Laha2018, Liu2021}. This correlation can also be described as the relation between the UV/optical and X-ray luminosities, i.e., the $L_{\rm 2keV}-L_{\rm 2500}$ relation \citep[e.g.,][]{Avni1982, Lusso2016, Lusso2017}. A well-calibrated non-linear $L_{\rm 2keV}-L_{\rm 2500}$ relation can also be used to estimate cosmological parameters \citep[e.g., ][]{Lusso2016, Lusso2017}.
Another relation between the hard X-ray photon index  $\Gamma$ and the Eddington ratio  $\leddR$ was extensively discussed, and the hard X-ray spectrum becomes softer as the Eddington ratio becomes larger
 \citep[e.g.,][]{Bian2003a, Wang2004,Yang2007, Brightman13,  Liu2015, Meyer2017, Trakhtenbrot2017, Qiao2018, Wang2019, Huang2020, Liu2021}.

In these studies, the Eddington ratio $\leddR$ is usually used to represent the accretion strength. The dimensionless accretion rate $\mdot$ is also used to represent the accretion strength instead of \leddR, where $\mdot\equiv \dot{M}/\dot{M}_{\rm Edd}$, $\dot{M}$ is the mass accretion rate, $\dot{M}_{\rm Edd}=L_{\rm Edd}/c^2$ \cite[e.g.,][]{Wang2013, Du2015, Kubota2018, Liu2021}. 
The relation between $\mdot$ and $\leddR$ depends on the conversion efficiency $\eta$ (from the accretion mass to the radiation), $\lb=\eta \dot{M} c^2$ \citep{Bian2003b, DL2011}. For both parameters $\mdot$ and $\leddR$, \mbh is the key parameter that needs to be determined.

For type 1 AGN, broad-line region (BLR) clouds can be used as a probe of the gravitational potential of the SMBH. The virial mass can be derived as follows \citep[e.g.,][]{Pe04, N2013}:
 \begin{equation}
 \label{eq1}
\mbh=f\times \frac{R_{\rm BLR}~(\Delta V)^2}{G}  .
\end{equation}
where  $\Delta V$ is the velocity dispersion of the BLR clouds,  $f$ is a corresponding virial factor, $R_{\rm BLR}$ is the distance from the SMBH to the BLR, and $G$ is the gravitational constant. $R_{\rm BLR}$  can be estimated from the reverberation mapping (RM) method \citep[e.g.,][]{BM82} or from the empirical $R_{\rm BLR}-\lv$ relation ($L_{\rm 5100}$ is the 5100 \AA\ monochromatic luminosity) \cite[e.g.,][]{Ka00, Be13, Du2019, Yu2020a}.  $\Delta V$ is  usually determined from the broad \hb Full-width at half-maximum (FWHM) or the line dispersion ($\shb$) measured from a single-epoch spectrum or a mean/rms spectrum \citep[e.g.,][]{Pe04, WS2019, Yu2020b}. The factor $f$ is a key quantity, which is usually calibrated through the \ms relation ($\sigma_*$ is the bulge stellar velocity dispersion) or other independent methods to derive the SMBH masses \citep[e.g.,][]{On04, Yu2019, Yu2020b}. For a sample of RM AGN, with respect to \mbh from $\sigma_*$ or $\sigma_{\rm \hb,rms}$, it was found that we can obtain \mbh from $\sigma_{\rm \hb,mean}$ with smaller scatter than from $\rm FWHM_{\hb}$.  \shb is preferred as a velocity tracer to calculate the single-epoch virial \mbh \citep{Yu2020b}.
\cite{Dalla2020} also suggested that the use of \shb gives better results, while the use of $\rm FWHM_{\hb}$ introduces a bias, stretching the mass scale such that high masses are overestimated and low masses are underestimated, although both velocity tracers are usable. 

The accretion rate can be derived from the disk model of \cite{SS73}, which has been extensively applied to fit the spectra of AGN \citep[e.g.,][]{DL2011, Mejia2018}. Considering that the radial distribution of the effective disk temperature is given by $T_{\rm eff} \propto R^{-3/4}$, \mdot is \citep[e.g.,][]{Du2016a}
 \begin{equation}
 \label{eq2}
\mdot \equiv \dot{M}/\dot{M}_{\rm Edd} = 20.1\left(\frac{l_{44}}{\cos\,\it{i}}\right)^{3/2} m_{7}^{-2}.
\end{equation}
where $l_{44} = \lv/10^{44}\ergs$ , $m_{7}=M_{\rm BH}/10^7 \msun$.  An average value of ${\rm cos}~ i = 0.75$ is adopted. \mdot is inversely proportional to \mbh to the second power. 
Recently, \mdot was used to search for super-Eddington accreting massive black holes in a  RM  project \citep[SEAMBHs,][]{Wang2013, Du2016a}. From the SEAMBH RM project, an extended empirical $R_{\rm BLR}-\lv$ relation  including \rfe (the flux ratio of the broad \hb to the optical \feii  within $4435-4685$ \AA) was  suggested \citep{Du2019, Yu2020a, Khadka2022}. This extended relation is possibly due to the dependence of the spectral energy distribution (SED) on the accretion rate \citep[e.g,][]{KE15, Yu2020a}. 


We have used a sample of 208 broad-line AGN to investigate the relations between the coronal properties
and the accretion process \citep{Wang2019}. 
That sample was selected from the Swift/BAT AGN Spectroscopic Survey (BASS) in the ultra-hard X-ray band (14–195 keV).
Most of the sources in that sample are nearby ($z<0.2$). 
There are 13 narrow-line Seyfert 1 galaxies (NLS1s) with $\rm FWHM_{\hb} < 2000 ~km/s$ \citep{Bian2003a}, which are usually thought to have small SMBHs with super-Eddington emissions \citep[see Figure 2 in ][]{Wang2019}.  
In that paper,  $\rm FWHM_{\hb}$ and the empirical \RL relation \citep{Be13} were used to calculate \mbh, and \leddR was used to describe the accretion strength.
In order to extend our sample to higher redshifts, a new sample of 113 quasars ($0.05 <z < 0.7$) drawing from the Sloan Digital Sky Survey (SDSS)  DR14 quasar catalogue with available hard X-ray $\Gamma$ (2-10 keV) are also used here \citep{Huang2020}.
From these two samples (Swift BASS and SDSS) with measured hard X-ray photon index, in this paper, we compile a large sample of 311 broad-line AGN ($z<0.7$)  to further investigate the coronal properties. 
Through spectral decomposition, we measure \shb and \rfe from their optical spectra.  
For determination of \mbh, we use \shb instead of FWHM, and use the extended empirical $R_{\rm BLR}-\lv$ relation. For the accretion strength, we use \mdot instead of \leddR \citep[see also][]{Liu2022}. This paper is organized as follows. Section 2 presents the sample. Section 3 describes the spectral decomposition and data analysis. Section 4 contains our results and discussion. Section 5 summarizes our conclusions. All of the cosmological calculations in this paper assume $\Omega_{\Lambda}=0.7$, $\Omega_{\rm M}=0.3$, and $H_{0}=70~ \kms {\rm Mpc}^{-1}$. 

\section{The AGN Sample with measured hard X-ray photon index}

\begin{figure}
\includegraphics[angle=0,width=3.4in]{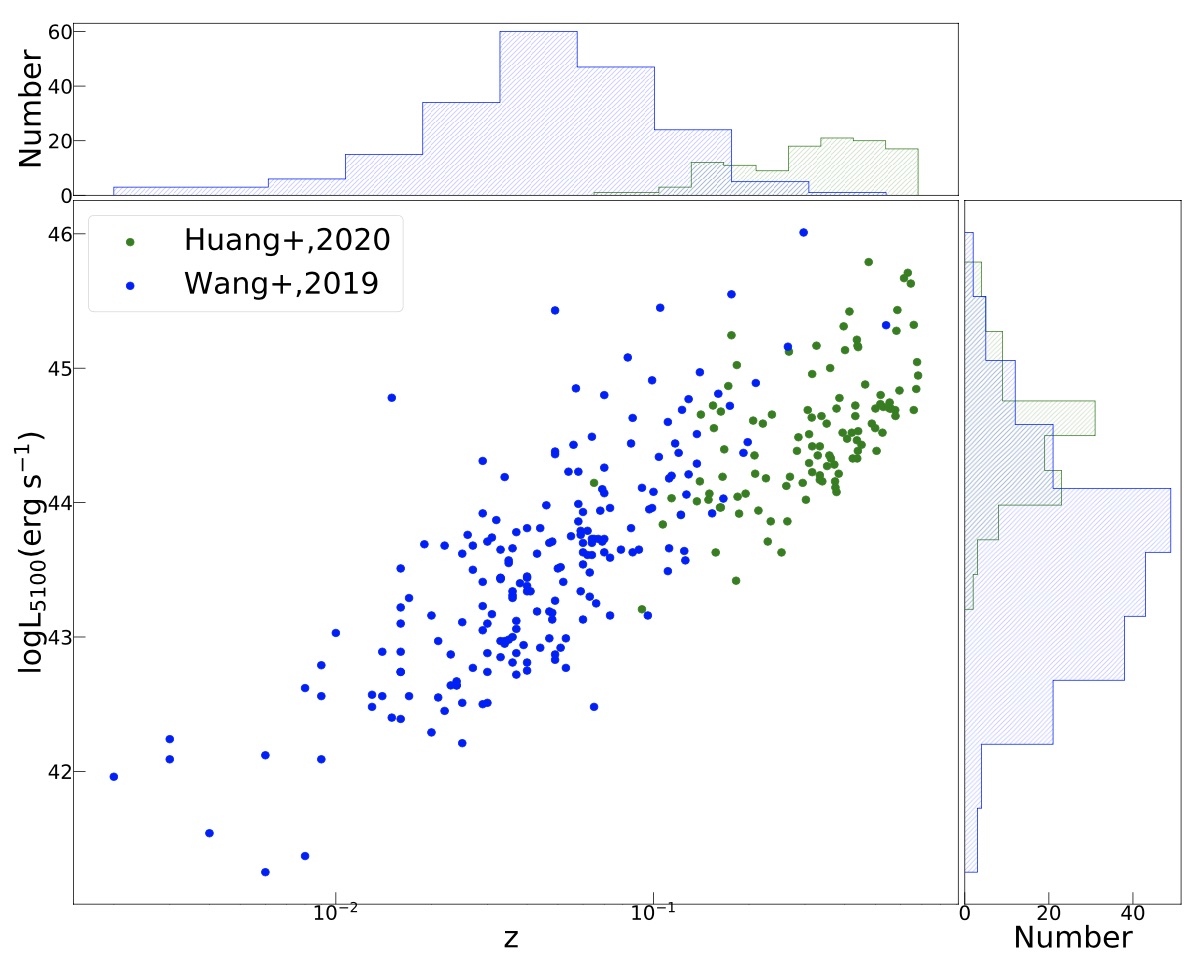}
\caption{The 5100\AA~  luminosity \lv versus redshift $z$. The blue dots denote AGN from Swift/BASS \citep{Wang2019}, the green dots denote AGN from SDSS \citep{Huang2020}. The top panel shows distributions of $z$ for the two subsamples, the right panel shows their distributions of \lv. }
\label{fig1}
\end{figure}

\begin{figure}
\includegraphics[angle=0,width=3.4in]{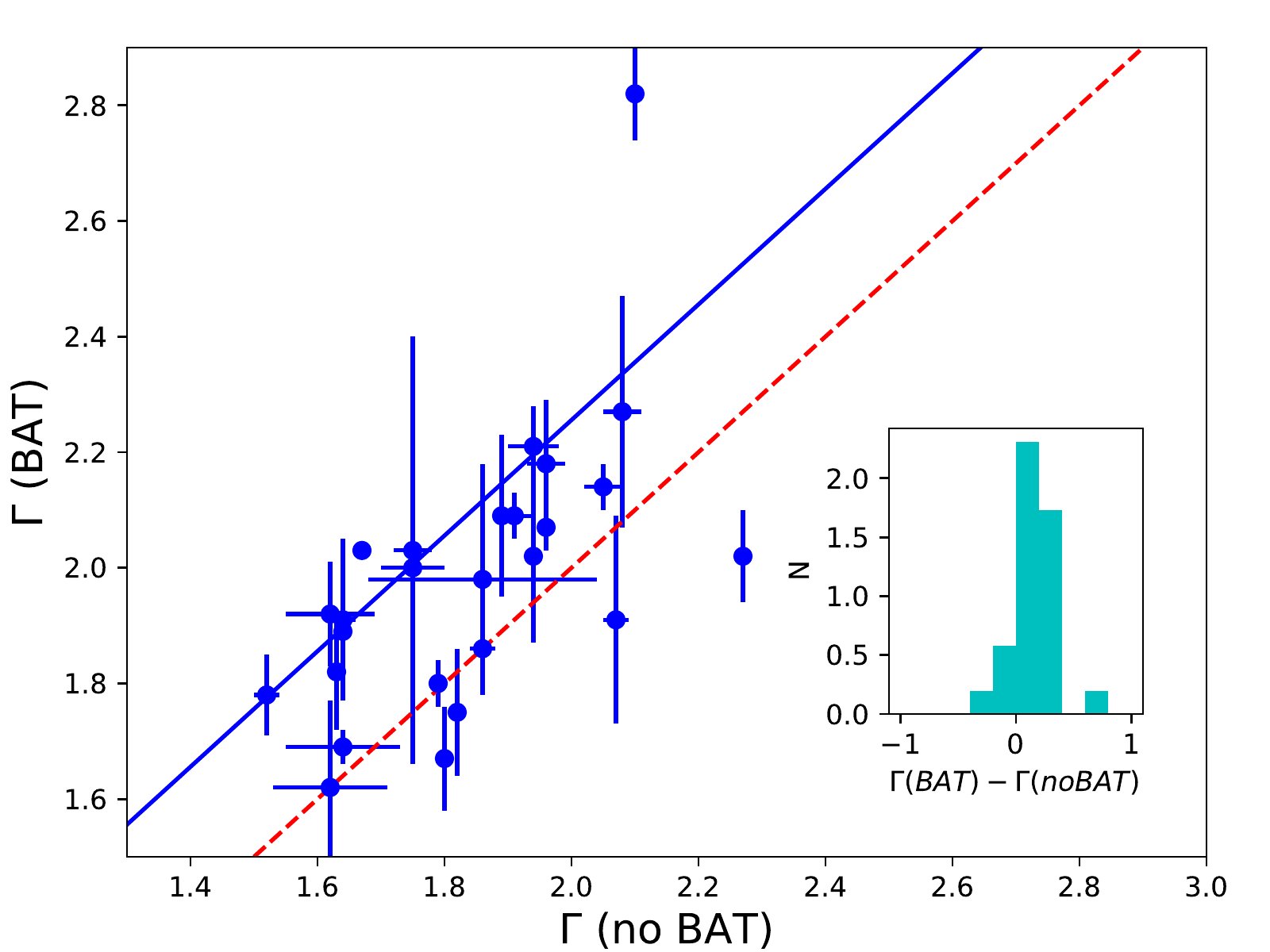}
\caption{$\Gamma$ of AGN with BAT data versus $\Gamma$ of AGN without BAT data for a sample by \cite{Liu2021}. The dashed red line is 1:1. The solid line is the best linear fit with a fixed slope of 1. The inset at the bottom right shows the distribution of their differences. }
\label{fig2}
\end{figure}


Our compiled sample of broad-line AGN ($z<0.7$) is selected from Swift BASS and SDSS, where 206 AGN \footnote{Two AGNs, i.e. SWIFT J1119.5+5132 and SWIFT J1313.6+3650A, have no $\Gamma$ shown in \citep{Ricci2017} because they were not observed in the 0.3–10 keV range.} selected from Swift BASS and 113 selected from SDSS \citep{Koss2017, Wang2019, Huang2020}. There is an AGN in common between the Swift BASS subsample and the SDSS subsample.  Our final compiled sample consists of 311 AGN excluding 7 AGN with $\rm S/N<3$. Figure \ref{fig1} shows the 5100 \AA\ monochromatic luminosity \lv versus the redshift $z$. $z$ is less than $0.7$. \lv covers from $\sim 10^{41}$ to $\sim 10^{46}$ \ergs. The distributions show that AGN from SDSS occupy ranges of higher $z$ and higher \lv with respect to AGN from Swift BASS. 

The Swift/BAT-detected AGN ($z<0.3$) used here were presented by \cite{Wang2019}, which were drawn from the Swift/BAT 70-month catalogue \citep{Baumgartner2013}. The Swift/BAT survey is an all-sky survey in the ultra-hard X-ray band (14-195 keV). The X-ray data and the analysis were presented by \cite{Ricci2017}, which we briefly summarize. For the Swift BASS subsample, the analysis covered the observed-frame energy range of $0.3-150$ keV, combining XMM-Newton, Swift/XRT, ASCA, Chandra, and Suzaku observations in the hard X-ray band ($\rm <10 keV$) with 70-month averaged Swift/BAT data in the ultra-hard X-ray band ($\rm >14keV$). The data were modelled with a set of models that rely on an absorbed power-law X-ray SED with a high-energy cutoff, and a reflection component, as well as additional components accounting for warm absorbers, soft excess, Fe $\rm K \alpha$ lines, and/or other spectral features. The typical uncertainty on the hard X-ray photon index $\Gamma$ is  less than 0.3 \citep{Ricci2017,Trakhtenbrot2017}.  

SDSS-selected AGN ($z<0.7$) were presented by \cite{Huang2020}, a sample which was drawn from the SDSS DR14 quasar catalogue and the Chandra/ACIS and XMM-Newton/EPIC archives. They selected AGN with more than 200 net source counts in the observed-frame $2/(1 + z)–7$keV band for Chandra/ACIS data and $2/(1 + z)–10$keV for XMM-Newton/EPIC data  (excluding the Fe K complex).  They used a redshifted single power-law model modified by Galactic absorption to fit the hard X-ray spectrum, and excluded objects with intrinsic absorption. The typical uncertainty on the $\Gamma$ values for the SDSS subsample is less than 0.2. For our entire sample, the Swift BASS subsample and the SDSS subsample,  the mean values of $\Gamma$ and standard deviations are $1.88\pm 0.23$, $1.83\pm 0.22$, and $1.98\pm 0.23$, respectively. The SDSS subsample has on average larger $\Gamma$ than for the Swift BASS subsample.

We compare $\Gamma$ of 26 Swift/BAT AGN measured with and without BAT from \cite{Liu2021} and \cite{Wang2019}.  \cite{Liu2021} used the same X-ray fitting model as for our SDSS subsample of \cite{Huang2020}. 
In Figure \ref{fig2}, we  show $\Gamma$ of AGN with BAT data versus $\Gamma$ of AGN fit without the BAT data for these same 26 AGN.  The dashed red line is 1:1. The solid line is the best linear fit with a fixed slope of 1 (considering errors in both coordinates) and the intercept is $0.15$. 
The mean value and the standard deviation of their difference is 0.15 and 0.19. 
$\Gamma$ of AGN including the BAT data is on average larger than that for the AGN excluding BAT, which arises from including a high-energy cutoff and the reflection
component in the model for the AGN including the BAT data. \cite{Ricci2017} also found that the large majority ($\sim 80\%$) of their sample show larger $\Gamma$ for AGN with BAT data  than $\Gamma$ of AGN with data only in the $0.3 -10$ keV band (ignoring the high-energy cutoff and the reflection component) (see their Figure 18). From simulations, they found that the value of $\Gamma$ for AGN with BAT data does not increase significantly for larger values of the reflection parameter, and increases significantly as the energy of the cutoff decreases (see their Figure 19). 
The mean value of the difference in $\Gamma$ for these 26 Swift/BAT AGN is less than the typical adopted error for the BASS subsample (0.3) or SDSS subsample (0.2). In order to be consistent with the $\Gamma$ of AGN with BAT data by \cite{Ricci2017}, a shift of 0.15 is used to correct $\Gamma$ for AGN without  BAT data given by \cite{Huang2020}. 

The hard X-ray photon indexes $\Gamma$ are adopted from Col. (7) in Table 1 in  \cite{Wang2019}  \citep[also see][]{Ricci2017}, and Col. (7) in Table 2 in \cite{Huang2020}.  \lxm for  Swift/BAT-detected AGN  is adopted from Col. (2) in Table 1 in \cite{Wang2019} \citep[also see][]{Koss2017}. $L_{2-10\rm keV}$ for SDSS-selected AGN is adopted from Col. (8) in Table 1 in \cite{Huang2020}. 

\cite{Koss2017} used a factor of $1/2.67$ to convert \lxm to $L_{2-10\rm keV}$, following \cite{Rigby2009}, which is based on scaling the templates to higher X-ray energies \citep{Marconi2004} .  
Considering the form of $f_{\nu} \propto \nu^{-\Gamma +1}$, the ratio of \lxm to $L_{2-10\rm keV}$ is 5.49, 1.64, 0.50, 0.17 for $\Gamma=1.5, 2, 2.5, 3$, respectively. The luminosity conversion factor is smaller for AGN with softer $\Gamma$, which suggests the importance of  the flux contribution from $2-10~ \rm keV$.   For our entire sample, the mean value of $\Gamma=1.88$ gives a ratio of 2.20.  Assuming the form of $f_{\nu} \propto \nu^{-\Gamma +1}$ and using the value of $\Gamma$ for each AGN,  we calculate $L_{2-195 \rm keV}$ from \lxm for the Swift BASS subsample, and from $L_{2-10\rm keV}$ for the SDSS subsample. 
Neglecting the complexity in the soft X-ray band \cite[e.g.,][]{N2013, Kubota2018}, we use $L_{2-195 \rm keV}$ instead of $L_{14-195 \rm keV}$ to calculate the fraction dissipated in the hot corona \citep[see also][]{Wang2019}.  
For the luminosity conversion factor, we don't consider the high-energy cutoff and the reflection component.
A smaller cutoff would lead to a smaller \lxm, and  $L_{2-10\rm keV}$ would be a larger fraction of $L_{2-195 \rm keV}$. A larger cutoff would have a smaller effect, especially for a cutoff larger than about 195 keV. 
For the Swift BASS subsample, ignoring the cutoff would overestimate \lxm, and underestimate the luminosity conversion factor of $L_{2-195 \rm keV}/L_{14-195\rm keV}$.  For the SDSS subsample, ignoring the cutoff would overestimate the luminosity conversion factor of $L_{2-195 \rm keV}/L_{2-10\rm keV}$.
The uncertainty of the luminosity conversion factor can be attributed to the uncertainty of $\Gamma$. 
Assuming a cut-off energy at 100 keV vs 300 keV, $\Gamma$ of AGN with data in the 0.3 - 10 keV band would be overestimated by about 0.3 versus 0.1. For a reflection parameter 0 or 1, the effect on $\Gamma$ is almost negligible \citep[see Fig. 19 in][]{Ricci2017}.
Considering the uncertainty in $\Gamma$ of 0.2 for the SDSS subsample, the uncertainty in the conversion factor of $\log L_{2-195 \rm keV}/L_{2-10\rm keV}$ is less than $\sim 50\%$ for $\Gamma$ between 1.2 and 2.5. That level of uncertainty would result in an error of 0.18 dex for $\log L_{2-195 \rm keV}$ for the SDSS subsample. 
For the Swift BASS subsample, the uncertainty of the conversion factor of $L_{2-195 \rm keV}/L_{14-195\rm keV}$ is also less than $\sim 50\%$, resulting in an error of 0.18 dex for $\log L_{2-195 \rm keV}$.


\section{Spectral decomposition and data Analysis}
\subsection{Spectral decomposition}

\begin{figure*}
\includegraphics[angle=0,width=8in]{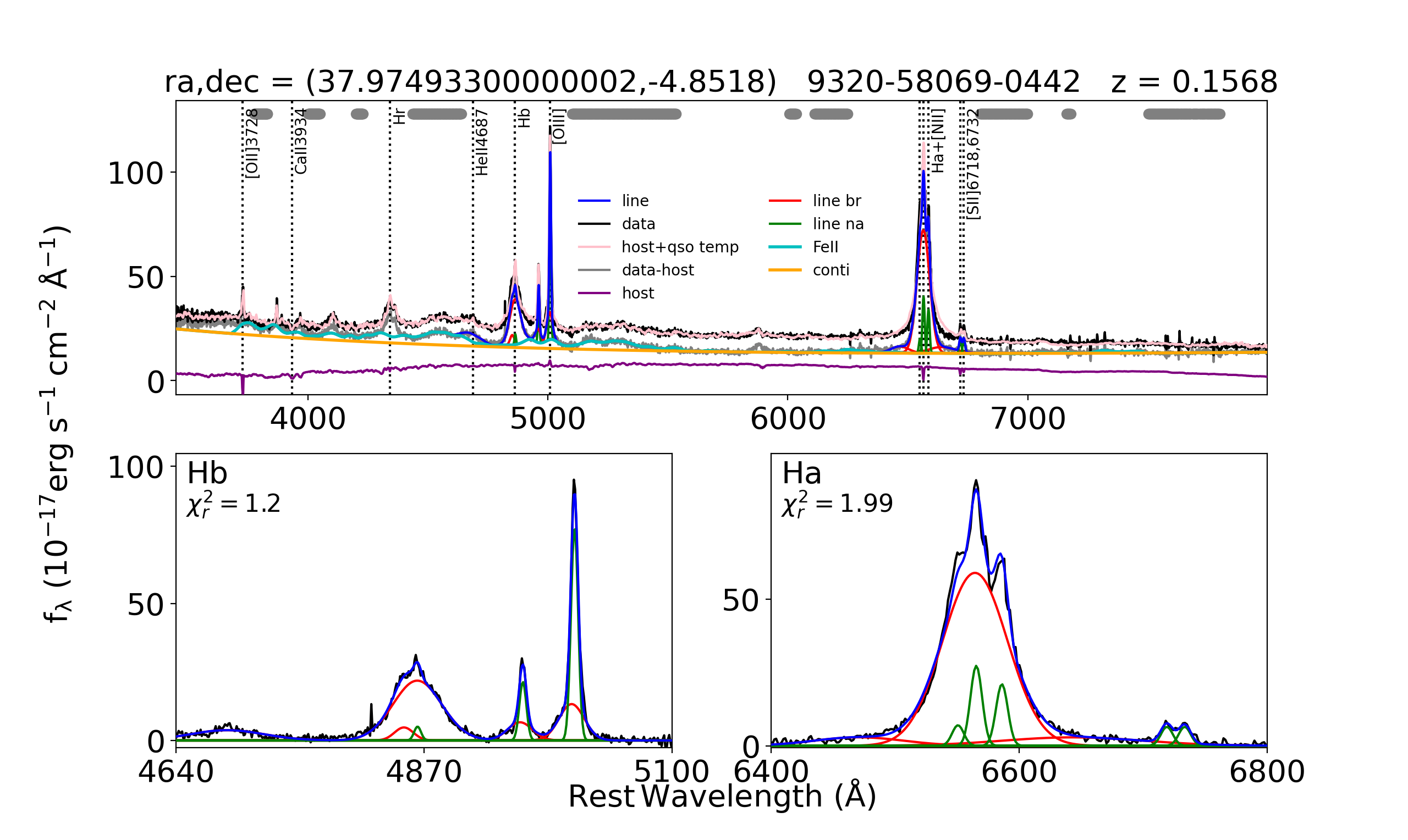}
\caption{An example of optical spectral decomposition. 
The pseudo-continuum including the host galaxy, \feii, a power-law, and a third-order polynomial is shown, as well as the narrow and broad profiles for \hb, \heii, \oiii, \ha,\nii, \sii. The bottom shows the line fitting around \hb (left), around \ha (right). 
 }
\label{fig3}
\end{figure*}

Using BLR gas as a probe to calculate single-epoch virial \mbh, we can obtain \mbh from  the broad \hb  component \shb as the BLR velocity tracer with smaller scatter than from $\rm FWHM_{\hb}$ \citep{Pe04, Yu2020b, Dalla2020}. At the same time, we need to include the relative \feii strength \rfe in the \RL relation  to estimate the BLR size. 
In order to measure \shb and \rfe, the spectra of the AGN in our sample need to be decomposed. The optical spectra for our sample are collected from the Swift BASS archives \citep{Koss2017} and the SDSS DR16 archives. There are 142 out of the 642 Swift/BASS AGN with spectra collected from SDSS \citep{Koss2017}. The spectra for the remaining 64 AGN were collected by other telescopes, which were given on the Swift BASS website \footnote{http://www.bass-survey.com/data.html}. We do the spectral decomposition using the Python code PyQSOFit \citep{Guo2018, Shen19}. The fitting is performed in the rest frame of the AGN using the redshift, after correcting for Galactic reddening using the dust map in \cite{Schlegel1998} and the extinction curve with $R_V=3.1$ from \cite{Cardelli1989}. The decomposition for the host and AGN contributions is applied, where the templates used are from \cite{Yip2004a, Yip2004b}. 
The optical \feii emission is fitted using empirical templates from \cite{BG92}. A pseudo-continuum including the optical \feii emission, a power law and a third-order polynomial is subtracted from the spectrum to form a line-only spectrum for which we measure emission-line properties. 

Considering the possible broad wing shown in \oiii \citep[e.g.,][]{Hu2008, WS2019}, two sets of two Gaussians are used to fit the \oiii~ $\lambda \lambda$4959, 5007 doublets, with constraints on their corresponding strengths, velocities and widths. The \oiii~ $\lambda \lambda$4959, 5007 lines are forced to have the same FWHM and no relative wavelength shift, and their intensity ratio fixed to the theoretical value of 3.0.  Three Gaussians are used to model the \hb profile, two Gaussians as the  broad component and a Gaussian as a narrow component.  The narrow \hb component is tied to the  core component of \oiii 5007 (same FWHM and no relative wavelength shift), which is coming from the narrow-line region (NLR). We also use two Gaussians to fit \heii~ $\lambda$4686, one is from the NLR and the other is from the BLR.  An example of fitting a spectrum is  shown in Figure \ref{fig3}. 
From the pseudo-continuum and the reconstructed profile from the two \hb broad components, we measure the optical integral \feii flux ($F_{\rm Fe{~\sc  II}}$) between 4434 \AA\ and  4684 \AA, \shb, FWHM, and integral flux ($F_{\rm \hb}$). Then we calculate the optical \feii relative strength, $\rfe= F_{\rm Fe {~\sc II}}/F_{\rm \hb}$. Based on the flux error, Monte Carlo simulation is performed for 20 trials to estimate the measurement uncertainties of the spectral properties. We correct the spectral resolution effect on the line width measurements following \cite{Pe04}, adopting the spectral resolutions presented by \cite{Koss2017}. 
The results are presented in Tables 1 and 2.


\begin{figure}
\centering
\includegraphics[angle=0,width=2.5in]{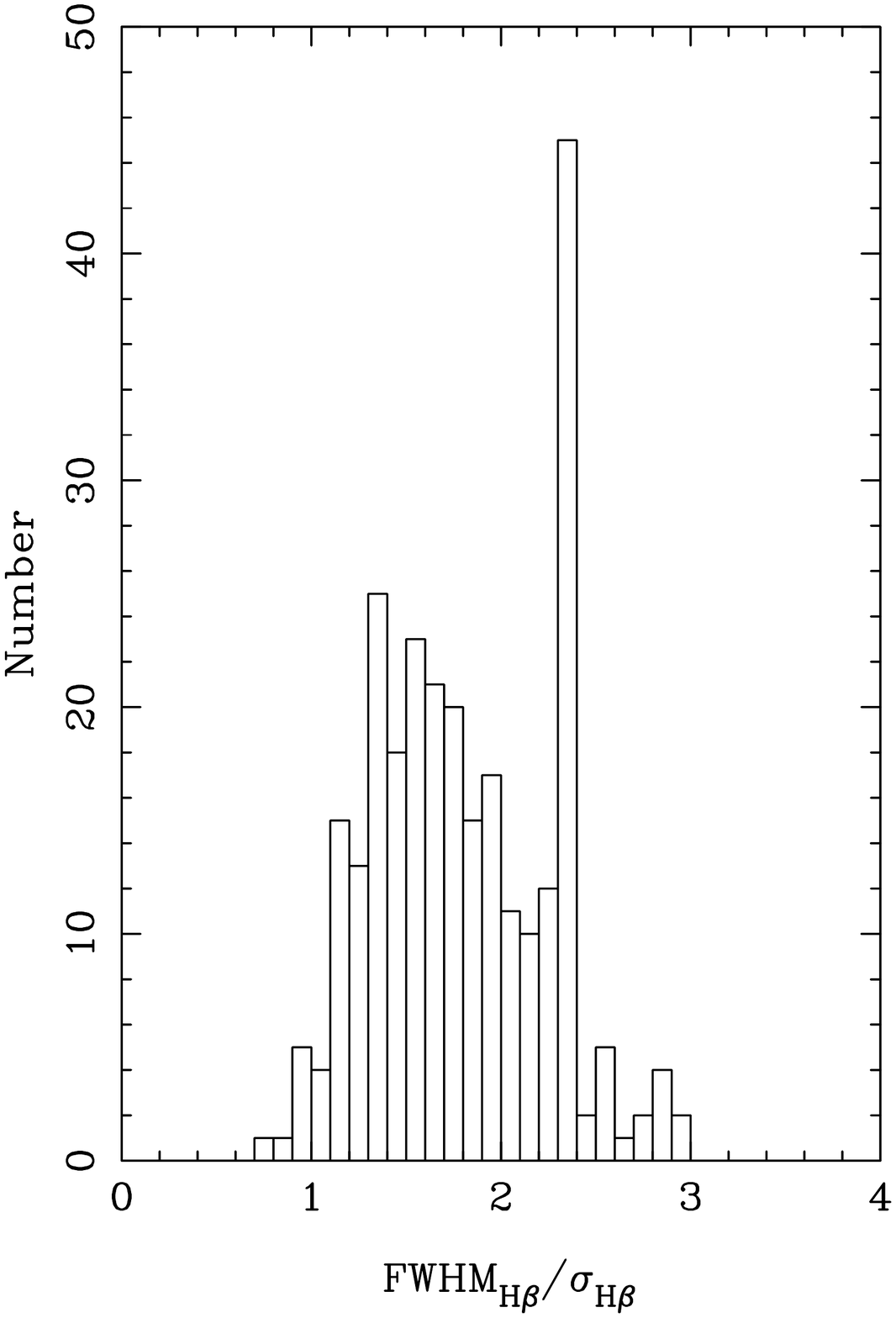}
\caption{The distribution of $\rm FWHM/\shb$. 
}
\label{fig4}
\end{figure}

For the SDSS subsample, \cite{Huang2020} fit the SDSS spectra  following the same procedure described in \cite{Hu2008, Hu2015}. They fit the complex pseudo-continuum together (power-law continuum, \feii template, Balmer continuum, the single simple stellar population model) and then fit the broad/narrow emission-line profiles from the pure emission-line spectrum \citep{Hu2008, Hu2015}. They didn't measure \shb. For the Swift subsample, \cite{Koss2017} used the continuum and line fitting procedure described in \cite{TN2012}, where a power-law continuum was fitted first, and then an \feii template was fitted and subtracted from the continuum-free spectrum, and finally a set of Gaussian profiles was used to fit the broad/narrow emission lines. They didn't measure \shb or \rfe. We compare our measured $\rm FWHM_{\hb}$ with that from theirs, as well as that from previous work \citep{Koss2017, Wang2019, Huang2020}.  
They are very well consistent; the difference in $\log (\rm FWHM_{\hb})$ is $0.009\pm 0.07$. In the following, we use our measured $\rm FWHM_{\hb}$.
In Figure \ref{fig4}, we show the distribution of $\rm FWHM_{\hb}/\shb$. There is a peak at 2.35, which is because the second broad \hb component is  weak in our fitting.  For the entire sample, the mean values and standard deviations of $\rm FWHM_{\hb}$, \shb, \rfe, $\rm FWHM_{\rm}/\shb$ are $5013.4\pm 2069.5$ \kms, $2799.7\pm 1231.9$ \kms, $0.53\pm 0.44$, $1.82 \pm 0.47$, respectively. 
The value of $\rm FWHM_{\rm}/\shb$ is 2.35 for a single Gaussian profile \citep{Collin2006, Du2016a}. For our sample, the  distribution of $\rm FWHM_{\rm}/\shb$ is $1.82\pm 0.47$, showing the deviation from the value of $2.35$ for a single Gaussian profile. For 76 RM AGN, the value is $1.94\pm 0.61$ \citep{Du2016a, Yu2020b}. For SDSS DR5 AGN,  the value is $1.78\pm 0.28$ \citep{Hu2008}.  
Our result is consistent with that for the sample of SDSS DR5 AGN, which is different from that for RM AGN measured in mean spectra. 

There is a relation between $\rm FWHM_{\hb}$ and \rfe. The Spearman correlation test gives a correlation coefficient $r = -0.466$ and a probability of the null hypothesis $p_{\rm null}=3.5\times 10^{-18}$. The diagram of $\rm FWHM_{\hb}$ versus \rfe shows a plane of Eigenvector 1 (EV 1), which is driven by the Eddington ratio or \mdot \citep{BG92, Sulentic2000, Boroson2002, Shen2014, Ge2016}.  It was suggested that there is substantial scatter between $\rm FWHM_{\hb}$ used to calculate \mbh and the actual virial velocity of BLRs \citep{Shen2014}. That is consistent with the bias in using $\rm FWHM_{\hb}$ as the tracer of  virial velocity in the \mbh calculation \citep[e.g.,][]{Dalla2020}. Recently, it was suggested there was	a selection bias for this relation \citep{Coffey2019}. For the project SDSS-IV/SPIDERS (optical spectroscopy of the counterparts to the X-ray selected sources detected in the ROSAT all-sky survey and the XMM-Newton slew survey in the footprint of the eBOSS), \cite{Coffey2019} did spectral fitting for simulated spectra and found the minimum detectable spectral S/N in the diagram of $\rm FWHM_{\hb}$ versus \rfe. They suggested that the detection of high \rfe and high FWHM needs high S/N, and that there exists a bias in this diagram because of the limitation of enough S/N. The limited S/N or selection effects  were suggested to be producing some of the strength of this correlation \citep{Coffey2019}.

\subsection{ \lv and  \lb}
The monochromatic luminosities at 5100 \AA\ in the rest frame, $L_{\rm 5100}$, are adopted from  Col. (3) in Table 1 in  \cite{Wang2019}, and Col. (7) in Table 1 in \cite{Huang2020}. The latter didn't make a correction for the host galaxy contribution. From the empirical formula by \cite{Ge2016}, we remove the host contribution in \lv given by \cite{Huang2020}. The host fraction $f^{\rm host}$ in the total
continuum luminosity at 5100 \AA\ ($L_{\rm 5100}^{total}$) is as follows \citep{Ge2016},
\begin{equation}
f^{\rm host} = (10.265\pm 2.92)-(0.225\pm0.07)\frac{\log L_{\rm 5100}^{total}}{\ergs}
\end{equation}
where we make the correction for AGN with $L_{\rm 5100}^{total} \leq 10^{45.6} \rm erg~ s^{-1}$. 
For the entire sample, the mean value of corrected $\log L_{\rm 5100}$ is 43.86 with a standard deviation of 0.85. 

Using the host-corrected luminosity $L_{\rm 5100}$, we calculate the bolometric luminosity \lb through the bolometric correction factor at 5100\AA\ \citep{Marconi2004, N2019}. The correction factor formula as a function of the bolometric luminosity is \citep{Marconi2004, Wang2019},
\begin{equation}
\begin{split}
\log(\lb/L_{5100})=0.837-0.067\ell+0.017\ell^{2} & \\
-0.0023\ell^{3}
\end{split}
\end{equation}
where $\ell$ = $\log (\lb/L_{\odot}) -12$.





\subsection{\mbh and  \mdot}

\begin{figure*}
\includegraphics[angle=0,width=2.5in]{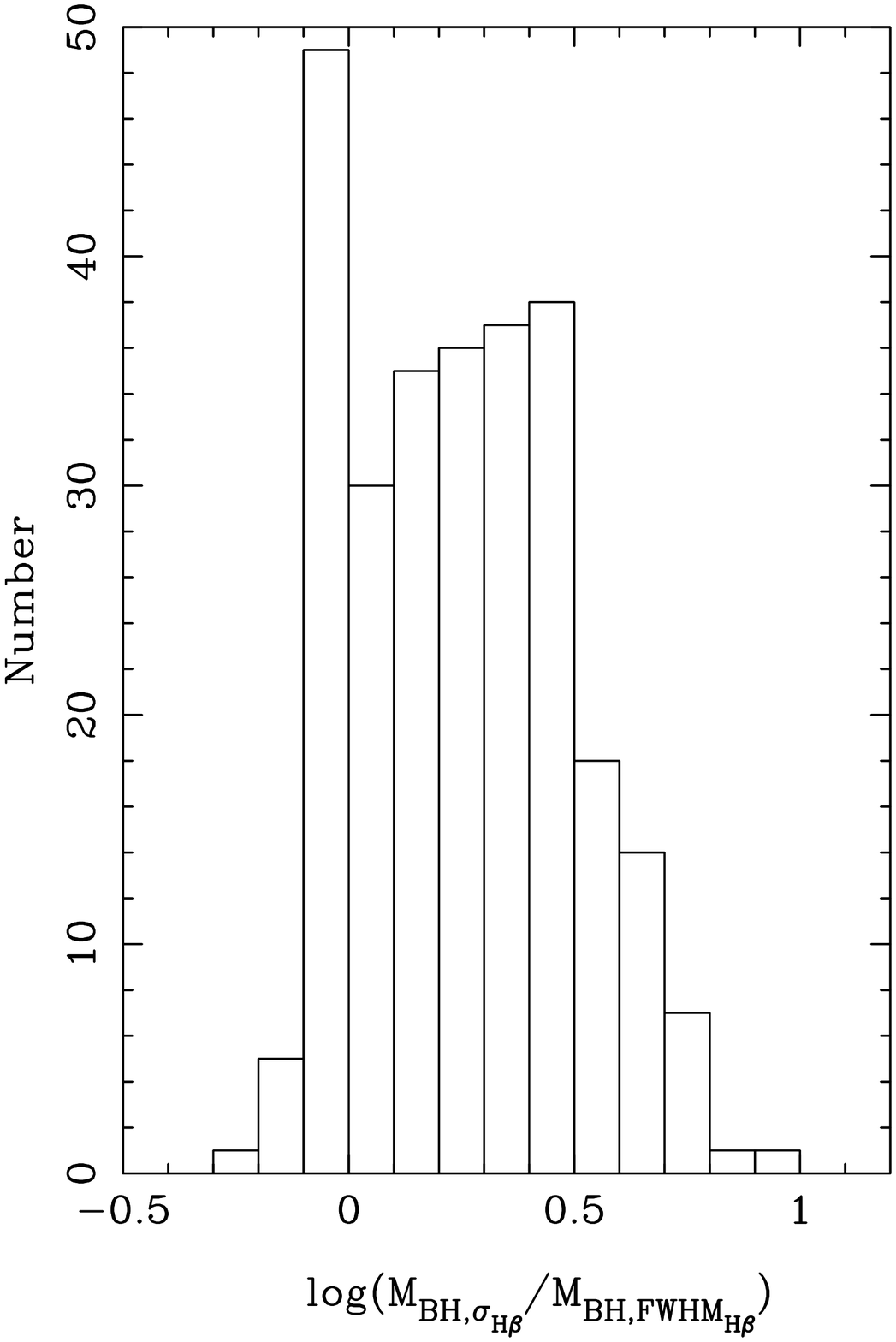}
\includegraphics[angle=0,width=2.5in]{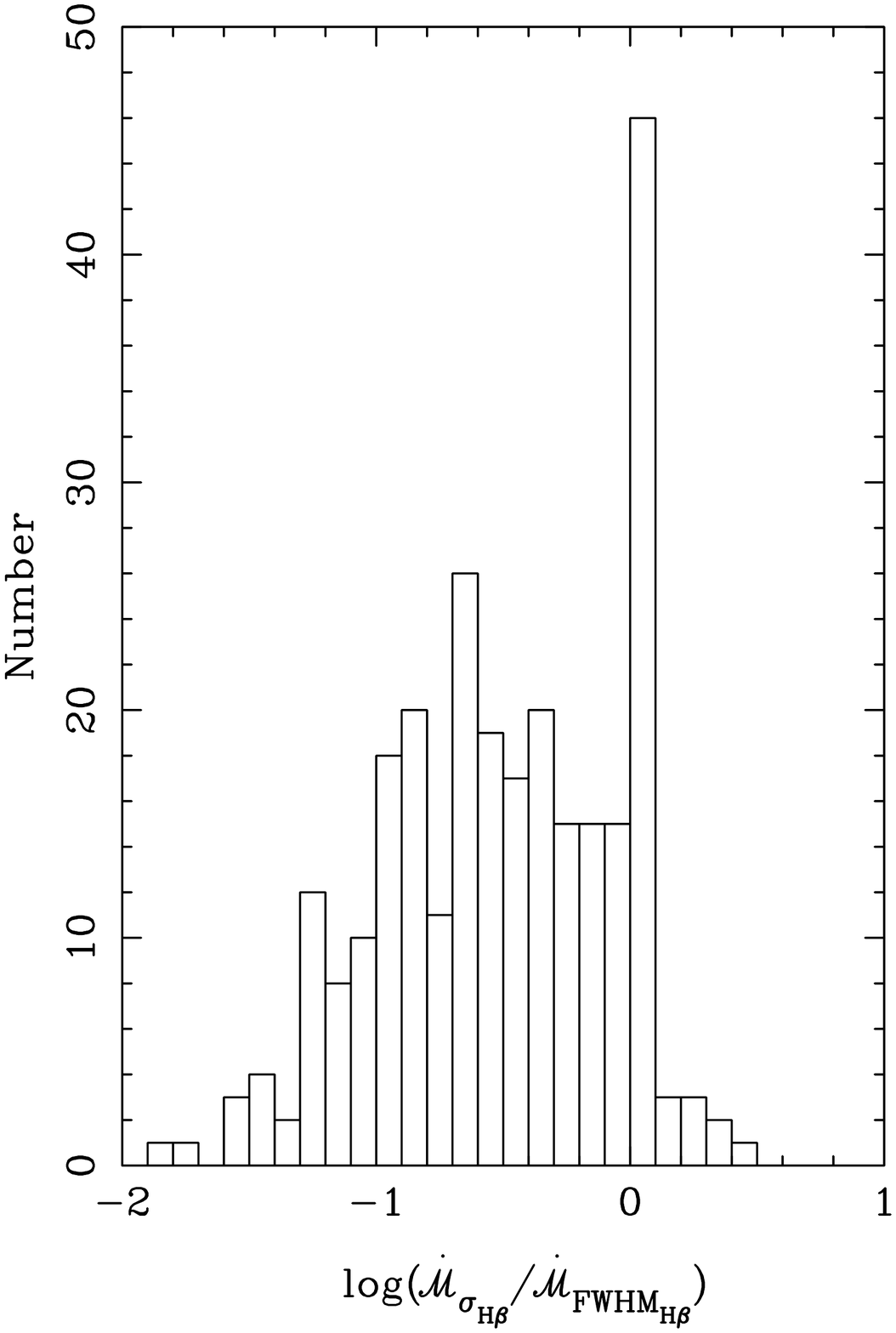}
\includegraphics[angle=0,width=2.5in]{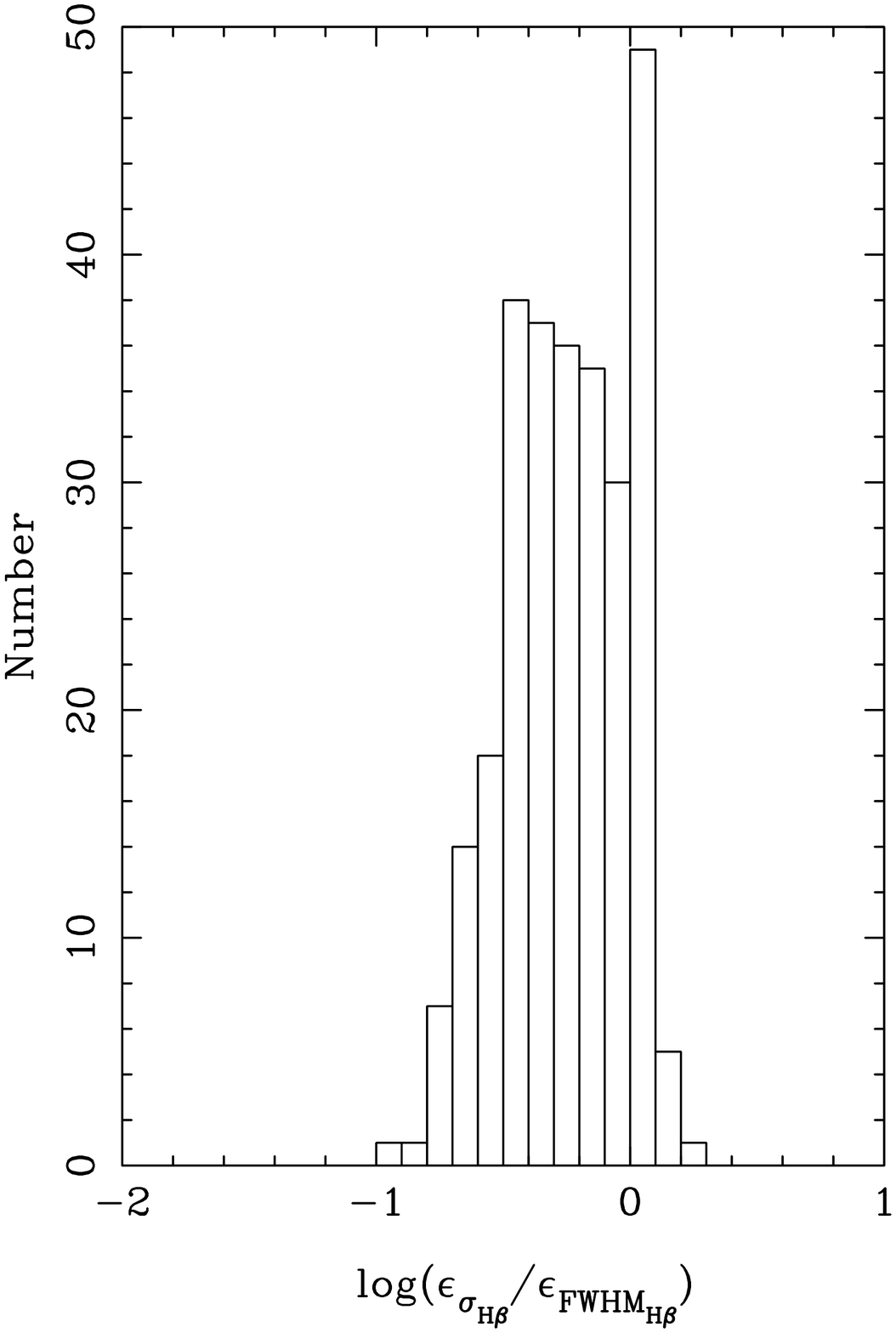}
\caption{Distributions of the differences of \mbh (left), \mdot (middle), and \leddR (right) based on \shb and $\rm FWHM_{\hb}$.  }
\label{fig5}
\end{figure*}

\begin{figure}
\includegraphics[angle=0,width=2.6in,angle=-90]{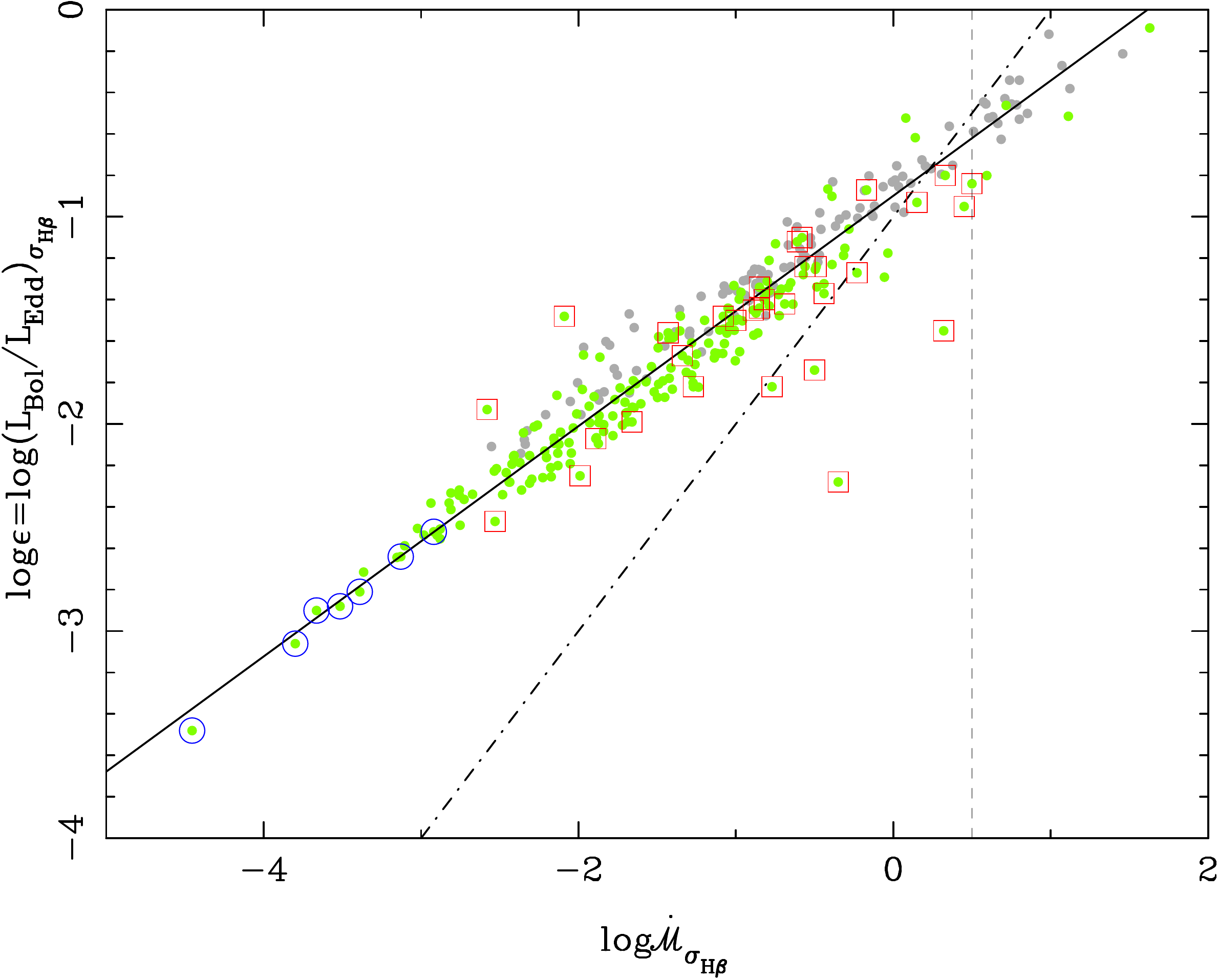}
\caption{\leddR versus \mdot. Red squares represent 32 reverberation-mapping sources in our sample. Blue circles denote 8 AGN with stellar velocity dispersion.  The solid line is our best fit showing a non-linear relation, $\leddR \propto \mdot^{0.56}$. The dash dot line is the relation assuming a constant coefficient of $\eta=0.1$. The dashed line is \mdot=3, which is the criterion distinguishing super-Eddington SMBH and sub-Eddington SMBH.}
\label{fig6}
\end{figure}

The SMBH masses of 311 broad-line AGN in our sample are estimated as follows:
 (1) for 32 RM AGN, $\shb$-based masses are preferentially adopted from Col. (9) in Table 1 in  \cite{Yu2020b}; (2) for the other 8 AGN with stellar velocity dispersions \citep{Koss2017}, their SMBH masses are calculated from the $\mbh-\sigma_*$ relation \citep{KH13, Koss2017}; (3) for the rest of the  AGN, we calculate their single-epoch SMBH masses from \shb and the following extended \RL relation. 

An extended empirical \RL relation including \rfe was found for the RM AGN sample \citep{Du2019, Yu2020a}:
\begin{equation}
\label{eq3}
\begin{split}
\log \frac{\rhb}{\rm light-days}=0.48\log l_{44}-0.38\rfe+1.67,
\end{split}
\end{equation}
With the extended \RL relation, when \shb is used as the tracer of the virial velocity, and $f$ is adopted as 5.5 \citep[e.g.,][]{Yu2020a, Yu2020b},
the single-epoch $\shb$-based \mbh can be expressed as
\begin{equation}
\label{eq4}
\begin{split}
\log \frac {M_{\rm BH,\shb}}{M_{\odot}} =7.7+2\log \frac{\rm \sigma_{H\rm \beta}}{\rm 1000~ \kms} & \\
+0.48\log l_{44} -0.38\rfe,
\end{split}
\end{equation}
Substituting Equation \ref{eq4} into Equation \ref{eq2}, the $\shb$-based \mdot formula is:
\begin{equation}
\label{eq5}
\begin{split}
\log (\mdot_{\shb}) = 0.54 \log l_{44}-4\log \frac{\shb}{\rm 1000~ \kms} & \\
+0.76\rfe+0.09. \\
\end{split}
\end{equation}

In order to compare \mbh, \mdot and \leddR from different velocity tracers,  we also give the formula from  $\rm FWHM_{\hb}$. Using $\rm FWHM_{\hb}$ as the tracer of the virial velocity, and adopting $f$ as 1 \citep[e.g.,][]{Du2019}, 
the single-epoch FWHM-based \mbh formula is:
\begin{equation}
\label{eq6}
\begin{split}
\log \frac {M_{\rm BH,FWHM_{\hb}}}{M_{\odot}} =6.96+2\log \frac{\rm FWHM_{\hb}}{\rm 1000~ \kms} & \\ 
+0.48\log l_{44} -0.38\rfe,
\end{split}
\end{equation}
and a similar FWHM-based \mdot formula is:
\begin{equation}
\label{eq7}
\begin{split}
\log (\mdot_{\rm FWHM_{\hb}}) = 0.54 \log l_{44}-4\log \frac{\rm FWHM_{H\rm \beta}}{\rm 1000~ \kms} & \\
+0.76\rfe+1.57. \\
\end{split}
\end{equation}

For our sample,  the mean value of  $\log (M_{\rm BH,\shb}/M_{\odot})$  is 8.30 with a standard deviation of 0.63 dex.
The mean value of  $\log (M_{\rm BH,FWHM_{\hb}}/M_{\odot})$  is 8.04 with a standard deviation of 0.68 dex. The $\shb$-based \mbh is on average larger than the $\rm FWHM_{\hb}$-based \mbh by 0.26 dex. It is smaller than the $0.5$ dex found for a NLS1s sample in \cite{Bian2008}. The mean value of  $\log \mdot_{\shb}$  is -1.15 with a standard deviation of 0.99 dex. The mean value of  $\log  \mdot_{\rm FWHM_{\hb}}$  is -0.64 with a standard deviation of 1.18 dex. The $\shb$-based \mdot is on average smaller than the $\rm FWHM_{\hb}$-based \mdot by 0.51 dex, which is consistent with the formula of Equation \ref{eq2}, $\mdot \propto \mbh^{-2}$. For \leddR, the $\shb$-based \leddR is on average smaller than the $\rm FWHM_{\hb}$-based \leddR by 0.26 dex. 

The distributions of the differences of \mbh, \mdot, and \leddR from \shb and $\rm FWHM_{\hb}$ are shown in Figure \ref{fig5}. 
A criterion of $\mdot > 3$ is adopted to select super-Eddington accreting AGN \citep[e.g.,][]{Du2016a}.  When using \shb to calculate the \mbh, \mdot would be smaller with respect to that using $\rm FWHM_{\hb}$, which would move the locations of some super-Eddington accreting AGN to the range below $\mdot<3$. With $\shb$-based \mbh, the number of super-Eddington sources becomes smaller.  Using $\shb$-based $\mdot> 3$, 22 (7\%) of the 311 AGN are super-Eddington accreting AGN (Figure \ref{fig6}). 

Hereinafter, in our relations analysis, we use \mbh, \mdot and \leddR derived from \shb.
Considering the typical errors of 0.1 dex for $\log \lv$ and the conversion factor of  $\log(\lb/\lv)$, the error propagation formula leads to a $\log \lb$ error of 0.14 dex. (If we think the conversion factor $\log(\lb/\lv)$ is larger, the  error of $\log \lb$ would be larger.) Considering the calibration scatter \citep[e.g.,][]{Yu2020b}, the typical error of \mbh is adopted as 0.3 dex. Adopting the uncertainties of $\log \lb$ and $\log \mbh$ as 0.14 dex and $0.30$ dex, the error propagation gives the typical errors of \mdot and \leddR as $0.62$ dex and $0.33$ dex, respectively. The error of $\fx=L_{\rm 2-195keV}/\lb$ is from $L_{\rm 2-195keV}$ and \lb. If the errors of $L_{\rm 2-195keV}$ and $\lb$ are 0.2 dex, the error of $\fx= L_{\rm 2-195keV}/\lb$ is 0.28 dex. Therefore, in our following analysis, we adopt 0.3 dex as the typical error in $\fx$.
For $\Gamma$, we use the error from the X-ray fitting presented by \cite{Ricci2017} and \cite{Huang2020}.  

We use $\shb$-based \mdot instead of \leddR to trace the accretion strength to investigate the disk-corona connection. In Figure \ref{fig6}, we show \leddR versus \mdot. The Spearman correlation test gives the correlation coefficient $r=0.95$ and the probability of the null hypothesis $p_{\rm null} = 1.1\times10^{-312}$. This large $r$ is due to the self-correlation with  \lv and \mbh. We use the bivariate correlated errors and scatter method \cite[BCES;][]{AB96} to perform the linear regression.  
Considering errors in both coordinates, the BCES($\rm Y|X$) best-fitting relation for our total sample shows $\leddR \propto \mdot^{0.56\pm 0.01}$, which is consistent with the slopes of $0.53$ found by \cite{DL2011} and $0.52$ by \cite{Huang2020}. The slight difference is due to the methods to obtain the bolometric luminosity and \mbh. 
Using Equations \ref{eq1} and \ref{eq2}, $\mdot \propto \lv^{1.5}\mbh^{-2}$, $\leddR \propto \lb \mbh^{-1} \propto \lv \mbh^{-1}$,  so $\leddR \propto \mdot^{0.5} \lv^{0.25}$. The dependence on \lv is weaker than on \mdot. The slope of $0.5$ is well consistent with our fitted slope of $0.56\pm 0.01$.  The non-linear relation between \leddR and \mdot would have an impact on the relations using \mdot instead of \leddR. 
Considering $\lb=\eta \dot{M}c^2$, where $\eta$ is the accretion efficiency, $\leddR=\frac{\eta \dot{M}c^2}{\dot{M}_{\rm Edd}c^2}=\eta \mdot$.  The difference between \leddR and \mdot comes from $\eta$. Adopting the non-linear relation $\leddR \propto \mdot ^{0.5} \lv^{0.25}$ means that $\eta \propto \mdot^{-0.5} \lv^{0.25} \propto (\leddR)^{-1}  \lv^{0.5}$. The non-linear relation between \leddR and \mdot implies that the conversion efficiency $\eta$ is anti-correlated with \mdot or \leddR ( including a relatively weaker dependence on \lv) , i.e., smaller efficiency for AGN with higher dimensionless accretion rate.

\section{Results and Discussion}
\subsection{The relation between $f$ and $\rm FWHM_{\hb}$}

\begin{figure}
\includegraphics[angle=0,width=2.6in,angle=-90]{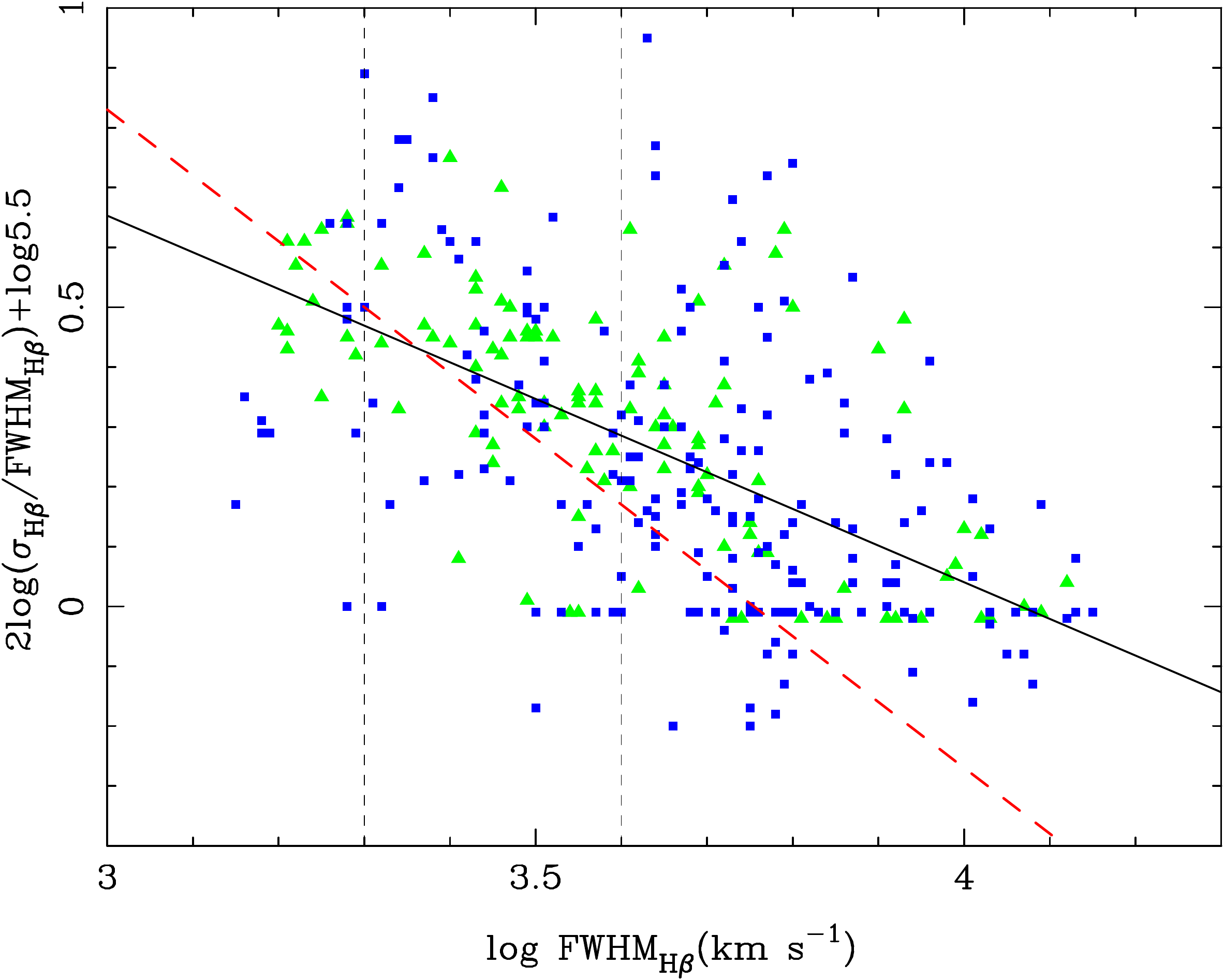}
\caption{The relation between $2\log(\shb/\rm FWHM_{\hb})+log 5.5$ and $\rm FWHM_{\hb}$, i.e., $f$ versus $\rm FWHM_{\hb}$.  The black solid line is our best fit with a slope of $-0.61$. The red dashed line is the result of \cite{Yu2020b} with a slope of $-1.1$. The two vertical dashed lines denote 2000 \kms and 4000 \kms. The blue squares denote AGN from \cite{Wang2019} and the green triangles denote AGN from  \cite{Huang2020}. }
\label{fig7}
\end{figure}

From Equation \ref{eq1}, we can calculate the single-epoch virial mass from different velocity tracers with the corresponding virial factor $f$, i.e., $\mbh=f_{\sigma}\times \shb^2 \rhb/G=f\times \rm FWHM_{\hb}^2 \rhb/G$. The difference between $\shb$-based \mbh and $\rm FWHM_{\hb}$-based \mbh is from adopting $f_{\sigma}=5.5$ for the former and $f=1$ for the latter. Considering  a variable $f$ to calculate the $\rm FWHM_{\hb}$-based \mbh means $f=f_{\sigma} (\shb / \rm FWHM_{\hb})^2$. 
In Figure \ref{fig7}, we show $\log f$ versus $\rm FWHM_{\hb}$ with $ f_{\sigma}=5.5$ as suggested by \cite{Yu2020b}.  The Spearman correlation test gives $r=-0.58, p_{\rm null}=2.4\times 10^{-28}$. The BCES($\rm Y|X$) best-fitting relation for our total sample shows $\log f =-(0.61\pm 0.05) \log \rm FWHM_{\hb}+(2.48\pm 0.18)$ (the black solid line in Figure \ref{fig7}). For RM AGN \citep{Yu2020b}, it was found that the relation is   $\log f =-(1.10\pm 0.4) \log \rm FWHM_{\hb}+(4.13\pm 0.11)$ (the red dashed line in Figure \ref{fig7}). Although our fitted slope is flatter than that of \cite{Yu2020b}, the trend is similar, i.e., larger $f$ for AGN with smaller $\rm FWHM_{\hb}$.  With our large number of AGN, the error of our slope is smaller, and our slope ($-0.61\pm 0.05$) is consistent with the slope ($-1.10\pm0.4$) for RM AGN considering their errors. 

$\dhb=\rm FWHM_{\hb}/\shb$ is the shape of the \hb broad-line profile. Therefore, the $\rm FWHM_{\hb}$-based $f$ follows $\log f=\log f_{\sigma}-2\times \log \dhb$. 
The $\rm FWHM_{\hb}$-based $f$ is proportional to $\dhb^{-2}$. 
The value of $\dhb$ is 2.35, 2.83, and 0 for a Gaussian, an edge-on rotating ring, and a Lorentzian profile, respectively \citep{Collin2006, Du2016a}. In Figure \ref{fig7}, the vertical axis is  $\log f_{\sigma}-2\times \log \dhb$, and the value of the y coordinate is  -0.0017 for $\dhb=2.35$. That result shows that the \hb profiles of most of AGN in our sample deviate from a Gaussian profile, implying the presence of a broad wing. For the extreme case of a Lorentzian profile, the broad wing leads to  $\dhb \rightarrow \infty$. The negative correlation in Figure \ref{fig7} shows that there is a positive correlation between $\dhb$ and $\rm FWHM_{\hb}$, i.e., a smaller $\dhb$ for AGN with smaller $\rm FWHM_{\hb}$, although with a large scatter. All the NLS1s but one in our sample show \dhb smaller than 2.35. In Figure \ref{fig7}, we show two vertical dashed lines denoting 2000 and 4000 \kms, which are used to distinguish NLS1s and population A AGN from other types of AGN \citep{Sulentic2000}. It was found that there is no significant correlation between \dhb and \rfe for a sample of 63 RM AGN \citep{Du2016a}. Using a compiled sample of 120 RM AGN, \cite{Yu2020a} found a relation between $\dhb$ and \mdot \citep[see also][]{Collin2006, Du2016a}, which suggested that the \hb broad-line profile is governed by the ionizing flux and hydrogen density related to the accretion rate. 
Therefore,  $\rm FWHM_{\hb}$-based $f$ is not a constant, but depends on \dhb or has a relation with $\rm FWHM_{\hb}$. In the following analysis, \shb instead $\rm FWHM_{\hb}$ is used to calculate virial \mbh and \mdot.

\subsection{The relation between $\fx$ and $\mdot$}

\begin{figure*}
\includegraphics[angle=0,width=2.6in,angle=-90]{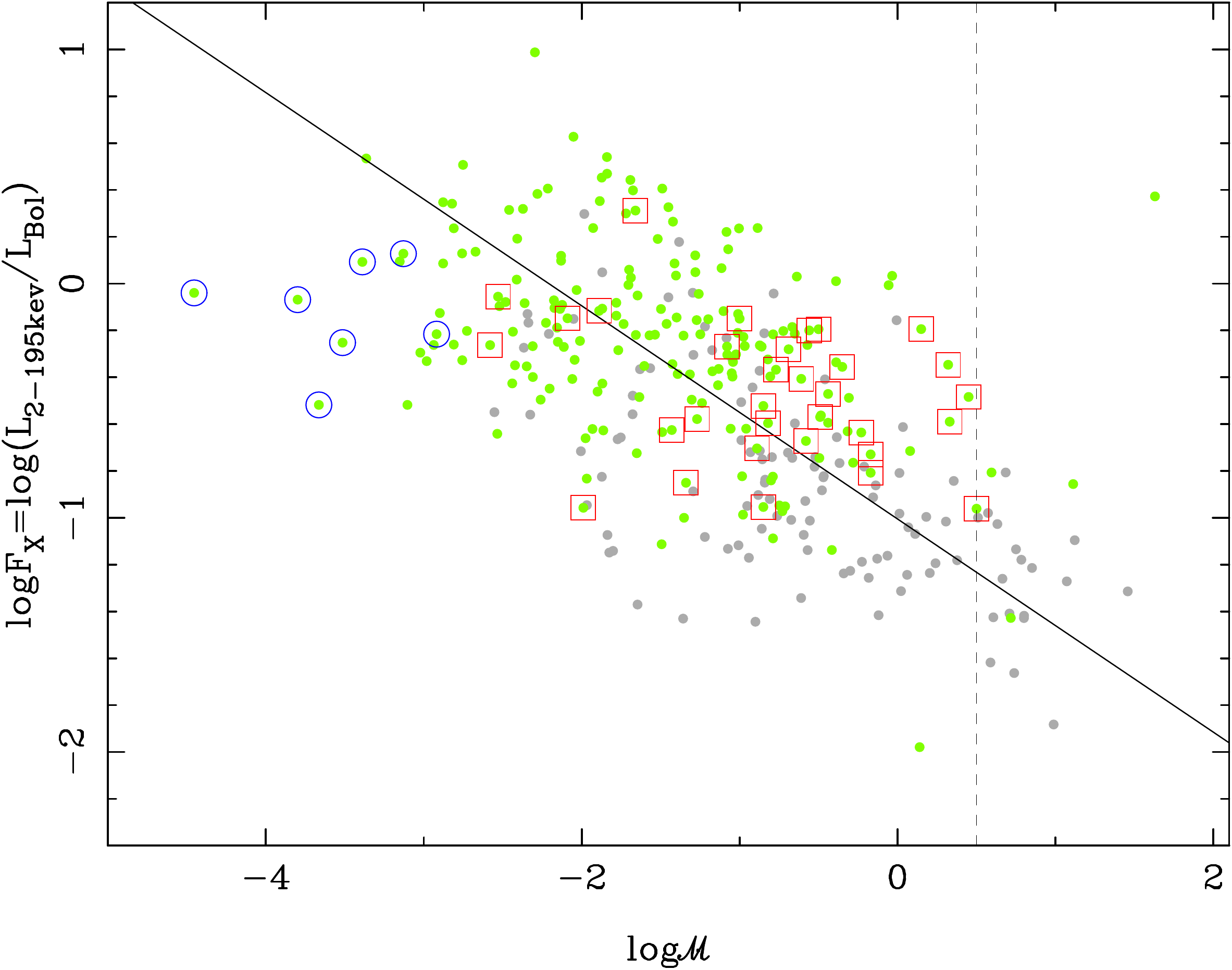}
\includegraphics[angle=0,width=2.6in,angle=-90]{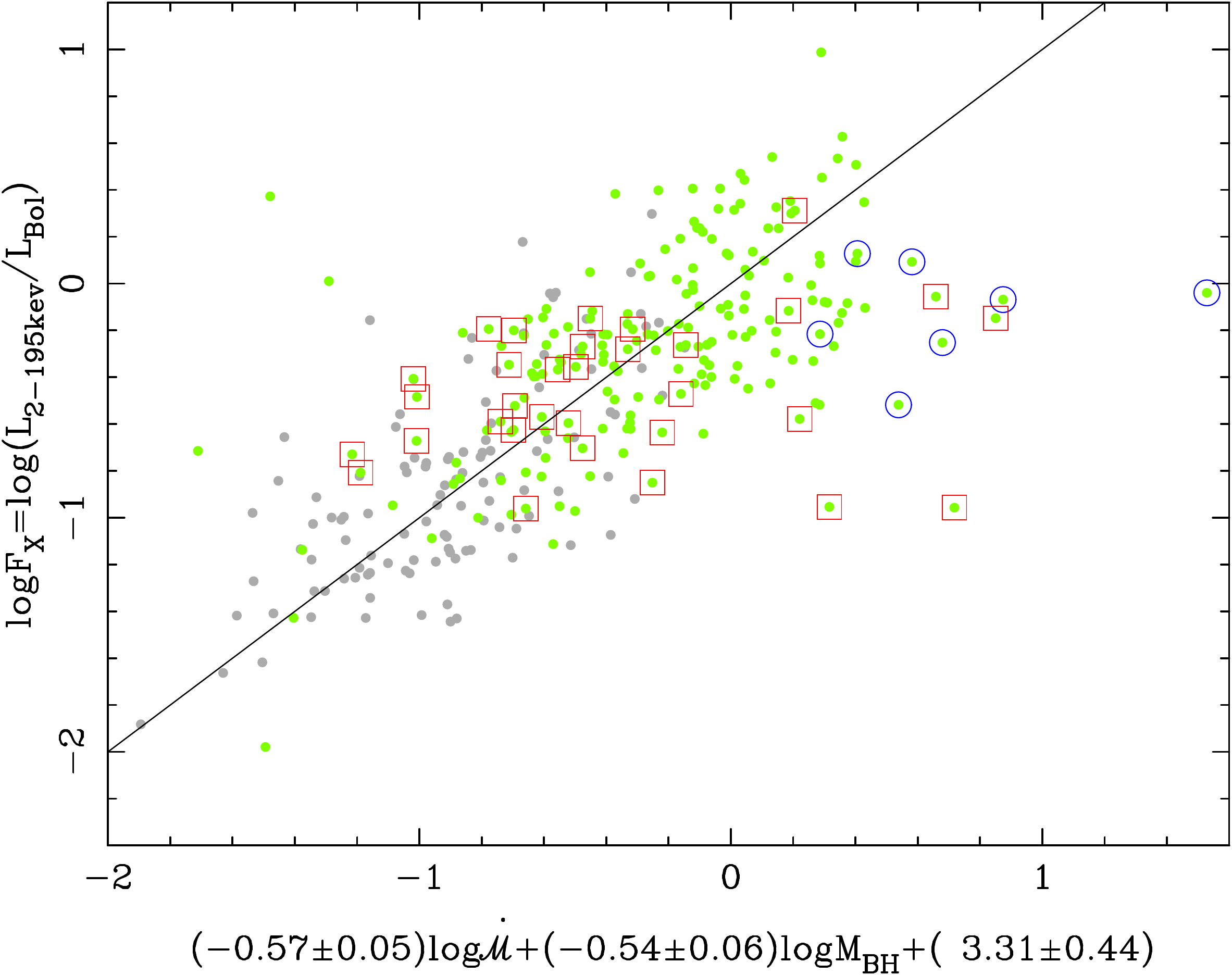}
\caption{Left: \fx versus \mdot. The green points denote  Swift/BAT-detected AGN, the grey points denote SDSS AGN.  Red squares represent 32 reverberation-mapping sources in our sample. Blue circles denote 8 AGN with stellar velocity dispersions. The solid black line is our best fit. The vertical dashed line is \mdot=3.  Right: Dependence of \fx on \mdot and \mbh. The solid line is 1:1.}
\label{fig8}
\end{figure*}

Using $2-195$ keV, we calculate the fraction of energy dissipated in the hot corona, i.e., $\fx= L_{\rm 2-195~keV}/\lb$. In the left panel in Figure \ref{fig8}, we show \fx versus $\shb$-based \mdot. The Spearman correlation test gives $r=-0.60$, $p_{\rm null} = 3.0\times10^{-32}$. The correlation is significant. 
Using the uncertainties of \fx, \mdot of $0.3$, $0.62$ dex, respectively, the BCES($\rm Y|X$)  best-fitting relation for our total sample gives $\log \fx=-(0.45\pm 0.05)\log \mdot -(1.01\pm 0.06)$. 
Comparing with the result by \cite{Wang2019}, our slope of $-0.45\pm 0.05$ is flatter than theirs, i.e., $-0.60\pm 0.1$.  The negative correlation shows that, with increasing dimensionless  accretion rate, the energy dissipated in the corona becomes smaller. 

It was suggested that  \fx depends  on \mbh \citep{Yang2007, Wang2019}.
The multivariate regression analysis technique is used to investigate the relation of \fx with \mdot and \mbh  \citep{Wang2019, Yu2020a, Yu2020b}.  
For the best linear fit in the form $ y = \alpha_{1}x_{1} + \alpha_{2}x_{2} + \beta_1$, the estimator of $\chi^2$ is used to find the best values for these fitting coefficients: $\chi^2 = \Sigma_{i} \frac{(y_{i} - \alpha_{1}x_{1i}- \alpha_{2}x_{2i}-\beta_1)^2}{\sigma_{\rm int}^{2}+\sigma_{y_{i}}^{2}+(\alpha_{1}\sigma_{x_{1i}})^2+(\alpha_{2}\sigma_{x_{2i}})^2}$. $y_{i}$ is the dependent variable. $x_{1i}$ and $x_{2i}$ are the independent variables. $\sigma_{y_{i}}$, $\sigma_{x_{1i}}$ and $\sigma_{x_{2i}}$ are the uncertainties of $y_{i}$, $x_{1i}$ and $x_{2i}$, respectively. $\sigma_{\rm int}$ is the intrinsic scatter. 
Considering the uncertainties of \fx, \mdot, \mbh of $0.3$, $0.62$, $0.3$ dex, respectively (see section 3.2), the fitting result is $\log \fx=-(0.57\pm 0.05)\log \mdot-(0.54\pm 0.06)\log \mbh+(3.31\pm 0.44)$ (see the right panel in Figure \ref{fig8}). 
The bootstrap method is used to estimate the error bars of the fitting coefficients. We re-sample the data 100 times and repeat the multiple linear regressions. Therefore, we can derive 100 values for each coefficient. For each coefficient, we sort these values, adopt the 16th and 84th values as end points of the 1 $\sigma$ confidence interval, and calculate its error \citep{Yu2020a}. 
Including \mbh in the relation between \fx and \mdot, the correlation becomes stronger, from $r=0.60, p_{\rm null}=3.0\times 10^{-32}$ to  $r=0.73, p_{\rm null}=1.0\times 10^{-308}$. 
The scatter reduces from 0.440 to 0.398. The $\chi^2$ decreases from 663.98 to 546.23. 
We compare the $\chi^2_N$, $\chi^2_M$, respectively for the single variable regression and multivariate regression, and use the $F$-test \citep[chap. 12.1]{Lupton1993} to calculate how significantly the multivariate regression model improves the fit. $f=(n-s)\frac{\chi^2_{N}-\chi^2_M}{\chi^2_N}$, n=311, s=2.  We found that the probability P that the multivariate regression model cannot improve the fit is $2.55\times 10^ {-12}$. Therefore, the significance of improving the fitting by the multivariate regression is greater than 3$\sigma$ ($1-P >99.73\%$).
With increasing \mbh and \mdot, \fx becomes smaller. The trend is consistent with the result by \cite{Wang2019}, where \leddR instead of \mdot was used. However, our fitting slope of $-0.57\pm 0.05$ is flatter than the value of $-0.74\pm 0.14$ by \cite{Wang2019}. 

For the SDSS subsample by \cite{Huang2020}, the X-ray data don't cover the band beyond $\rm 10~keV$. We also do the analysis only for the Swift BASS subsample by \cite{Wang2019}. For the relation of \fx with \mdot, the slope is $-0.30\pm 0.07$ for the Swift BASS subsample. 
Considering the uncertainties of \fx, \mdot, \mbh of $0.3$, $0.62$, $0.3$ dex, respectively (see section 3.2), multivariate regression analysis shows that $\fx \propto \mdot^{-0.55\pm 0.05} \mbh^{-0.52\pm 0.05}$ with $r=0.55, p_{\rm null}=2.3\times 10^{-17}$ for the Swift BASS subsample, which is well consistent with  the entire sample. Considering the uncertainty, the slope of $-0.55\pm 0.05$ is very well consistent with  $-0.57\pm 0.05$ as shown in the right panel in Figure \ref{fig8}. The slope can be used to constrain the shear stress tensor of the accretion disk of AGN.

The fraction of the energy transported by magnetic buoyancy is $\fx=P_{\rm mag}v_{p}/Q$, where $P_{\rm mag} = B^{2}/8\pi$,  $Q=-(3/2)c_{\rm s}t_{\rm r\phi}$ (see Section 1). The viscous stress is $t_{r\phi}=-k_0 P_{\rm mag}$, $\fx= C\sqrt{\frac{(-t_{r\phi})}{P_{\rm tot}}}$.
If the magnetic stress $t_{r\phi}$ is assumed, we can get \fx at every radius and for different accretion rates. A global value of $<\fx>$ can then be obtained by integrating over all the disc area for different magnetic-stress tensors \citep{Svensson1994, Merloni2002, Wang2004, Yang2007}.  
Theoretical curves were presented for the relation between $\fx$ and \mdot for six distinct types of $t_{r \phi}$, which have relations with $P_{gas}, P_{rad}, and P_{total}$ \citep{Wang2004,Yang2007}. The stress tensors of $t_{r\phi}$ are $-\alpha P_{tot}$, $-\alpha P_{gas}$, $-\alpha P_{rad}$, $-\alpha \sqrt{P_{rad}P_{tot}}$, $-\alpha \sqrt{P_{gas}P_{tot}}$, and $-\alpha \sqrt{P_{gas}P_{rad}}$ for models 1 to 6, respectively, where $\alpha$ is the viscosity.  Different models have different slopes for the \fx-\mdot relation. For model 1, the slope is zero. For models 3 and 4, the slope is positive. For model 6, the slope changes from a positive value to a negative value and has a peak at $\mdot \sim 0.1$. For models 2 and 5, the slope is negative, but their slope values are different \citep[see Figure  5 in][]{Yang2007}.  
For their model 2 of magnetic stress tensor $t_{r \phi} \propto P_{gas}$, $\fx \propto \mdot^{-0.77}$. For their model 5 of magnetic stress tensor $t_{r \phi} \propto \sqrt{P_{gas}P_{total}}$, $\fx \propto \mdot^{-0.44}$. 
Using $\shb$-based \mdot instead of \leddR for the accretion strength, the flat slope of $-0.45\pm 0.05$ or $-0.57\pm 0.05$ favors model 5, which suggests the shear stress tensor being proportional to $\sqrt{P_{gas}P_{tot}}$. 


Considering the relation of $\eta \propto \mdot^{-0.5}  \lv^{0.25}$,  the negative relation between $\fx$ and \mdot (i.e., $\fx \propto \mdot^{-0.57}$) implies that there is a positive correlation between \fx and $\eta$,  i.e., $\fx\propto \eta^{1.14}\lv^{-0.285}$.
\fx increases more slowly than expected from the fitting solid line in Figure \ref{fig8} for decreasing \mdot, implying a limit on $\eta$ for smaller \mdot for the faint AGN with large \lv. 
The accretion efficiency $\eta$ depends on the innermost radius of the accreting matter, which is related to the SMBH spin \citep[e.g.,][]{N2013}.  The fraction of energy dissipated in the hot corona \fx has a correlation with the accretion efficiency $\eta$. It implies that AGN with SMBHs that spin faster (smaller \mdot) would have larger \fx.

\subsection{The relation between $\Gamma$ and $\mdot$}
\begin{figure*}
\includegraphics[angle=0,width=2.6in,angle=-90]{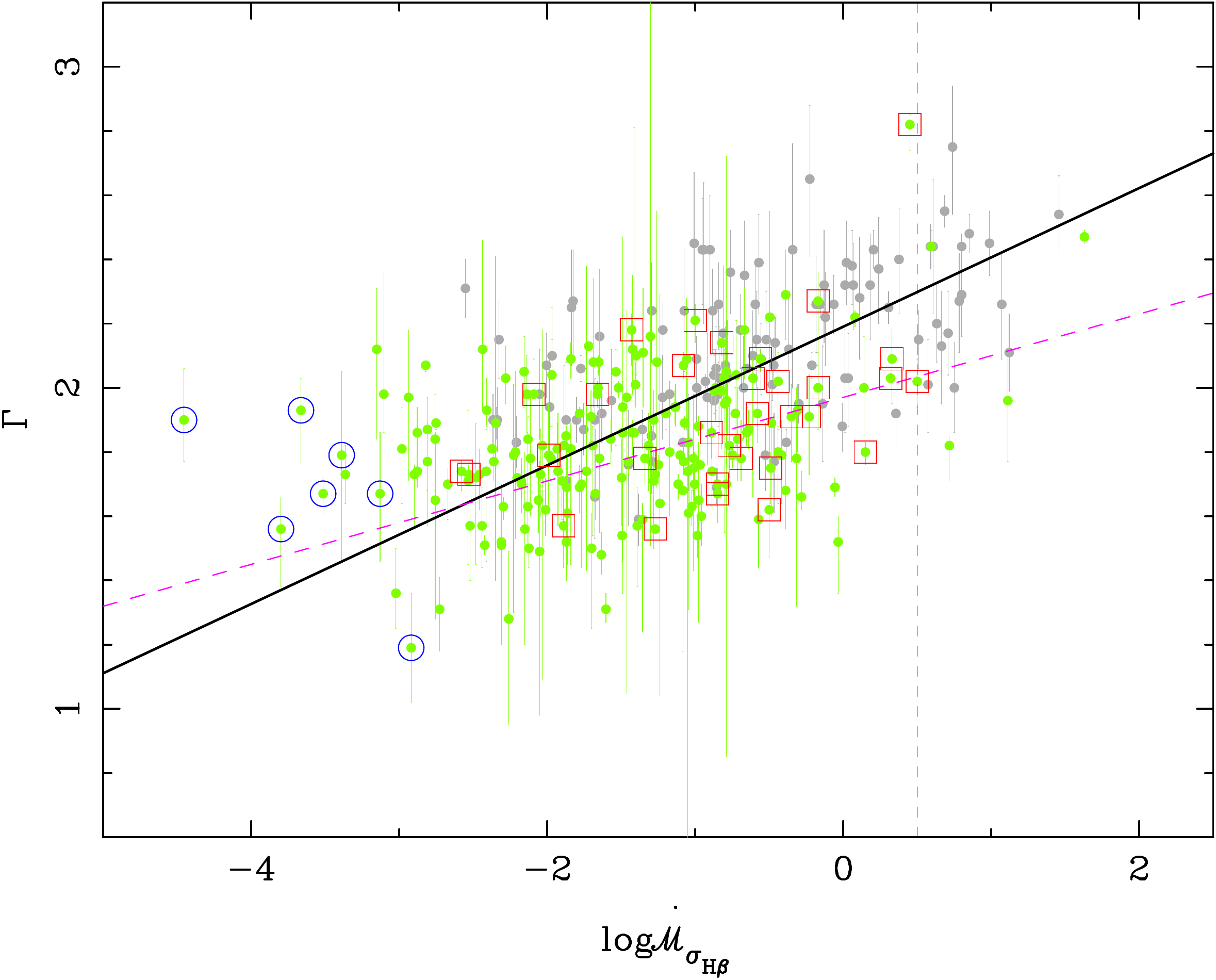}
\includegraphics[angle=0,width=2.6in,angle=-90]{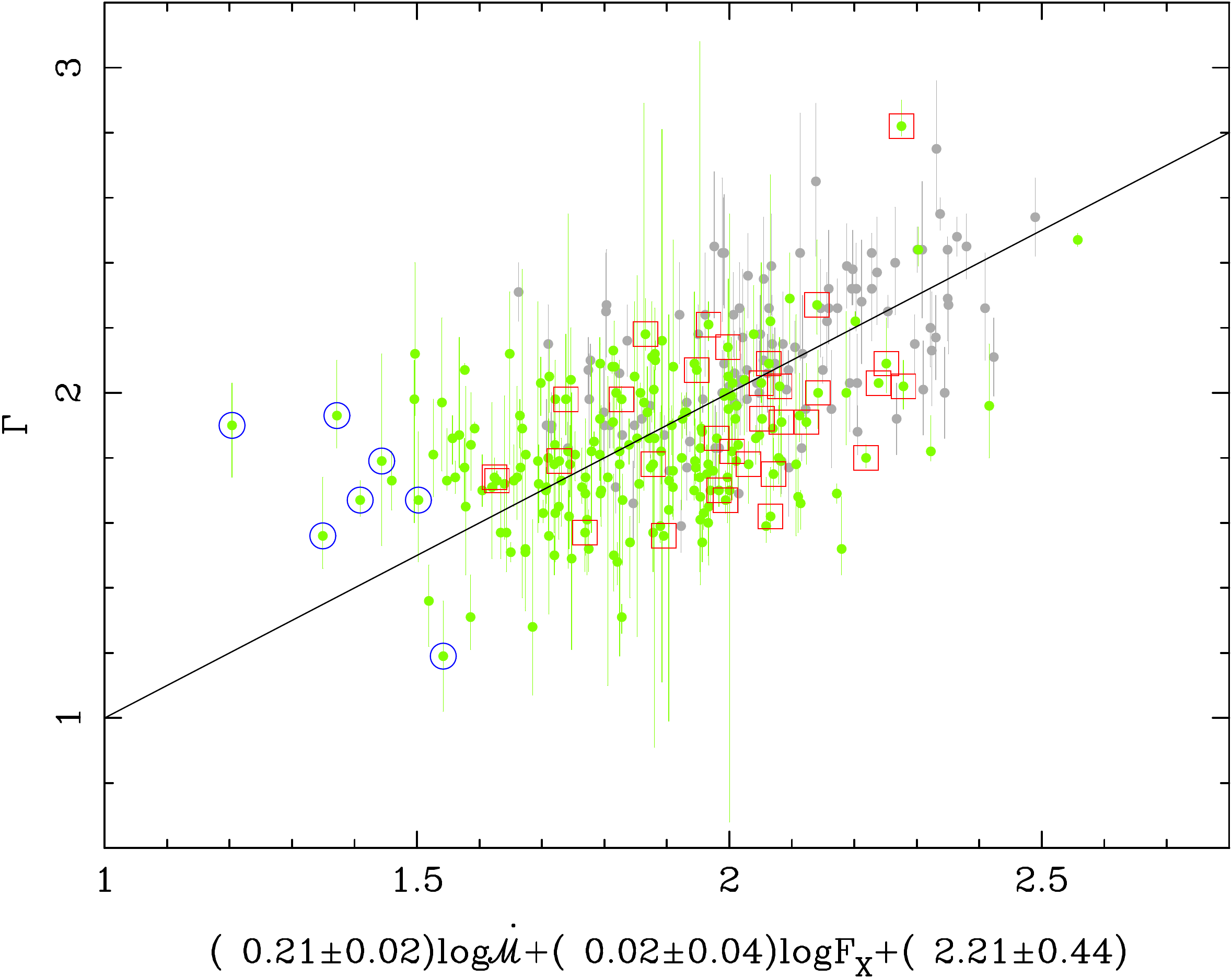}
\caption{ 
Left: $\Gamma$ versus \mdot. The solid black line is our best fit. The purple line is the result by \cite{Huang2020}. Right: Dependence of $\Gamma$ on \mdot and \fx. The solid line is 1:1. The symbols are the same as in Figure \ref{fig8}. 
}
\label{fig9}
\end{figure*}
In the left panel in Figure \ref{fig9}, we show the hard X-ray photon index $\Gamma$ versus \mdot. The Spearman correlation test gives  $r=0.52$, $p_{\rm null} = 1.1\times10^{-22}$. The correlation is significant. Considering errors in both coordinates, the BCES($\rm Y|X$) best-fitting relation for our total sample shows $\Gamma=(0.22\pm 0.02)\log \mdot +(2.19\pm 0.03)$. 
For 113 SDSS AGN by \cite{Huang2020}, $r=0.52, p_{\rm null}=3.1\times 10^{-9}$, $\Gamma \propto (\log \mdot)^{0.23\pm 0.04}$.  For 198 Swift-BASS AGN by \cite{Wang2019}, $r=0.31, p_{\rm null}=1.1\times 10^{-5}$, $\Gamma \propto (\log \mdot)^{0.14\pm 0.03}$. 
The slope ($0.23\pm 0.04$) for the SDSS subsample is slightly larger than the slope ($0.14\pm 0.03$) for the Swift-BASS subsample. It is possibly due to the high-luminosity for the SDSS subsample and/or lacking AGN with small \mdot in the SDSS subsample.
For 32 RM AGN in our sample, the Spearman correlation test gives  $r=0.4$, $p_{\rm null} = 0.024$. 
For 22 super-Eddington accreting AGN with $\mdot \geq 3$, we find that they are following the linear fitting (solid line the left panel in Figure \ref{fig9} ), although with a large uncertainty in \mdot ($\sim 0.62~$ dex). 



For our sample, the relation between $\Gamma$ and \leddR is also investigated. The Spearman correlation test gives  $r=0.56$, $p_{\rm null} = 6.7\times10^{-27}$. The correlation is also significant. Considering errors in both coordinates, the BCES($\rm Y|X$) best-fitting relation for our total sample shows $\Gamma=(0.39\pm 0.04)\log \leddR +(2.54\pm 0.06)$. Using $\rm FWHM_{\hb}$-based \mbh to calculate \leddR, the slopes are $0.21$ by \cite{Wang2019} and $0.23$ by \cite{Huang2020}, which are smaller than ours of $0.39$ . Considering the relation of $\leddR \propto (\mdot)^{0.56}$, $\Gamma \propto (\log \leddR)^{0.39}$ is consistent with $\Gamma \propto (\log \mdot)^{0.22}$. Using $\shb$-based \leddR, the larger slope of $0.39\pm 0.04$  will relieve the need for large viscosity in a model of the condensation of the corona by \cite{Qiao2018}. We also find that the relation between $\Gamma$ and \mbh is very weak with $r_s=-0.066$, and $p_{\rm null}= 0.247$. 

The correlation between the hard X-ray photon index $\Gamma$ and \leddR is usually interpreted in  the scenario of the disk–corona connection. For AGN with a  larger \leddR, the enhanced UV/optical emission from the accretion disk leads to more effective Compton cooling in the corona. This decreases the temperature and/or optical depth in the corona, which leads to the softening of the X-ray spectrum (a larger value of $\Gamma$) \citep[e.g., ][]{Pounds1995, Fabian2015, Qiao2018, Cheng2020} .

Considering the uncertainties of \fx, \mdot of $0.3$, $0.62$ dex and the error of $\Gamma$,  multivariate regression analysis
shows that  $\Gamma=(0.21\pm 0.02) \log \mdot+(0.02\pm 0.04)\log \fx+(2.21\pm 0.44)$. 
Including \fx in the relation between $\Gamma$ and \mdot, the correlation becomes slightly stronger than the single variable regression, from $r=0.52, p_{\rm null}=1.1\times 10^{-22}$ to  $r=0.53, p_{\rm null}=4.7\times 10^{-24}$. 
The scatter decreases slightly  from 0.237 to 0.234. The $\chi^2$ decreases from 7109 to 6756. Using the F-test, from the $\chi^2$ difference, we found that the probability P that the multivariate regression model cannot improve the fit is $2.18\times 10^ {-4}$. Therefore, the significance of improving the fitting by the multivariate regression is greater than 3$\sigma$ ($1-P>99.73\%$).
\cite{Kubota2018} developed a spectral model for the AGN SED, including an outer standard disc to produce UV/optical emission, an inner warm Comptonizing region to produce the soft X-ray excess and a hot corona to produce hard X-rays. They suggested that the corona size scale has to decrease with increasing Eddington fraction, as well as the steeper spectral index with increasing \mdot. Considering the reprocessed emission produced by the hot corona illuminating the warm Comptonization and standard disc regions,  they predicted a decreasing amount of optical variability with increasing \mdot. They also predicted that with increasing \fx , the curve of $\Gamma-\mdot$ relation is lower (see their Figure 5). With increasing of \fx, the curve of the $\Gamma-\mdot$ relation is lower and the X-ray spectrum becomes harder. 
For our compiled sample, we find that the hard X-ray spectrum becomes softer with increasing of \fx, although the scatter is large ($0.02\pm 0.04$). It is not consistent with the result for the spectral model suggested by \cite{Kubota2018}. We caution that this contradiction is partly due to the large uncertainties in these parameters.

\section{Conclusions}
A sample of 311 broad-line AGN ($z<0.7$) is collected from Swift BASS and SDSS with hard X-ray photon indexes, and \lv covering from $\sim 10^{41}$ to $\sim 10^{46}$ \ergs.  Through optical spectral decomposition, \shb and \rfe are measured, which are used to calculate the $\shb$-based virial \mbh. Using the standard disk model, the dimensionless accretion rate \mdot can be derived from \lv and the $\shb$-based \mbh. The main conclusions can be summarized as follows:
\begin{itemize}
\item  For our entire sample, with respect to  $\rm FWHM_{\hb}$, the mean value of $\shb$-based \mbh is  on average larger by $0.26$ dex,  the mean value of $\shb$-based \mdot is on average smaller by $0.51$ dex.  

\item There exists a non-linear relationship between the Eddington ratio  ($\leddR$) and the dimensionless accretion  (\mdot), $\leddR \propto \mdot^{0.56}$. From the accretion efficiency $\eta$, $\leddR=\eta \mdot$. The difference between \leddR and \mdot comes from $\eta$. The non-linear relationship implies smaller efficiency for AGN with higher dimensionless accretion rate, i.e., $\eta \propto \mdot^{-0.5}  \lv^{0.25}$.

\item The difference between $\rm FWHM_{\hb}$-based and $\shb$-based \mbh comes from the corresponding proportionality factor $f$. For a variable $\rm FWHM_{\hb}$-based $f$, we find a relation between $f$ and $\rm FWHM_{\hb}$,  $\log f =-(0.61\pm 0.05) \log \rm FWHM_{\hb}+(2.48\pm 0.18)$. It has a similar trend to that suggested by \cite{Yu2020b}.

\item We find a strong bivariate correlation of the fraction of energy released in the corona \fx with $\shb$-based \mdot and $\shb$-based \mbh,  $\fx\propto \mdot^{-0.57\pm 0.05} M_{\rm BH}^{-0.54\pm 0.06}$. The flat slope of $-0.57\pm 0.05$ favors the shear stress tensor being proportional to the geometric mean of gas pressure and total pressure. It implies AGN with SMBHs that spin faster (smaller \mdot) would have larger \fx.

\item  It is found that the hard X-ray photon index  $\Gamma$ has a significant correlation with $\shb$-based \mdot, with a  shallow slope with respect to that with \leddR. The relation between $\Gamma$ and \mbh is very weak with $r_s=-0.066, p_{\rm null}=0.247$. Including \fx, we find a slightly stronger bivariate relation of $\Gamma$ with $\shb$-based \mdot and \fx. We find that the hard X-ray spectrum becomes softer with increasing of \fx, although the scatter is large ($0.02\pm 0.04$).

\end{itemize}

\acknowledgments

We are very grateful to the anonymous referee for her/his instructive comments which significantly improved the content of the paper. We are very grateful to C. Hu  for his suggestion. We are very grateful to R. F. Green for his carefully revising our manuscript.
This work has been supported by the National Science Foundations of China (Nos. 11973029 and 11873032). This work is supported by the National Key Research and Development Program of China (No. 2017YFA0402703).

\newpage

\begin{thebibliography}{}
\bibitem[Akritas \& Bershady (1996)]{AB96} Akritas, M. G., \& Bershady, M. A. 1996, ApJ, 470, 706
\bibitem[Avni \& Tananbaum (1982)]{Avni1982} Avni, Y., \& Tananbaum, H. 1982, ApJL, 262, L17

\bibitem[Batiste et al., (2017)]{Ba17} Batiste, M., Bentz, M. C., Manne-Nicholas, E. R., Onken, C. A., Bershady, M. A. 2017, ApJ, 838, L10
\bibitem[Baumgartner et al. (2013)]{Baumgartner2013} Baumgartner, W. H., Tueller, J., Markwardt, C. B., et al. 2013, ApJS, 207, 19
\bibitem[Bentz et al. (2013)]{Be13} Bentz, M. C.,  et al. 2013, \apj, 767, 149
\bibitem[Bian \& Zhao (2002)]{Bian2002} Bian, W. H., \& Zhao, Y. H. 2002, A\&A, 395, 465
\bibitem[Bian \& Zhao (2003a)]{Bian2003a} Bian, W. H., \& Zhao, Y. H., 2003a, MNRAS, 343, 164
\bibitem[Bian \& Zhao (2003b)]{Bian2003b} Bian, W. H., \& Zhao, Y. H., 2003b, PASJ, 55, 599
\bibitem[Bian et al. (2008)]{Bian2008} Bian, W. H., et al. 2008, MNRAS, 390, 752
\bibitem[Blandford \& McKee (1982)]{BM82} Blandford, R., McKee, C. 1982, \apj, 255, 419
\bibitem[Boroson \& Green (1992)]{BG92}  Boroson, T. A. \& Green, R. F., 1992, \apjs, 80, 109
\bibitem[Boroson(2002)]{Boroson2002} Boroson T. A., 2002, ApJ, 565, 78
\bibitem[Brightman et al. (2013)]{Brightman13} Brightman, M., et al. 2013, MNRAS, 433, 2485
\bibitem[Cardelli et al. (1989)]{Cardelli1989} Cardelli, J. A., Clayton, G. C., \& Mathis, J. S. 1989, ApJ, 345, 245
\bibitem[Cheng et  al. (2020)]{Cheng2020} Cheng, H., Liu, B. F., Liu, J., et al. 2020, MNRAS, 495, 1158

\bibitem[Coffey et al. (2019)]{Coffey2019} Coffey, D., et al. 2019, A\& A, 625, 123   
\bibitem[Collin et al. (2006)]{Collin2006} Collin, S., Kawaguchi, T., Peterson, B. M., \& Vestergaard, M. 2006, A\&A,
456, 75
\bibitem[Dalla Bont\`{a} et al. (2020)] {Dalla2020} Dalla Bont\`{a}, E., Peterson B. M., Bentz, M. C., et al. 2020, \apj, 903, 112

\bibitem[Davis \& Laor (2011)]{DL2011}Davis, S. W., \& Laor, A. 2011, \apj, 728, 98
\bibitem[Du et al. (2015)]{Du2015} Du, P., et al. 2015, \apj, 806, 22
\bibitem[Du et al. (2016a)]{Du2016a} Du, P., et al. 2016a, \apj, 818, L14
\bibitem[Du et al. (2016b)]{Du2016b} Du, P., et al. 2016b, \apj, 825, 126
\bibitem[Du et al. (2018)]{Du2018} Du, P., et al. 2018, \apj, 856, 6
\bibitem[Du \& Wang (2019)]{Du2019} Du, P., Wang, J.-M. 2019, \apj, 886, 42, arxiv: 1909.06735
\bibitem[Fabian et al. (2015)]{Fabian2015} Fabian, A. C., Lohfink, A., Kara, E., et al. 2015, MNRAS, 451, 4375

\bibitem[Ge et al. (2016)]{Ge2016} Ge, X., Bian, W. H., Jiang, X. L., et al., 2016, MNRAS, 462, 966
\bibitem[Guo et al. (2018)]{Guo2018} Guo, H., Shen, Y., \& Wang S.,  2018,  ascl: 1809:008
\bibitem[Haardt \& Maraschi (1991)]{Haardt1991} Haardt F., Maraschi L., 1991, ApJ, 380, L51
\bibitem[Haardt et al. (1994)]{Haardt1994} Haardt F., Maraschi L., Ghisellini G., 1994, ApJL, 432, L95
\bibitem[Ho \& Kim (2014)]{HK14} Ho, L., Kim, M. 2014, \apj, 789, 17
\bibitem[Hu et al. (2008)]{Hu2008} Hu, C., Wang, J.-M., Ho, L. C., et al. 2008, ApJ, 687, 78
\bibitem[Hu et al. (2015)]{Hu2015} Hu, C., Du, P., Lu, K. X., et al. 2015, ApJ, 804, 138
\bibitem[Huang et al. (2020)]{Huang2020} Huang, J. et al. 2022, ApJ, 895, 114
\bibitem[Jin et al. (2012)]{Jin2012} Jin, C., Ward, M., \& Done, C. 2012, MNRAS, 425, 907

\bibitem[Kaspi et al. (2000)]{Ka00} Kaspi, S., et al. 2000, \apj, 533, 631
\bibitem[Kilerci Eser et al. (2015)]{KE15} Kilerci, Eser, E., Vestergaard, M., Peterson, B. M., Denney, K. D., Bentz, M. C. 2015, \apj, 801, 8
\bibitem[Khadka et al. (2022)]{Khadka2022} Khadka, N., Mart{\'\i}nez-Aldama, M. L., Zaja\v{c}ek, M., et al. 2022, MNRAS, 513, 1985
\bibitem[Kormendy \& Ho (2013)]{KH13} Kormendy, J., \& Ho, L. C. 2013,  ARA\&A, 51, 511
\bibitem[Koss et al. (2017)]{Koss2017} Koss, M., Trakhtenbrot, B., Ricci, C., et al., 2017, ApJ, 850, 74
\bibitem[Kubota \& Done (2018)]{Kubota2018} Kubota, A., \& Done, C., 2018, MNRAS, 480, 1247
\bibitem[Laha et al. (2018)]{Laha2018} Laha, S., Ghosh, R., Guainazzi, M., \& Markowitz, A. G. 2018, MNRAS, 480, 1522

\bibitem[Liang et al. (1979)]{Liang1979} Liang, E. et al. 1979, ApJL, 231, L111
\bibitem[Liu et al. (2015)]{Liu2015} Liu, B.-F., et al. 2015, ApJ, 806, 223
\bibitem[Liu et al. (2021)]{Liu2021} Liu, H. Z., et al. 2021, ApJ, 910, 103
\bibitem[Liu \& Bian (2022)]{Liu2022} Liu, Y. S.,  \& Bian, W. H. 2022, ApJ, in press
\bibitem[Lupton (1993)]{Lupton1993} Lupton, R. H. 1993, Statistics in Theory and Practice (Princeton: Princeton Univ. Press)
\bibitem[Lusso \& Risaliti (2016)]{Lusso2016} Lusso, E., \& Risaliti, G. 2016, ApJ, 819, 154
\bibitem[Lusso \& Risaliti (2017)]{Lusso2017} Lusso, E., \& Risaliti, G. 2017, A\&A, 602, A79

\bibitem[Marconi et al. (2004)]{Marconi2004} Marconi, A., Risaliti, G., Gilli, R. et al. 2004, MNRAS, 351, 169
\bibitem[Mej{\'\i}a-Restrepo et al. (2018)]{Mejia2018} Mej{\'\i}a-Restrepo, J. E., et al. 2018, Nature Astronomy, 2, 63, arXiv: 1709.05345

\bibitem[Meyer-Hofmeister et al. (2017)]{Meyer2017} Meyer-Hofmeister, E., et al. 2017, A\&A, 607, 94
\bibitem[Merloni \& Fabian (2002)]{Merloni2002} Merloni A., \& Fabian A. C., 2002, MNRAS, 332, 165 (MF02)
\bibitem[Netzer (2013)]{N2013} Netzer, H. 2013, The Physics and Evolution of Active Galactic Nuclei
\bibitem[Netzer (2019)]{N2019} Netzer, H. 2019, \mnras, 488, 5185 (arxiv: 1907.09534)
\bibitem[Onken et al. (2004)]{On04} Onken, C. A., et al. 2004, \apj, 615, 645
\bibitem[Peterson et al. (2004)]{Pe04} Peterson, B. M., et al. 2004,  \apj, 613, 682
\bibitem[Pounds et al. (1995)]{Pounds1995} Pounds, K. A., Done, C., \& Osborne, J. P. 1995, MNRAS, 277, L5
\bibitem[Qiao \& Liu (2018)]{Qiao2018} Qiao, E. \&  Liu, B. F. 2018, MNRAS, 477, 210
\bibitem[Rigby et al. (2009)]{Rigby2009} Rigby, J. R., Diamond-Stanic, A. M., \& Aniano, G. 2009, ApJ, 700, 1878
\bibitem[Ricci et al. (2017)]{Ricci2017} Ricci, C., Trakhtenbrot, B., Koss, M. et al. 2017, ApJS, 233, 17
\bibitem[Schlegel et al. (1998)]{Schlegel1998} Schlegel, D. J.,  Finkbeiner, D. P.,  \& Davis, M. 1998, ApJ,  500, 525
\bibitem[Shakura \& Sunyaev (1973)]{SS73}Shakura, N. I., \& Sunyaev, R. A. 1973, A\&A, 24, 337
\bibitem[Shen et al. (2011)]{Sh11} Shen, Y., et al. 2011, \apjs, 194, 45
\bibitem[Shen \& Ho (2014)]{Shen2014} Shen, Y., \&  Ho, L. C. 2014, Nature, 513, 210
\bibitem[Shen et al. (2019)]{Shen19} Shen, Y., et al. 2019, \apjs, 241,34
\bibitem[Stella \& Rosner (1984)]{Stella1984} Stella L., Rosner R., 1984, ApJ, 277, 312
\bibitem[Sturm et al. (2018)]{Sturm2018} Sturm, E., et al. 2018, Nature, 563, 657
\bibitem[Sulentic et al. (2000)]{Sulentic2000} Sulentic, J. W., Zwitter, T., Marziani, P., \& Dultzin-Hacyan, D. 2000, ApJ, 536, L5
\bibitem[Svensson \& Zdziarski (1994)]{Svensson1994} Svensson, R., \& Zdziarski, A. A., 1994, ApJ, 436, 599
\bibitem[Trakhtenbrot \& Netzer (2012)]{TN2012} Trakhtenbrot, B., \& Netzer, H. 2012, MNRAS, 427, 3081
\bibitem[Trakhtenbrot et al.  (2017)]{Trakhtenbrot2017}Trakhtenbrot, B., Ricci, C., Koss, M. J., et al. 2017, MNRAS, 470, 800
\bibitem[Wang et al. (2013)]{Wang2013} Wang, J. M., et al. 2013, PhRvL, 110, 081301
\bibitem[Wang et al.  (2004)]{Wang2004} Wang, J. M., Watarai, K. Y., Mineshige, S., 2004, ApJ, 607, L107
\bibitem[Wang et al. (2019a)]{Wang2019} Wang, C., Yu, L. M., Bian, W. H., Zhao, B. X., 2019a, \mnras, 487, 2463
\bibitem[Wang et al. (2019b)]{WS2019} Wang, S., Shen, Y., Jiang, L. H., et al. 2019b, \apj, 882, 4
\bibitem[Williams et al. (2018)]{Williams2018} Williams, P. R., Pancoast, A., Treu, T., et al. 2018, \apj, 866, 75
\bibitem[Yang et al.  (2007)]{Yang2007} Yang, F., Hu, C., Chen, Y. et al. 2007, ChJAA, 7, 353
\bibitem[Yip et al. (2004a)]{Yip2004a} Yip, C. W., et al. 2004a, AJ, 128, 585
\bibitem[Yip et al. (2004b)]{Yip2004b} Yip, C. W., et al. 2004b, AJ, 128, 2603
\bibitem[Yu et al. (2019)]{Yu2019} Yu, L. M., Wang, C., Bian, W. H., Zhao, B. X.,  Ge, X. 2019, \mnras, 488. 1519
\bibitem[Yu et al. (2020a)]{Yu2020a} Yu, L. M., et al. 2020a, MNRAS, 491, 5881
\bibitem[Yu et al. (2020b)]{Yu2020b} Yu, L. M., et al. 2020b, ApJ, 901, 133
\end{thebibliography}

\begin{table*}
\centering
\caption{Properties of 113 SDSS AGN.\\
Col. (1): Name; Col. (2): hard X-ray $\Gamma$\citep{Huang2020}; Col. (3): $\log L_{\rm X} (2-10 \rm keV)$ in units of $\ergs$; Col. (4-5): $\rm FWHM_{\hb}$, $\shb$, in units of $\kms$~; Col. (6): $\rfe$; Col. (7-8): host-corrected $\lv$, $\lb$, in units of $\ergs$; Col. (9): $\shb$-based \mbh. \\
}
\label{table1}
\small
\begin{lrbox}{\tablebox}
\begin{tabular}{llllllllll}
\hline
   Name  & $\Gamma$& $\log L_{\rm X}$ & $\rm FWHM_{\hb}$  &$\shb$ & $\rfe$ & $\lv$ & $\lb$ & $\mbh$ \\
 (J2000) & & $\rm erg~s^{-1}$ &$\kms$ & $\kms$ &  & $\ergs$ & $\ergs$ &$\rm M_\odot$ \\
 (1)&(2)&(3)&(4)&(5)&(6)&(7)&(8)& (9)\\
\hline
002233.27-003448.4 &$  2.22^{+0.17}_{-0.16}$ & 43.79   &$  1900.1\pm    82.1$&$  1689.2\pm    71.8$ &$    1.12\pm    0.03$&    44.38        & 45.25 & 7.91       \\
004319.74+005115.4 &$  1.74^{+0.05}_{-0.05}$ & 44.50   &$ 12239.0\pm   188.6$&$  5129.6\pm    79.1$ &$    0.32\pm    0.02$&    44.29        & 45.16 & 9.14       \\
005709.94+144610.1 &$  1.92^{+0.10}_{-0.10}$ & 44.57   &$ 10462.8\pm   629.4$&$  5114.0\pm  1904.4$ &$    0.36\pm    0.02$&    44.87        & 45.70 & 9.40       \\
012549.97+020332.2 &$  1.68^{+0.13}_{-0.13}$ & 44.73   &$  6881.1\pm    47.7$&$  2883.2\pm    20.0$ &$    0.38\pm    0.03$&    44.70        & 45.54 & 8.81       \\
013418.19+001536.7 &$  1.62^{+0.16}_{-0.15}$ & 44.51   &$  4347.7\pm   120.9$&$  2628.5\pm    32.7$ &$    0.67\pm    0.04$&    45.13        & 45.95 & 8.83       \\
014959.27+125658.0 &$  2.28^{+0.43}_{-0.33}$ & 44.01   &$  3367.5\pm    99.9$&$  2074.0\pm   129.0$ &$    0.64\pm    0.03$&    44.64        & 45.49 & 8.40       \\
015950.24+002340.8 &$  2.23^{+0.12}_{-0.11}$ & 44.02   &$  2674.2\pm    70.3$&$  1968.3\pm   124.1$ &$    1.03\pm    0.02$&    44.68        & 45.52 & 8.22       \\
020011.52-093126.2 &$  1.82^{+0.13}_{-0.13}$ & 44.27   &$  8368.6\pm   156.6$&$  3506.6\pm  1296.1$ &$    0.43\pm    0.06$&    45.00        & 45.82 & 9.11       \\
020039.15-084554.9 &$  2.29^{+0.21}_{-0.21}$ & 43.90   &$  1792.3\pm    11.9$&$  1579.9\pm    25.1$ &$    1.21\pm    0.03$&    44.72        & 45.56 & 7.98       \\
020354.68-060844.0 &$  2.09^{+0.19}_{-0.19}$ & 44.26   &$  5914.3\pm    64.8$&$  2782.3\pm    14.1$ &$    0.18\pm    0.01$&    44.88        & 45.71 & 8.94       \\
020840.66-062716.7 &$  2.02^{+0.21}_{-0.21}$ & 42.90   &$  3509.6\pm   178.5$&$  1782.1\pm    90.0$ &$    0.70\pm    0.03$&    43.21        & 44.18 & 7.55       \\
021318.29+130643.9 &$  1.88^{+0.15}_{-0.14}$ & 44.28   &$  3532.5\pm   234.5$&$  1485.0\pm  1885.6$ &$    0.49\pm    0.10$&    44.48        & 45.33 & 8.08       \\
021657.78-032459.4 &$  2.21^{+0.13}_{-0.13}$ & 43.64   &$  2981.8\pm    46.1$&$  2144.3\pm    20.4$ &$    0.61\pm    0.00$&    44.01        & 44.90 & 8.13       \\
022014.57-072859.3 &$  1.79^{+0.15}_{-0.15}$ & 44.64   &$  8827.0\pm   111.9$&$  3698.9\pm    46.9$ &$    0.30\pm    0.12$&    43.94        & 44.84 & 8.69       \\
022024.78-050231.8 &$  1.75^{+0.13}_{-0.13}$ & 43.56   &$  4481.2\pm   486.2$&$  3198.1\pm   659.0$ &$    0.60\pm    0.06$&    43.63        & 44.56 & 8.31       \\
023153.98-045106.4 &$  2.17^{+0.14}_{-0.13}$ & 43.23   &$  3105.4\pm    65.9$&$  1341.1\pm   313.1$ &$    0.90\pm    0.04$&    43.63        & 44.56 & 7.43       \\
030639.57+000343.1 &$  1.86^{+0.05}_{-0.05}$ & 43.41   &$  2842.8\pm   149.2$&$  1598.3\pm  1209.8$ &$    0.44\pm    0.06$&    43.84        & 44.74 & 7.86       \\
073309.20+455506.2 &$  1.64^{+0.11}_{-0.10}$ & 44.11   &$  3920.8\pm    27.8$&$  2255.2\pm    20.9$ &$    0.58\pm    0.00$&    44.66        & 45.50 & 8.50       \\
075112.18+174351.7 &$  1.83^{+0.04}_{-0.04}$ & 44.07   &$  3143.6\pm    97.9$&$  2248.6\pm   103.4$ &$    0.25\pm    0.01$&    43.92        & 44.82 & 8.27       \\
080101.41+184840.7 &$  2.33^{+0.06}_{-0.06}$ & 43.61   &$  1778.2\pm    16.4$&$  1131.5\pm     3.2$ &$    1.17\pm    0.01$&    44.16        & 45.04 & 7.44       \\
080608.13+244421.0 &$  2.50^{+0.24}_{-0.23}$ & 43.86   &$  2663.2\pm    51.8$&$  1803.5\pm    46.7$ &$    0.68\pm    0.01$&    44.35        & 45.22 & 8.12       \\
080908.28+202420.4 &$  1.82^{+0.21}_{-0.20}$ & 44.02   &$  3267.7\pm    33.5$&$  2051.6\pm    46.1$ &$    0.28\pm    0.01$&    44.04        & 44.93 & 8.24       \\
081331.13+040949.4 &$  1.85^{+0.13}_{-0.13}$ & 44.25   &$  5628.0\pm   361.0$&$  2833.5\pm    23.8$ &$    0.59\pm    0.03$&    44.52        & 45.37 & 8.63       \\
081422.12+514839.4 &$  1.68^{+0.15}_{-0.15}$ & 44.64   &$  4438.4\pm   176.7$&$  2898.4\pm    75.1$ &$    0.44\pm    0.01$&    44.70        & 45.54 & 8.79       \\
081441.91+212918.5 &$  2.17^{+0.05}_{-0.05}$ & 43.77   &$  1670.3\pm    13.9$&$  1373.9\pm    33.4$ &$    0.65\pm    0.02$&    43.96        & 44.86 & 7.71       \\
081840.29+210208.5 &$  2.30^{+0.23}_{-0.22}$ & 43.42   &$  4107.6\pm   218.0$&$  2558.3\pm   175.3$ &$    0.80\pm    0.08$&    43.86        & 44.77 & 8.14       \\
084339.24+053125.2 &$  2.09^{+0.13}_{-0.13}$ & 44.41   &$  7055.4\pm     0.8$&$  2956.1\pm     0.3$ &$    0.63\pm    0.03$&    44.80        & 45.63 & 8.79       \\
090900.42+105934.8 &$  1.75^{+0.08}_{-0.08}$ & 44.06   &$  6285.9\pm   297.9$&$  4752.3\pm   101.5$ &$    0.36\pm    0.02$&    43.96        & 44.86 & 8.90       \\
091029.03+542718.9 &$  1.94^{+0.07}_{-0.07}$ & 44.47   &$  4947.1\pm   291.3$&$  2919.7\pm  1253.1$ &$    0.66\pm    0.20$&    44.52        & 45.37 & 8.63       \\
091557.29+292618.3 &$  1.88^{+0.17}_{-0.15}$ & 44.13   &$  3264.2\pm   126.7$&$  1961.0\pm    56.1$ &$    0.93\pm    0.05$&    44.71        & 45.55 & 8.27       \\
091737.53+284046.2 &$  1.61^{+0.19}_{-0.17}$ & 44.51   &$  5696.9\pm   435.4$&$  3096.5\pm  1518.7$ &$    0.32\pm    0.10$&    44.33        & 45.19 & 8.72       \\
092140.83+293809.6 &$  1.91^{+0.13}_{-0.13}$ & 43.59   &$  3828.3\pm   317.9$&$  2083.3\pm  1003.5$ &$    0.30\pm    0.06$&    44.18        & 45.06 & 8.31       \\
095048.38+392650.4 &$  1.91^{+0.04}_{-0.04}$ & 44.37   &$  5219.5\pm   100.5$&$  4267.4\pm    60.7$ &$    0.43\pm    0.03$&    44.61        & 45.46 & 9.09       \\
095240.16+515249.9 &$  2.11^{+0.09}_{-0.09}$ & 44.11   &$  3549.1\pm   130.2$&$  2233.5\pm    69.4$ &$    1.10\pm    0.07$&    44.74        & 45.58 & 8.34       \\
100025.24+015852.0 &$  1.77^{+0.09}_{-0.09}$ & 44.16   &$  5101.9\pm   604.8$&$  3213.2\pm  1375.9$ &$    0.29\pm    0.06$&    44.16        & 45.04 & 8.68       \\
100055.71+314001.2 &$  2.39^{+0.12}_{-0.12}$ & 43.44   &$  1613.7\pm    69.6$&$  1130.6\pm    94.4$ &$    2.03\pm    0.06$&    44.07        & 44.95 & 7.07       \\
100057.50+684231.0 &$  2.11^{+0.13}_{-0.13}$ & 44.02   &$  4915.5\pm   210.4$&$  2630.6\pm  2029.7$ &$    0.92\pm    0.17$&    44.55        & 45.40 & 8.46       \\
100324.56+021831.3 &$  2.28^{+0.18}_{-0.17}$ & 43.89   &$  4147.8\pm   123.0$&$  2849.2\pm   169.0$ &$    0.57\pm    0.04$&    44.73        & 45.57 & 8.75       \\
101850.52+411508.1 &$  2.28^{+0.17}_{-0.16}$ & 44.34   &$  4211.1\pm   362.7$&$  2824.2\pm    93.1$ &$    0.51\pm    0.03$&    44.69        & 45.53 & 8.74       \\
103935.75+533038.7 &$  1.81^{+0.16}_{-0.16}$ & 43.79   &$  4850.7\pm   345.7$&$  2817.3\pm    66.9$ &$    0.39\pm    0.03$&    43.71        & 44.63 & 8.31       \\
104741.75+151332.2 &$  2.17^{+0.17}_{-0.16}$ & 44.34   &$  2862.9\pm    43.4$&$  1808.7\pm    37.7$ &$    0.92\pm    0.05$&    44.78        & 45.61 & 8.24       \\
105104.54+625159.3 &$  1.86^{+0.15}_{-0.14}$ & 44.60   &$  2071.0\pm   723.0$&$  1708.1\pm    47.2$ &$    1.00\pm    0.07$&    45.21        & 46.02 & 8.37       \\
105143.89+335926.7 &$  1.81^{+0.04}_{-0.04}$ & 44.06   &$  3676.1\pm    27.9$&$  2307.2\pm    44.3$ &$    0.36\pm    0.01$&    44.40        & 45.26 & 8.48       \\
105237.24+240627.3 &$  2.01^{+0.11}_{-0.11}$ & 44.39   &$  8487.8\pm   115.9$&$  5316.2\pm   177.2$ &$    0.60\pm    0.01$&    45.31        & 46.12 & 9.55       \\
111006.95+612521.3 &$  2.17^{+0.18}_{-0.17}$ & 43.55   &$  2332.6\pm    66.2$&$  1714.2\pm    86.0$ &$    0.85\pm    0.02$&    44.12        & 45.01 & 7.90       \\
111117.65+133054.0 &$  1.80^{+0.18}_{-0.18}$ & 43.74   &$  2670.0\pm    66.4$&$  2155.3\pm   208.5$ &$    1.30\pm    0.08$&    44.21        & 45.09 & 7.98       \\
111135.77+482945.4 &$  1.99^{+0.08}_{-0.08}$ & 44.76   &$  4036.3\pm   147.9$&$  2163.2\pm    85.5$ &$    0.54\pm    0.08$&    44.70        & 45.54 & 8.50       \\
111830.28+402554.0 &$  2.28^{+0.06}_{-0.06}$ & 43.93   &$  1909.0\pm    10.4$&$  1727.3\pm    19.4$ &$    1.01\pm    0.02$&    44.55        & 45.40 & 8.06       \\
111908.67+211917.9 &$  2.11^{+0.04}_{-0.04}$ & 44.83   &$  2959.4\pm   175.9$&$  2244.4\pm   119.0$ &$    0.64\pm    0.01$&    45.25        & 46.05 & 8.76       \\
113233.55+273956.3 &$  2.02^{+0.13}_{-0.13}$ & 44.01   &$  1711.3\pm    42.2$&$  1479.2\pm    69.5$ &$    1.04\pm    0.07$&    44.95        & 45.77 & 8.10       \\
114820.83+012634.2 &$  1.80^{+0.06}_{-0.06}$ & 43.77   &$  5273.5\pm   120.5$&$  2515.3\pm    72.8$ &$    1.14\pm    0.05$&    44.64        & 45.49 & 8.38       \\
114933.87+222227.0 &$  1.85^{+0.14}_{-0.14}$ & 43.95   &$  4176.9\pm   311.2$&$  1850.3\pm    91.7$ &$    1.78\pm    0.08$&    44.70        & 45.54 & 7.90       \\
115937.87+554622.9 &$  1.73^{+0.08}_{-0.07}$ & 44.86   &$  3500.0\pm   138.7$&$  1468.8\pm  1093.6$ &$    0.25\pm    0.04$&    44.71        & 45.55 & 8.28       \\
120051.01+341702.4 &$  2.25^{+0.17}_{-0.16}$ & 43.53   &$  1614.0\pm    40.2$&$  1393.5\pm    93.8$ &$    1.08\pm    0.05$&    44.08        & 44.96 & 7.62       \\
120442.10+275411.8 &$  1.70^{+0.03}_{-0.03}$ & 44.47   &$  5010.4\pm    52.0$&$  2742.3\pm    67.8$ &$    0.34\pm    0.01$&    44.19        & 45.07 & 8.54       \\
121425.55+293612.6 &$  1.83^{+0.14}_{-0.14}$ & 44.74   &$  9622.9\pm   144.6$&$  4328.0\pm   121.1$ &$    0.17\pm    0.05$&    44.51        & 45.36 & 9.15       \\
121509.22+330955.1 &$  1.68^{+0.04}_{-0.04}$ & 45.23   &$  4515.7\pm   301.9$&$  2621.5\pm   187.4$ &$    0.39\pm    0.02$&    45.67        & 46.46 & 9.19       \\
121549.90+670621.7 &$  1.88^{+0.18}_{-0.18}$ & 44.29   &$  3016.0\pm    62.4$&$  1872.9\pm    16.2$ &$    0.38\pm    0.03$&    43.86        & 44.77 & 8.03       \\
121640.56+071224.3 &$  1.85^{+0.16}_{-0.16}$ & 44.77   &$  7200.0\pm   187.9$&$  3186.8\pm   716.6$ &$    0.63\pm    0.06$&    45.43        & 46.23 & 9.16       \\

\hline
\end{tabular}
\end{lrbox}
\scalebox{0.8}{\usebox{\tablebox}}
\\
(This table is available in its entirety in machine-readable form.)
\end{table*}

\begin{table*}
\setcounter{table}{0}
\caption{--continu}
\centering
\label{table1}
\small
\begin{lrbox}{\tablebox}
\begin{tabular}{llllllllll}
\hline
   Name  & $\Gamma$& $\log L_{\rm X}$ & $\rm FWHM_{\hb}$  &$\shb$ & $\rfe$ & $\lv$ & $\lb$ & $\mbh$ \\
 (J2000) & & $\rm erg~s^{-1}$ &$\kms$ & $\kms$ &  & $\ergs$ & $\ergs$ &$\rm M_\odot$ \\
 (1)&(2)&(3)&(4)&(5)&(6)&(7)&(8)& (9)\\
\hline
122004.37-002539.0 &$  1.44^{+0.08}_{-0.08}$ & 44.78   &$  4432.0\pm   924.3$&$  2746.9\pm   133.0$ &$    0.00\pm    0.00$&    44.52        & 45.37 & 8.83       \\
122018.43+064119.6 &$  2.00^{+0.12}_{-0.12}$ & 44.42   &$  7978.9\pm   261.7$&$  5592.2\pm   137.3$ &$    0.41\pm    0.11$&    44.49        & 45.34 & 9.27       \\
122106.50+114625.4 &$  2.10^{+0.18}_{-0.18}$ & 43.65   &$  4065.6\pm   382.2$&$  3577.2\pm  1262.8$ &$    0.26\pm    0.11$&    44.16        & 45.04 & 8.78       \\
122210.00+271902.4 &$  1.95^{+0.15}_{-0.14}$ & 44.23   &$  5670.4\pm   219.4$&$  2789.5\pm    37.7$ &$    0.60\pm    0.09$&    45.16        & 45.97 & 8.92       \\
122757.19+130232.8 &$  2.16^{+0.09}_{-0.09}$ & 44.69   &$ 13314.6\pm   860.9$&$  5954.0\pm   175.1$ &$    0.11\pm    0.07$&    44.69        & 45.53 & 9.54       \\
122950.08+033444.8 &$  2.13^{+0.19}_{-0.19}$ & 43.85   &$  2104.9\pm    41.2$&$  1484.2\pm   106.0$ &$    0.68\pm    0.03$&    44.35        & 45.22 & 7.95       \\
123054.11+110011.2 &$  1.87^{+0.02}_{-0.02}$ & 44.34   &$  4905.3\pm   178.0$&$  2609.6\pm   140.7$ &$    0.38\pm    0.01$&    44.66        & 45.50 & 8.70       \\
123325.78+093123.3 &$  2.12^{+0.17}_{-0.16}$ & 44.77   &$  6143.3\pm   332.1$&$  5416.7\pm    18.1$ &$    0.33\pm    0.01$&    45.42        & 46.22 & 9.72       \\
123356.11+074755.9 &$  1.62^{+0.10}_{-0.10}$ & 44.23   &$  3739.2\pm   117.8$&$  2421.9\pm    82.1$ &$    0.00\pm    0.00$&    44.28        & 45.15 & 8.60       \\
123604.02+264135.9 &$  1.94^{+0.11}_{-0.11}$ & 43.68   &$  3154.5\pm   135.1$&$  2276.6\pm   416.1$ &$    0.88\pm    0.04$&    44.21        & 45.09 & 8.18       \\
123800.91+621336.0 &$  2.07^{+0.15}_{-0.14}$ & 43.50   &$  2875.2\pm   140.1$&$  2735.9\pm   511.5$ &$    1.75\pm    0.14$&    44.38        & 45.25 & 8.09       \\
125455.09+084653.9 &$  1.77^{+0.11}_{-0.11}$ & 44.63   &$  2382.2\pm   373.7$&$  1704.8\pm   100.8$ &$    0.74\pm    0.04$&    45.17        & 45.98 & 8.44       \\
125553.04+272405.2 &$  1.92^{+0.12}_{-0.12}$ & 44.13   &$  6507.5\pm   133.8$&$  2726.7\pm    56.0$ &$    0.83\pm    0.03$&    44.42        & 45.28 & 8.46       \\
130112.91+590206.6 &$  2.30^{+0.10}_{-0.10}$ & 44.46   &$  2875.3\pm   520.7$&$  2212.4\pm    65.7$ &$    1.72\pm    0.10$&    45.79        & 46.58 & 8.59       \\
130926.28+534130.6 &$  2.03^{+0.09}_{-0.09}$ & 43.83   &$  3654.7\pm   210.4$&$  2032.5\pm    27.6$ &$    0.31\pm    0.04$&    43.42        & 44.37 & 7.92       \\
130946.99+081948.2 &$  1.54^{+0.05}_{-0.06}$ & 44.13   &$  4469.9\pm   179.8$&$  2496.3\pm   482.3$ &$    0.33\pm    0.02$&    44.72        & 45.56 & 8.72       \\
131046.77+271644.7 &$  1.56^{+0.08}_{-0.08}$ & 44.29   &$  8056.9\pm   231.0$&$  3376.1\pm    96.8$ &$    0.19\pm    0.06$&    44.02        & 44.91 & 8.70       \\
131217.75+351521.0 &$  1.72^{+0.06}_{-0.06}$ & 43.87   &$  4927.0\pm   112.2$&$  3793.2\pm    24.2$ &$    0.41\pm    0.01$&    45.02        & 45.84 & 9.19       \\
132101.41+340657.9 &$  2.09^{+0.16}_{-0.15}$ & 43.84   &$  5294.0\pm   319.5$&$  3452.1\pm   311.6$ &$    0.89\pm    0.04$&    44.17        & 45.05 & 8.52       \\
132447.65+032432.6 &$  2.05^{+0.12}_{-0.12}$ & 44.16   &$  3456.0\pm   243.7$&$  1454.2\pm    72.7$ &$    1.08\pm    0.05$&    44.69        & 45.53 & 7.95       \\
133806.59-012412.8 &$  2.11^{+0.14}_{-0.13}$ & 44.13   &$  4593.9\pm   406.0$&$  2775.2\pm   182.3$ &$    0.80\pm    0.10$&    44.43        & 45.29 & 8.49       \\
134749.85+582109.4 &$  1.95^{+0.05}_{-0.05}$ & 45.08   &$  8445.3\pm   302.6$&$  6261.4\pm  1037.1$ &$    0.33\pm    0.04$&    45.63        & 46.42 & 9.95       \\
134848.24+262219.2 &$  2.11^{+0.17}_{-0.16}$ & 44.09   &$  2169.4\pm   662.2$&$  1354.3\pm   122.2$ &$    1.39\pm    0.13$&    44.83        & 45.67 & 7.84       \\
135516.56+561244.6 &$  2.40^{+0.05}_{-0.05}$ & 43.81   &$  1632.6\pm     8.6$&$  1184.2\pm    19.9$ &$    1.23\pm    0.05$&    43.91        & 44.81 & 7.33       \\
141146.62+140807.6 &$  1.91^{+0.12}_{-0.12}$ & 44.09   &$  2810.4\pm   359.9$&$  1961.5\pm  2233.2$ &$    0.40\pm    0.09$&    44.23        & 45.10 & 8.24       \\
141213.61+021202.1 &$  1.75^{+0.10}_{-0.10}$ & 44.34   &$ 10657.4\pm   470.9$&$  4466.4\pm   197.5$ &$    0.12\pm    0.05$&    44.15        & 45.03 & 9.02       \\
141500.37+520658.5 &$  2.24^{+0.16}_{-0.15}$ & 43.80   &$  3563.1\pm   105.7$&$  2281.9\pm    31.0$ &$    0.78\pm    0.04$&    44.33        & 45.19 & 8.28       \\
141700.82+445606.3 &$  2.10^{+0.05}_{-0.05}$ & 43.59   &$  2795.4\pm    41.8$&$  1633.4\pm    36.1$ &$    1.38\pm    0.02$&    44.03        & 44.92 & 7.62       \\
142052.43+525622.4 &$  2.24^{+0.13}_{-0.13}$ & 44.30   &$  3534.0\pm   171.0$&$  2271.8\pm   246.5$ &$    1.04\pm    0.06$&    45.05        & 45.87 & 8.52       \\
142455.53+421407.6 &$  2.11^{+0.15}_{-0.15}$ & 44.22   &$  3076.9\pm    37.1$&$  2233.9\pm    94.5$ &$    0.80\pm    0.02$&    44.96        & 45.78 & 8.55       \\
142557.63+334626.2 &$  2.12^{+0.16}_{-0.15}$ & 43.95   &$  2671.4\pm    32.4$&$  1588.4\pm    34.9$ &$    1.55\pm    0.02$&    44.59        & 45.44 & 7.79       \\
142817.81+354021.9 &$  1.75^{+0.18}_{-0.18}$ & 43.73   &$ 11783.9\pm   557.5$&$  5000.4\pm   234.9$ &$    0.95\pm    0.16$&    44.20        & 45.08 & 8.83       \\
142943.07+474726.2 &$  1.92^{+0.04}_{-0.04}$ & 44.24   &$  2529.6\pm    41.5$&$  1785.3\pm    41.2$ &$    0.51\pm    0.01$&    44.59        & 45.44 & 8.29       \\
143025.78+415956.6 &$  2.28^{+0.23}_{-0.21}$ & 43.73   &$  3711.3\pm   244.3$&$  2754.3\pm   310.2$ &$    0.76\pm    0.06$&    44.27        & 45.14 & 8.42       \\
143118.42+325131.6 &$  1.98^{+0.18}_{-0.17}$ & 43.64   &$  1749.4\pm    23.3$&$  1345.2\pm    44.4$ &$    1.14\pm    0.03$&    44.42        & 45.28 & 7.73       \\
144404.50+291412.2 &$  2.60^{+0.21}_{-0.19}$ & 44.32   &$  3017.5\pm   313.1$&$  1924.1\pm    80.3$ &$    1.41\pm    0.15$&    45.32        & 46.13 & 8.37       \\
144414.66+063306.7 &$  1.89^{+0.08}_{-0.08}$ & 44.41   &$  3742.7\pm   106.6$&$  2161.8\pm    41.7$ &$    0.24\pm    0.01$&    44.35        & 45.22 & 8.45       \\
144645.94+403505.7 &$  2.29^{+0.07}_{-0.07}$ & 44.09   &$  2697.3\pm   243.7$&$  2126.3\pm    64.8$ &$    1.58\pm    0.05$&    45.12        & 45.94 & 8.29       \\
145006.93+581456.9 &$  2.00^{+0.09}_{-0.09}$ & 44.11   &$  1918.0\pm    42.0$&$  1368.8\pm    31.2$ &$    0.82\pm    0.02$&    44.63        & 45.48 & 7.97       \\
145108.76+270926.9 &$  2.29^{+0.04}_{-0.04}$ & 43.37   &$  1591.5\pm    11.8$&$  1159.4\pm    30.6$ &$    1.17\pm    0.02$&    44.15        & 45.03 & 7.46       \\
145459.47+184452.4 &$  1.68^{+0.08}_{-0.08}$ & 44.65   &$  9716.5\pm   716.8$&$  4509.2\pm   273.0$ &$    0.00\pm    0.00$&    44.59        & 45.44 & 9.29       \\
151526.17+415612.1 &$  1.83^{+0.10}_{-0.10}$ & 44.32   &$  5472.2\pm   256.0$&$  2292.8\pm   107.3$ &$    0.59\pm    0.06$&    44.11        & 45.00 & 8.25       \\
153132.00+353439.5 &$  1.72^{+0.08}_{-0.08}$ & 44.05   &$ 10415.4\pm   343.6$&$  4364.9\pm   144.1$ &$    0.67\pm    0.05$&    44.38        & 45.25 & 8.91       \\
153159.10+242047.1 &$  1.51^{+0.13}_{-0.13}$ & 45.23   &$  9994.5\pm  1311.3$&$  4953.7\pm   602.2$ &$    0.12\pm    0.02$&    45.71        & 46.50 & 9.87       \\
153633.55+161530.8 &$  1.75^{+0.07}_{-0.07}$ & 44.42   &$  6065.6\pm    56.3$&$  5101.9\pm   262.1$ &$    0.33\pm    0.03$&    45.28        & 46.09 & 9.61       \\
155638.22+442853.3 &$  1.97^{+0.18}_{-0.17}$ & 44.23   &$  3065.9\pm   226.8$&$  2194.5\pm   282.5$ &$    0.82\pm    0.06$&    44.53        & 45.38 & 8.33       \\
162704.32+142124.7 &$  1.92^{+0.06}_{-0.06}$ & 43.67   &$  2874.8\pm    15.8$&$  1977.7\pm    11.5$ &$    0.81\pm    0.02$&    44.02        & 44.91 & 8.00       \\
162720.71+452519.7 &$  1.88^{+0.17}_{-0.16}$ & 44.22   &$  5732.7\pm   528.2$&$  2717.3\pm   131.4$ &$    0.61\pm    0.05$&    44.64        & 45.49 & 8.65       \\
163821.78+285902.3 &$  2.20^{+0.19}_{-0.17}$ & 44.66   &$  5416.5\pm   105.0$&$  2269.3\pm   174.8$ &$    0.28\pm    0.06$&    44.85        & 45.68 & 8.71       \\
163833.72+295128.8 &$  2.00^{+0.16}_{-0.15}$ & 44.22   &$  3328.9\pm   206.8$&$  2380.0\pm   125.3$ &$    0.85\pm    0.05$&    44.46        & 45.32 & 8.35       \\
171207.44+584754.4 &$  2.00^{+0.15}_{-0.15}$ & 43.78   &$  2327.2\pm   194.8$&$  1953.7\pm  1587.7$ &$    0.51\pm    0.06$&    44.19        & 45.07 & 8.18       \\
223607.68+134355.3 &$  2.14^{+0.13}_{-0.13}$ & 44.27   &$  1948.1\pm    56.0$&$  1345.4\pm    50.4$ &$    0.78\pm    0.02$&    45.17        & 45.98 & 8.22       \\
231928.72+081706.7 &$  2.03^{+0.12}_{-0.11}$ & 44.12   &$  2489.7\pm   275.4$&$  2503.1\pm    98.5$ &$    0.83\pm    0.06$&    44.33        & 45.19 & 8.34       \\
232259.99-005359.2 &$  1.96^{+0.12}_{-0.12}$ & 43.47   &$  2556.5\pm    43.8$&$  1190.7\pm    54.5$ &$    1.71\pm    0.03$&    44.07        & 44.95 & 7.23       \\
\hline
\end{tabular}
\end{lrbox}
\scalebox{0.8}{\usebox{\tablebox}}
\\(This table is available in its entirety in machine-readable form.)
\end{table*}

\begin{table*}
\centering
\caption{Properties of 198 Swift/BAT-detected broad-line AGN.\\
Col. (1): Name; Col. (2) $\Gamma$ \citep{Ricci2017}; Col. (3): $\log L_{\rm X} (14-195 \rm keV)$ in units of $\ergs$; Col. (4-5): $\rm FWHM_{\hb}$, $\shb$, in untis of $\kms$; Col. (6), $\rfe$; Col. (7-8): host-corrected $\lv$, $\lb$ in units of $\ergs$; Col. (9): $\shb$-based $\mbh$; Col. (10):Note, 1: RM $\shb$-based mass; 2: $M_{\rm BH}-\sigma_*$ relation, $\sigma_*$ from \cite{Koss2017}; 0: our measured $\shb$-based $\mbh$.\\
}
\label{table2}
\begin{lrbox}{\tablebox}
\begin{tabular}{lllllllllllll}
\hline
    Name & $\Gamma$ & $\log L_{\rm X}$ & $\rm FWHM_{\hb}$ & $\shb$ & \rfe & \lv & \lb & \mbh & Notes\\
                 &                    &  \ergs                   &\kms                          & \kms     &       & \ergs & \ergs &$\rm M_\odot$ & \\
 (1)&(2)&(3)&(4)&(5)&(6)&(7)&(8)&(9) & (10) \\
\hline
SWIFT J0006.2+2012  &$  2.82^{+0.08}_{-0.03}$ & 43.45  &$  1885.7\pm   103.6$&$  1673.7\pm    85.4$ &$    0.71\pm    0.14$   & 43.76      & 44.67 & 7.52       & 1     \\
SWIFT J0029.2+1319  &$  2.00^{+0.13}_{-0.11}$ & 44.82  &$  2571.5\pm   767.0$&$  2133.2\pm   633.1$ &$    0.52\pm    0.05$   & 44.97      & 45.79 & 8.56       & 1     \\
SWIFT J0051.9+1724  &$  1.81^{+0.36}_{-0.08}$ & 44.41  &$  5609.8\pm   159.1$&$  2405.4\pm   372.4$ &$    0.08\pm    0.03$   & 43.13      & 44.11 & 8.01       & 0     \\
SWIFT J0054.9+2524  &$  2.03^{+0.37}_{-0.06}$ & 44.98  &$  4873.1\pm   548.2$&$  2745.7\pm   118.3$ &$    0.19\pm    0.05$   & 44.81      & 45.64 & 8.66       & 1    \\
SWIFT J0059.4+3150  &$  1.92^{+0.03}_{-0.02}$ & 43.19  &$  4324.9\pm   475.9$&$  2061.2\pm   132.0$ &$    0.33\pm    0.03$   & 42.40      & 43.48 & 7.44       & 0     \\
SWIFT J0113.8+1313  &$  1.68^{+0.50}_{-0.22}$ & 43.97  &$  3096.7\pm   237.7$&$  1868.9\pm   121.4$ &$    0.69\pm    0.08$   & 42.87      & 43.88 & 7.44       & 0     \\
SWIFT J0113.8-1450  &$  1.82^{+0.28}_{-0.06}$ & 44.26  &$  4795.5\pm   301.6$&$  2725.2\pm    26.7$ &$    0.66\pm    0.02$   & 42.99      & 43.99 & 7.83       & 0     \\
SWIFT J0123.9-5846  &$  2.09^{+0.04}_{-0.03}$ & 44.39  &$  4657.4\pm  1083.5$&$  2793.5\pm   157.7$ &$    0.45\pm    0.03$   & 43.98      & 44.87 & 8.01       & 1     \\
SWIFT J0149.9-5019  &$  1.52^{+0.16}_{-0.18}$ & 43.41  &$  6828.7\pm   760.9$&$  2887.5\pm   204.0$ &$    0.16\pm    0.17$   & 42.74      & 43.77 & 7.96       & 0     \\
SWIFT J0157.2+4715  &$  1.93^{+0.13}_{-0.08}$ & 43.93  &$  2949.4\pm    69.8$&$  1599.8\pm    73.0$ &$    0.76\pm    0.06$   & 43.71      & 44.63 & 7.68       & 0     \\
SWIFT J0206.2-0019  &$  1.84^{+0.08}_{-0.06}$ & 44.15  &$  4114.3\pm   548.1$&$  2670.8\pm    42.2$ &$    0.33\pm    0.06$   & 43.62      & 44.55 & 8.25       & 0     \\
SWIFT J0208.5-1738  &$  1.97^{+0.26}_{-0.21}$ & 44.60  &$ 12270.3\pm   221.5$&$  6380.4\pm    93.8$ &$    0.10\pm    0.02$   & 44.21      & 45.09 & 9.37       & 0     \\
SWIFT J0214.6-0049  &$  1.78^{+0.08}_{-0.06}$ & 43.43  &$  7009.5\pm   509.0$&$  2938.9\pm   383.3$ &$    0.71\pm    0.08$   & 43.50      & 44.44 & 8.01       & 1     \\
SWIFT J0226.4-2821  &$  1.83^{+0.26}_{-0.14}$ & 44.05  &$  4348.9\pm    73.2$&$  2286.9\pm    17.7$ &$    0.66\pm    0.02$   & 43.63      & 44.56 & 7.99       & 0     \\
SWIFT J0228.1+3118  &$  2.01^{+0.07}_{-0.07}$ & 43.56  &$  3649.8\pm   302.7$&$  1889.4\pm    93.8$ &$    0.38\pm    0.06$   & 42.74      & 43.77 & 7.50       & 0     \\
SWIFT J0230.2-0900  &$  2.02^{+0.08}_{-0.07}$ & 42.87  &$  1432.8\pm   259.9$&$   918.8\pm   110.8$ &$    1.48\pm    0.10$   & 43.10      & 44.08 & 6.82       & 1     \\
SWIFT J0234.6-0848  &$  2.12^{+0.14}_{-0.23}$ & 44.14  &$  4792.4\pm    55.6$&$  2010.8\pm    26.8$ &$    0.18\pm    0.02$   & 43.19      & 44.17 & 7.85       & 0     \\
SWIFT J0244.8+6227  &$  1.63^{+0.04}_{-0.03}$ & 44.72  &$  9175.9\pm  1416.8$&$  3846.6\pm   415.4$ &$    0.77\pm    0.11$   & 42.83      & 43.85 & 8.02       & 0     \\
SWIFT J0300.0-1048  &$  1.98^{+0.12}_{-0.38}$ & 43.64  &$ 13368.7\pm   168.2$&$  5604.5\pm    92.2$ &$    0.13\pm    0.01$   & 43.44      & 44.39 & 8.88       & 0     \\
SWIFT J0311.5-2045  &$  1.94^{+0.14}_{-0.08}$ & 44.41  &$  6419.9\pm   210.3$&$  2878.9\pm   124.9$ &$    0.87\pm    0.03$   & 43.25      & 44.22 & 7.93       & 0     \\
SWIFT J0311.9+5032  &$  1.80^{+0.20}_{-0.16}$ & 43.99  &$  4889.0\pm   144.4$&$  2304.7\pm   763.8$ &$    0.52\pm    0.06$   & 43.61      & 44.53 & 8.04       & 0     \\
SWIFT J0333.3+3720  &$  1.70^{+0.26}_{-0.11}$ & 44.24  &$  4479.1\pm   250.6$&$  2938.2\pm   371.1$ &$    0.20\pm    0.06$   & 43.75      & 44.66 & 8.44       & 0     \\
SWIFT J0354.2+0250  &$  1.31^{+0.04}_{-0.05}$ & 43.57  &$  3087.1\pm   106.3$&$  2518.9\pm   274.8$ &$    0.61\pm    0.02$   & 42.98      & 43.98 & 7.78       & 0     \\
SWIFT J0414.8-0754  &$  1.97^{+0.92}_{-0.27}$ & 43.89  &$  4805.6\pm    97.8$&$  2681.2\pm    17.3$ &$    0.68\pm    0.02$   & 43.34      & 44.29 & 7.98       & 0     \\
SWIFT J0418.3+3800  &$  1.76^{+0.22}_{-0.10}$ & 44.83  &$  5993.0\pm   112.4$&$  2069.4\pm    35.8$ &$    0.00\pm    0.00$   & 44.38      & 45.25 & 8.51       & 0     \\
SWIFT J0426.2-5711  &$  1.92^{+0.05}_{-0.06}$ & 44.84  &$  3241.3\pm   365.6$&$  2461.3\pm    26.0$ &$    0.12\pm    0.02$   & 44.34      & 45.20 & 8.60       & 0     \\
SWIFT J0428.2-6704B &$  1.84^{+0.13}_{-0.16}$ & 43.52  &$  3962.1\pm   947.5$&$  1790.6\pm   278.4$ &$    0.47\pm    0.12$   & 43.73      & 44.65 & 7.90       & 0     \\
SWIFT J0429.6-2114  &$  1.81^{+0.07}_{-0.08}$ & 44.16  &$  8500.1\pm   141.5$&$  4250.9\pm    31.4$ &$    0.60\pm    0.01$   & 44.07      & 44.95 & 8.76       & 0     \\
SWIFT J0436.3-1022  &$  1.78^{+0.46}_{-0.22}$ & 43.69  &$  2553.5\pm   292.7$&$  1394.9\pm    26.0$ &$    0.55\pm    0.06$   & 43.55      & 44.48 & 7.57       & 0     \\
SWIFT J0438.2-1048  &$  2.12^{+0.19}_{-0.34}$ & 43.75  &$  8537.9\pm   291.9$&$  3579.2\pm   511.6$ &$    0.09\pm    0.10$   & 43.29      & 44.25 & 8.43       & 0     \\
SWIFT J0440.9+2741  &$  1.81^{+0.17}_{-0.13}$ & 43.85  &$  6246.8\pm   527.7$&$  6265.6\pm  1486.0$ &$    0.58\pm    0.06$   & 43.40      & 44.35 & 8.79       & 0     \\
SWIFT J0451.4-0346  &$  1.70^{+0.09}_{-0.08}$ & 43.23  &$  6326.2\pm  1570.8$&$  2652.7\pm   432.9$ &$    0.40\pm    0.37$   & 42.39      & 43.47 & 7.62       & 0     \\
SWIFT J0452.2+4933  &$  1.75^{+0.09}_{-0.07}$ & 44.08  &$  7043.1\pm   239.9$&$  2952.9\pm    59.0$ &$    1.45\pm    0.05$   & 43.41      & 44.36 & 7.81       & 0     \\
SWIFT J0503.0+2302  &$  1.66^{+0.04}_{-0.08}$ & 44.21  &$  4077.0\pm   434.2$&$  2226.2\pm   164.4$ &$    1.18\pm    0.10$   & 44.23      & 45.10 & 8.06       & 0     \\
SWIFT J0510.7+1629  &$  1.77^{+0.04}_{-0.02}$ & 43.79  &$  3890.9\pm   276.7$&$  1633.4\pm    18.4$ &$    0.12\pm    0.07$   & 42.56      & 43.62 & 7.39       & 0     \\
﻿SWIFT J0516.2-0009  &$  2.07^{+0.04}_{-0.04}$ & 44.23  &$  6320.1\pm   384.9$&$  2886.1\pm    40.4$ &$    0.68\pm    0.04$   & 43.87      & 44.77 & 8.15       & 1     \\
SWIFT J0501.9-3239  &$  1.73^{+0.02}_{-0.01}$ & 43.24  &$  4519.0\pm   168.1$&$  1524.5\pm    93.9$ &$    0.25\pm    0.13$   & 42.48      & 43.54 & 7.24       & 0     \\
SWIFT J0524.1-1210  &$  1.54^{+0.08}_{-0.17}$ & 44.02  &$  7339.0\pm   195.8$&$  3268.1\pm   314.4$ &$    0.37\pm    0.11$   & 44.36      & 45.23 & 8.76       & 0     \\
SWIFT J0532.9+1343  &$  1.79^{+0.49}_{-0.22}$ & 43.35  &$  8059.4\pm  2543.0$&$  3580.8\pm   307.0$ &$    0.83\pm    0.23$   & 42.64      & 43.68 & 7.84       & 0     \\
SWIFT J0535.4+4013  &$  1.80^{+0.16}_{-0.08}$ & 43.21  &$  4924.0\pm   445.5$&$  2065.7\pm     2.6$ &$    1.69\pm    0.21$   & 42.97      & 43.97 & 7.19       & 0     \\
SWIFT J0543.9-2749  &$  1.64^{+0.60}_{-0.65}$ & 42.58  &$  4100.7\pm   326.7$&$  2333.5\pm   218.1$ &$    1.55\pm    0.12$   & 42.09      & 43.21 & 6.93       & 0     \\
SWIFT J0554.8+4625  &$  1.89^{+0.02}_{-0.02}$ & 44.09  &$  3968.0\pm   271.2$&$  1665.7\pm   778.5$ &$    0.23\pm    0.04$   & 43.16      & 44.14 & 7.65       & 0     \\
SWIFT J0557.9-3822  &$  1.70^{+0.05}_{-0.03}$ & 43.90  &$  2779.7\pm    80.7$&$  2003.2\pm    42.0$ &$    0.73\pm    0.01$   & 42.97      & 43.97 & 7.53       & 0     \\
SWIFT J0559.8-5028  &$  2.47^{+0.02}_{-0.02}$ & 45.04  &$  1518.1\pm    48.1$&$   922.0\pm    28.5$ &$    1.63\pm    0.05$   & 44.29      & 45.16 & 7.15       & 0     \\
SWIFT J0602.2+2829  &$  1.50^{+0.06}_{-0.06}$ & 44.21  &$  4636.1\pm   150.9$&$  3342.3\pm    91.6$ &$    0.24\pm    0.01$   & 43.44      & 44.39 & 8.39       & 0     \\
SWIFT J0606.0-2755  &$  1.82^{+0.73}_{-0.36}$ & 44.38  &$  5805.3\pm   347.7$&$  3333.2\pm   286.6$ &$    0.21\pm    0.04$   & 43.65      & 44.58 & 8.50       & 0     \\
SWIFT J0634.7-7445  &$  1.61^{+0.30}_{-0.20}$ & 44.55  &$  4825.1\pm   136.3$&$  2685.0\pm   243.8$ &$    0.64\pm    0.15$   & 44.18      & 45.06 & 8.40       & 0     \\
SWIFT J0651.9+7426  &$  1.54^{+0.07}_{-0.06}$ & 43.69  &$  6168.8\pm   648.9$&$  2586.9\pm   437.3$ &$    0.98\pm    0.09$   & 43.69      & 44.61 & 8.01       & 0     \\
SWIFT J0714.3+4541  &$  1.70^{+0.85}_{-1.02}$ & 43.87  &$  4688.5\pm   240.2$&$  2497.1\pm    87.6$ &$    0.98\pm    0.04$   & 43.93      & 44.83 & 8.09       & 0     \\
SWIFT J0736.9+5846  &$  2.09^{+0.22}_{-0.19}$ & 43.50  &$  3914.3\pm   151.9$&$  2144.5\pm  3725.7$ &$    0.62\pm    0.18$   & 43.45      & 44.40 & 7.86       & 0     \\
SWIFT J0742.5+4948  &$  1.86^{+0.08}_{-0.12}$ & 43.71  &$  4211.4\pm   192.7$&$  2405.1\pm    43.9$ &$    0.18\pm    0.01$   & 43.68      & 44.60 & 7.95       & 1     \\
SWIFT J0743.3-2546  &$  2.08^{+0.14}_{-0.26}$ & 43.43  &$  3105.9\pm   137.7$&$  2351.4\pm   171.8$ &$    0.58\pm    0.16$   & 42.64      & 43.68 & 7.57       & 0     \\
SWIFT J0747.5+6057  &$  1.93^{+0.17}_{-0.10}$ & 43.47  &$  3806.7\pm  1261.1$&$  2771.4\pm    78.3$ &$    0.88\pm    0.06$   & 43.23      & 44.20 & 9.00       & 2     \\
SWIFT J0752.2+1937  &$  1.28^{+0.33}_{-0.21}$ & 44.75  &$ 12031.9\pm   595.9$&$  4437.3\pm    75.9$ &$    0.00\pm    0.00$   & 44.44      & 45.30 & 9.21       & 0     \\
SWIFT J0759.8-3844  &$  1.51^{+0.03}_{-0.03}$ & 44.28  &$  7198.3\pm   338.5$&$  4297.4\pm   132.1$ &$    0.16\pm    0.03$   & 43.81      & 44.72 & 8.81       & 0     \\
SWIFT J0803.4+0842  &$  1.50^{+0.25}_{-0.11}$ & 43.96  &$  5799.2\pm   257.3$&$  2432.0\pm   102.4$ &$    0.39\pm    0.04$   & 42.99      & 43.99 & 7.84       & 0     \\
SWIFT J0810.9+7602  &$  2.27^{+0.20}_{-0.09}$ & 44.56  &$  3295.8\pm   140.6$&$  2953.9\pm   323.9$ &$    0.79\pm    0.03$   & 44.91      & 45.74 & 8.51       & 1     \\
SWIFT J0814.3+0423  &$  1.78^{+0.23}_{-0.59}$ & 43.75  &$  2227.5\pm   192.5$&$  2337.9\pm    81.5$ &$    0.40\pm    0.10$   & 42.95      & 43.96 & 7.78       & 0     \\
SWIFT J0830.1+4154  &$  1.80^{+0.27}_{-0.25}$ & 44.74  &$  8799.0\pm   223.8$&$  3297.9\pm   199.2$ &$    0.00\pm    0.00$   & 43.57      & 44.50 & 8.53       & 0     \\
SWIFT J0832.5+3703  &$  1.87^{+0.30}_{-0.10}$ & 44.55  &$ 13296.0\pm   529.4$&$  5531.7\pm   225.4$ &$    0.02\pm    0.02$   & 44.11      & 45.00 & 9.23       & 0     \\
SWIFT J0839.6-1213  &$  2.03^{+0.08}_{-0.10}$ & 45.44  &$  4364.0\pm   528.6$&$  4504.2\pm   204.9$ &$    0.00\pm    0.00$   & 44.45      & 45.31 & 9.22       & 0     \\
SWIFT J0845.0-3531  &$  1.89^{+0.49}_{-0.52}$ & 44.81  &$  9073.2\pm   586.5$&$  5095.5\pm  1086.2$ &$    0.15\pm    0.03$   & 44.51      & 45.36 & 9.30       & 0     \\
SWIFT J0904.3+5538  &$  1.84^{+0.07}_{-0.07}$ & 43.69  &$  4665.0\pm   460.3$&$  2412.4\pm    63.3$ &$    0.00\pm    0.00$   & 42.72      & 43.75 & 7.85       & 0     \\
SWIFT J0917.2-6221  &$  2.11^{+0.17}_{-0.20}$ & 44.39  &$  4742.2\pm   891.1$&$  3600.7\pm   283.8$ &$    0.43\pm    0.05$   & 44.85      & 45.68 & 9.06       & 0     \\

   \hline
\end{tabular}
\end{lrbox}
\scalebox{0.75}{\usebox{\tablebox}}
\\(This table is available in its entirety in machine-readable form.)

\end{table*}

\begin{table*}
\setcounter{table}{1}
\caption{--continu}
\centering
\label{table2}
\begin{lrbox}{\tablebox}
\begin{tabular}{lllllllllllll}
\hline
    Name & $\Gamma$ & $\log L_{\rm X}$ & $\rm FWHM_{\hb}$ & $\shb$ & \rfe & \lv & \lb & \mbh & Notes\\
                 &                    &  \ergs                   &\kms                          & \kms     &       & \ergs & \ergs &$\rm M_\odot$ & \\
 (1)&(2)&(3)&(4)&(5)&(6)&(7)&(8)&(9) & (10) \\
\hline

SWIFT J0918.5+1618  &$  2.08^{+0.04}_{-0.02}$ & 43.82  &$  6274.5\pm   159.8$&$  3134.0\pm   166.6$ &$    0.38\pm    0.06$   & 43.92      & 44.82 & 8.51       & 0     \\
SWIFT J0923.7+2255  &$  1.76^{+0.06}_{-0.03}$ & 44.01  &$  3176.5\pm   114.9$&$  2354.8\pm    28.2$ &$    0.60\pm    0.05$   & 43.43      & 44.38 & 7.94       & 0     \\
SWIFT J0925.0+5218  &$  1.80^{+0.04}_{-0.03}$ & 44.22  &$  2986.9\pm   321.1$&$  1947.0\pm    14.0$ &$    0.25\pm    0.02$   & 43.66      & 44.58 & 7.41       & 1     \\
SWIFT J0926.1+6931  &$  2.16^{+0.65}_{-1.05}$ & 43.53  &$  6141.5\pm   522.6$&$  2575.5\pm   274.3$ &$    0.77\pm    0.27$   & 43.38      & 44.34 & 7.93       & 0     \\
SWIFT J0926.2+1244  &$  1.82^{+0.14}_{-0.10}$ & 43.48  &$  2494.9\pm   414.9$&$  2137.9\pm   296.5$ &$    0.56\pm    0.07$   & 43.05      & 44.04 & 7.69       & 0     \\
SWIFT J0927.3+2301  &$  1.67^{+0.06}_{-0.05}$ & 43.42  &$ 10261.1\pm   421.8$&$  3619.5\pm   134.2$ &$    0.47\pm    0.04$   & 42.77      & 43.80 & 8.58       & 2     \\
SWIFT J0935.5+2616  &$  1.71^{+0.21}_{-0.09}$ & 44.72  &$  3939.7\pm   126.3$&$  2421.3\pm   109.3$ &$    0.28\pm    0.02$   & 43.91      & 44.81 & 8.32       & 0     \\
SWIFT J0942.2+2344  &$  1.65^{+0.38}_{-0.18}$ & 43.17  &$  3942.5\pm   143.9$&$  1789.1\pm   117.3$ &$    1.03\pm    0.05$   & 42.45      & 43.52 & 7.07       & 0     \\
SWIFT J0947.7+0726  &$  1.77^{+0.09}_{-0.12}$ & 44.64  &$  8054.4\pm   641.8$&$  4734.1\pm    67.5$ &$    0.00\pm    0.00$   & 43.63      & 44.56 & 8.87       & 0     \\
SWIFT J0959.9+1303  &$  1.89^{+0.29}_{-0.10}$ & 43.24  &$  1916.2\pm    47.1$&$  1422.7\pm   101.3$ &$    0.76\pm    0.04$   & 43.00      & 44.00 & 7.24       & 0     \\
SWIFT J1020.5-0237B &$  1.72^{+0.12}_{-0.12}$ & 43.70  &$  8674.3\pm   162.9$&$  3622.5\pm    51.3$ &$    0.39\pm    0.02$   & 43.34      & 44.29 & 8.35       & 0     \\
SWIFT J1023.5+1952  &$  1.56^{+0.01}_{-0.01}$ & 42.66  &$  4005.2\pm    95.1$&$  2177.4\pm    34.5$ &$    0.43\pm    0.09$   & 42.24      & 43.34 & 7.06       & 1     \\
SWIFT J1031.9-1418  &$  1.86^{+0.07}_{-0.04}$ & 44.77  &$  5926.0\pm   349.8$&$  2832.1\pm   145.6$ &$    0.07\pm    0.02$   & 44.44      & 45.30 & 8.79       & 0     \\
SWIFT J1038.8-4942  &$  1.31^{+0.13}_{-0.10}$ & 44.36  &$  6988.6\pm   598.4$&$  4675.2\pm   193.0$ &$    0.03\pm    0.03$   & 43.70      & 44.62 & 8.88       & 0     \\
SWIFT J1043.4+1105  &$  1.69^{+0.08}_{-0.07}$ & 43.89  &$  6372.8\pm   200.2$&$  2465.3\pm   603.3$ &$    0.18\pm    0.02$   & 43.18      & 44.16 & 8.02       & 0     \\
SWIFT J1100.9+1104  &$  1.71^{+0.06}_{-0.05}$ & 43.62  &$  6073.9\pm   214.7$&$  2547.0\pm   542.1$ &$    0.00\pm    0.00$   & 42.81      & 43.83 & 7.94       & 0     \\
SWIFT J1105.7+5854B &$  1.57^{+0.19}_{-0.12}$ & 44.75  &$  5306.4\pm   117.0$&$  3138.2\pm   281.7$ &$    0.40\pm    0.02$   & 44.37      & 45.24 & 8.72       & 0     \\
SWIFT J1106.5+7234  &$  1.57^{+0.13}_{-0.12}$ & 43.59  &$  4988.1\pm   532.2$&$  2246.5\pm    24.9$ &$    0.61\pm    0.17$   & 42.79      & 43.81 & 7.78       & 1     \\
﻿SWIFT J1113.6+0936  &$  1.56^{+0.18}_{-0.10}$ & 43.39  &$  5950.0\pm     2.0$&$  2303.2\pm     0.8$ &$    0.34\pm    0.16$    & 42.50      & 43.56 & 8.52       & 2     \\
SWIFT J1125.6+5423  &$  1.62^{+0.15}_{-0.05}$ & 43.29  &$  3563.2\pm   107.7$&$  1710.9\pm    76.3$ &$    0.22\pm    0.15$   & 42.55      & 43.60 & 7.24       & 1     \\
SWIFT J1132.9+1019A &$  1.79^{+0.33}_{-0.26}$ & 43.85  &$  5317.6\pm   344.8$&$  4947.8\pm   899.8$ &$    0.29\pm    0.07$   & 42.92      & 43.92 & 8.63       & 2     \\
SWIFT J1132.9+1019B &$  1.68^{+0.12}_{-0.11}$ & 46.01  &$  2682.0\pm    69.7$&$  2313.0\pm   166.8$ &$    0.35\pm    0.01$   & 45.32      & 46.13 & 8.93       & 0     \\
SWIFT J1136.0+2132  &$  2.04^{+0.22}_{-0.17}$ & 43.43  &$  2722.8\pm   135.9$&$  1516.9\pm    49.0$ &$    0.68\pm    0.03$   & 42.88      & 43.89 & 7.27       & 0     \\
SWIFT J1139.1+5913  &$  1.63^{+0.06}_{-0.05}$ & 44.28  &$  4344.8\pm    87.5$&$  2122.7\pm    17.6$ &$    0.40\pm    0.06$   & 43.79      & 44.70 & 8.10       & 0     \\
SWIFT J1139.0-3743  &$  1.98^{+0.04}_{-0.04}$ & 43.69  &$  3141.2\pm   378.5$&$  1106.9\pm    52.9$ &$    1.34\pm    0.19$   & 42.56      & 43.61 & 7.50       & 1     \\
SWIFT J1141.3+2156  &$  1.86^{+0.04}_{-0.08}$ & 44.25  &$  3355.4\pm   126.4$&$  1409.6\pm    13.2$ &$    0.00\pm    0.00$   & 43.73      & 44.65 & 7.87       & 0     \\
SWIFT J1144.1+3652  &$  1.72^{+0.27}_{-0.17}$ & 43.73  &$  7543.2\pm   186.9$&$  3162.4\pm    78.3$ &$    0.00\pm    0.00$   & 42.94      & 43.95 & 8.19       & 0     \\
SWIFT J1143.7+7942  &$  1.69^{+0.03}_{-0.03}$ & 42.38  &$  1526.1\pm    10.5$&$   912.8\pm    33.0$ &$    1.55\pm    0.08$   & 41.25      & 42.52 & 5.71       & 0     \\
SWIFT J1145.6-1819  &$  1.77^{+0.04}_{-0.05}$ & 44.12  &$  2074.3\pm    69.1$&$  1852.3\pm   109.6$ &$    0.12\pm    0.02$   & 43.65      & 44.58 & 8.02       & 0     \\
SWIFT J1148.3+0901  &$  1.51^{+0.31}_{-0.18}$ & 44.14  &$  7348.7\pm   244.2$&$  3639.2\pm   114.1$ &$    0.00\pm    0.00$   & 43.71      & 44.63 & 8.68       & 0     \\
SWIFT J1149.3+5307  &$  1.67^{+0.21}_{-0.19}$ & 44.14  &$ 10598.5\pm   792.4$&$  4371.5\pm   375.4$ &$    0.35\pm    0.03$   & 43.16      & 44.14 & 8.68       & 2     \\
SWIFT J1152.1-1122  &$  1.99^{+0.04}_{-0.11}$ & 43.95  &$  2716.6\pm   218.4$&$  1799.0\pm   136.9$ &$    0.42\pm    0.03$   & 43.51      & 44.45 & 7.81       & 0     \\
SWIFT J1201.2-0341  &$  1.98^{+0.20}_{-0.11}$ & 43.00  &$  2336.8\pm   150.5$&$  1264.0\pm    35.2$ &$    0.08\pm    0.05$   & 42.29      & 43.38 & 6.76       & 1     \\
SWIFT J1203.0+4433  &$  1.70^{+0.11}_{-0.05}$ & 42.01  &$  1414.9\pm   490.2$&$   730.1\pm   176.5$ &$    1.23\pm    0.32$   & 41.96      & 43.10 & 6.44       & 1     \\
SWIFT J1204.8+2758  &$  1.93^{+0.05}_{-0.10}$ & 44.90  &$  5288.9\pm   271.8$&$  4348.1\pm    55.4$ &$    0.05\pm    0.01$   & 44.03      & 44.92 & 8.97       & 0     \\
SWIFT J1205.8+4959  &$  1.19^{+0.17}_{-0.17}$ & 44.00  &$  8305.0\pm   773.7$&$  3820.7\pm   183.4$ &$    0.58\pm    0.03$   & 43.30      & 44.26 & 8.68       & 2     \\
SWIFT J1210.5+3924  &$  1.73^{+0.03}_{-0.03}$ & 43.01  &$  5646.5\pm   562.6$&$  2368.1\pm   182.7$ &$    0.00\pm    0.00$   & 42.09      & 43.21 & 7.58       & 1     \\
SWIFT J1210.7+3819  &$  1.73^{+0.13}_{-0.14}$ & 43.41  &$  6059.8\pm   231.9$&$  2406.6\pm    99.9$ &$    0.00\pm    0.00$   & 42.87      & 43.88 & 7.92       & 0     \\
SWIFT J1217.3+0714  &$  1.90^{+0.13}_{-0.16}$ & 42.38  &$ 11152.8\pm  2784.6$&$  4351.3\pm   899.2$ &$    0.98\pm    0.20$   & 41.37      & 42.62 & 8.00       & 2     \\
SWIFT J1218.5+2952  &$  2.02^{+0.15}_{-0.09}$ & 42.90  &$  2176.2\pm  1280.4$&$  2086.5\pm    98.1$ &$    1.74\pm    0.14$   & 42.57      & 43.62 & 6.89       & 1     \\
SWIFT J1222.4+7520  &$  2.05^{+0.10}_{-0.09}$ & 44.28  &$  4974.0\pm   303.7$&$  2614.5\pm   133.7$ &$    0.47\pm    0.07$   & 44.80      & 45.63 & 8.74       & 0     \\
SWIFT J1223.7+0238  &$  1.77^{+0.06}_{-0.04}$ & 43.47  &$  5433.9\pm   203.2$&$  3136.2\pm   383.2$ &$    0.33\pm    0.08$   & 42.67      & 43.71 & 7.93       & 0     \\
SWIFT J1232.1+2009  &$  2.21^{+0.07}_{-0.05}$ & 44.13  &$  3908.8\pm   698.2$&$  2328.6\pm    62.9$ &$    0.20\pm    0.01$   & 43.70      & 44.62 & 8.02       & 1     \\
SWIFT J1232.0-4219  &$  1.62^{+0.08}_{-0.10}$ & 44.60  &$  5906.9\pm   163.3$&$  3639.3\pm   622.9$ &$    0.13\pm    0.04$   & 44.08      & 44.96 & 8.81       & 0     \\
SWIFT J1239.6-0519  &$  2.03^{+0.01}_{-0.01}$ & 43.06  &$  3677.0\pm   110.8$&$  1544.0\pm    26.9$ &$    0.54\pm    0.03$   & 42.62      & 43.66 & 7.11       & 1     \\
SWIFT J1241.6-5748  &$  1.73^{+0.16}_{-0.03}$ & 43.82  &$  8883.4\pm  1623.5$&$  4567.2\pm   245.0$ &$    0.17\pm    0.01$   & 43.11      & 44.09 & 8.53       & 0     \\
SWIFT J1252.3-1323  &$  1.91^{+0.18}_{-0.13}$ & 42.77  &$  1898.7\pm    71.4$&$   813.1\pm    24.8$ &$    0.00\pm    0.00$   & 42.56      & 43.61 & 6.78       & 1     \\
SWIFT J1255.0-2657  &$  1.48^{+0.06}_{-0.07}$ & 44.13  &$  6031.1\pm   446.9$&$  2529.3\pm   348.3$ &$    0.00\pm    0.00$   & 43.79      & 44.70 & 8.41       & 0     \\
SWIFT J1302.9+1620  &$  1.59^{+0.15}_{-0.05}$ & 44.28  &$  3183.3\pm   348.5$&$  1346.1\pm   125.7$ &$    0.00\pm    0.00$   & 43.73      & 44.65 & 7.83       & 0     \\
SWIFT J1303.8+5345  &$  1.85^{+0.03}_{-0.03}$ & 43.84  &$  5322.5\pm   448.4$&$  2232.4\pm   704.8$ &$    0.31\pm    0.10$   & 42.51      & 43.57 & 7.56       & 0     \\
SWIFT J1306.4-4025A &$  2.00^{+0.10}_{-0.12}$ & 43.26  &$  2409.8\pm   522.5$&$  2746.7\pm   381.6$ &$    1.95\pm    0.20$   & 42.74      & 43.77 & 7.23       & 0     \\
SWIFT J1313.1-1108  &$  1.94^{+0.06}_{-0.07}$ & 43.50  &$  3139.9\pm   425.6$&$  1978.8\pm    49.8$ &$    0.59\pm    0.03$   & 43.10      & 44.08 & 7.64       & 0     \\
SWIFT J1316.9-7155  &$  1.36^{+0.11}_{-0.14}$ & 44.29  &$  9150.5\pm  3636.7$&$  6280.8\pm   253.4$ &$    0.30\pm    0.10$   & 43.73      & 44.65 & 9.05       & 0     \\
SWIFT J1341.2-1439  &$  1.96^{+0.20}_{-0.07}$ & 43.95  &$  2767.9\pm   307.6$&$  1700.8\pm   924.8$ &$    0.46\pm    0.06$   & 43.44      & 44.39 & 7.72       & 0     \\
SWIFT J1341.9+3537  &$  1.79^{+0.18}_{-0.16}$ & 41.63  &$  5611.7\pm   284.3$&$  1898.6\pm   104.4$ &$    0.79\pm    0.22$   & 41.54      & 42.75 & 6.90       & 1     \\
SWIFT J1346.4+1924  &$  1.78^{+0.18}_{-0.21}$ & 44.29  &$  5234.3\pm   364.7$&$  3561.2\pm   601.4$ &$    0.14\pm    0.03$   & 43.81      & 44.72 & 8.66       & 0     \\
SWIFT J1349.3-3018  &$  1.89^{+0.02}_{-0.01}$ & 44.21  &$  6197.9\pm   337.9$&$  4779.3\pm   622.1$ &$    0.62\pm    0.10$   & 42.89      & 43.90 & 8.29       & 0     \\
SWIFT J1349.7+0209  &$  1.72^{+0.15}_{-0.12}$ & 43.61  &$  3367.3\pm   227.7$&$  1745.3\pm    88.7$ &$    0.00\pm    0.00$   & 42.85      & 43.86 & 7.63       & 0     \\
SWIFT J1352.8+6917  &$  1.75^{+0.11}_{-0.05}$ & 43.91  &$  7381.1\pm   248.3$&$  5917.0\pm   468.3$ &$    0.84\pm    0.12$   & 43.71      & 44.63 & 7.77       & 1     \\
SWIFT J1356.1+3832  &$  1.74^{+0.05}_{-0.05}$ & 44.12  &$  5872.8\pm    52.7$&$  4183.7\pm   414.1$ &$    0.13\pm    0.05$   & 42.92      & 43.92 & 8.37       & 0     \\
SWIFT J1416.9-1158  &$  1.57^{+0.17}_{-0.08}$ & 44.66  &$ 11406.6\pm   540.6$&$  4781.7\pm   315.4$ &$    0.17\pm    0.07$   & 43.96      & 44.86 & 8.98       & 0     \\
SWIFT J1417.9+2507  &$  1.82^{+0.10}_{-0.15}$ & 43.71  &$  7392.7\pm   319.9$&$  3446.3\pm    63.8$ &$    0.17\pm    0.01$   & 43.29      & 44.25 & 7.97       & 1     \\
SWIFT J1419.0-2639  &$  2.13^{+0.01}_{-0.01}$ & 43.68  &$  3219.2\pm   114.8$&$  2189.6\pm   106.2$ &$    0.37\pm    0.04$   & 42.64      & 43.68 & 7.59       & 0     \\
SWIFT J1421.4+4747  &$  2.05^{+0.23}_{-0.07}$ & 44.38  &$  5759.1\pm   110.6$&$  2719.1\pm   140.3$ &$    0.17\pm    0.01$   & 43.96      & 44.86 & 8.48       & 0     \\
SWIFT J1427.5+1949  &$  1.78^{+0.12}_{-0.10}$ & 44.63  &$  8290.6\pm   352.1$&$  4555.0\pm   196.2$ &$    0.32\pm    0.03$   & 44.60      & 45.45 & 9.18       & 0     \\
﻿SWIFT J1429.2+0118  &$  2.18^{+0.11}_{-0.05}$ & 44.52  &$  5784.8\pm   206.9$&$  4385.2\pm   230.4$ &$    0.25\pm    0.02$    & 44.63      & 45.47 & 8.93       & 1     \\
SWIFT J1434.9+4837  &$  2.00^{+0.06}_{-0.05}$ & 43.59  &$  4330.8\pm   211.8$&$  2201.8\pm    32.9$ &$    0.17\pm    0.02$   & 43.06      & 44.05 & 7.87       & 0     \\
SWIFT J1436.4+5846  &$  2.14^{+0.04}_{-0.03}$ & 43.76  &$  5420.1\pm   181.4$&$  2273.3\pm    33.8$ &$    0.69\pm    0.05$   & 43.74      & 44.66 & 7.96       & 1     \\
SWIFT J1446.7-6416  &$  1.82^{+0.29}_{-0.25}$ & 43.98  &$  5074.3\pm   384.7$&$  2128.8\pm    65.4$ &$    0.00\pm    0.00$   & 42.77      & 43.80 & 7.77       & 0     \\

   \hline
\end{tabular}
\end{lrbox}
\scalebox{0.75}{\usebox{\tablebox}}
\\(This table is available in its entirety in machine-readable form.)

\end{table*}

\begin{table*}
\setcounter{table}{1}
\caption{--continu}
\centering
\label{table2}
\begin{lrbox}{\tablebox}
\begin{tabular}{lllllllllllll}
\hline
    Name & $\Gamma$ & $\log L_{\rm X}$ & $\rm FWHM_{\hb}$ & $\shb$ & \rfe & \lv & \lb & \mbh & Notes\\
                 &                    &  \ergs                   &\kms                          & \kms     &       & \ergs & \ergs &$\rm M_\odot$ & \\
 (1)&(2)&(3)&(4)&(5)&(6)&(7)&(8)&(9) & (10) \\
\hline

SWIFTJ1448.7-4009   &$  1.86^{+0.50}_{-0.61}$ & 44.71  &$  7115.5\pm  1149.0$&$  3583.4\pm   509.4$ &$    0.35\pm    0.06$   & 44.69      & 45.53 & 9.01       & 0     \\
SWIFT J1453.3+2558  &$  1.86^{+0.07}_{-0.08}$ & 44.13  &$ 10128.9\pm   347.6$&$  4582.8\pm   192.1$ &$    0.09\pm    0.01$   & 43.27      & 44.23 & 8.64       & 0     \\
SWIFT J1454.9-5133  &$  1.96^{+0.19}_{-0.16}$ & 43.11  &$  2135.4\pm   170.0$&$  1107.5\pm   192.3$ &$    2.13\pm    0.24$   & 43.22      & 44.19 & 6.60       & 0     \\
SWIFT J1504.2+1025  &$  2.00^{+0.03}_{-0.02}$ & 44.05  &$  6230.1\pm   151.5$&$  2289.9\pm    49.3$ &$    0.40\pm    0.03$   & 43.12      & 44.10 & 7.84       & 0     \\
SWIFT J1506.7+0353A &$  1.65^{+0.10}_{-0.10}$ & 43.76  &$  5478.4\pm   105.2$&$  3430.7\pm   226.0$ &$    0.46\pm    0.05$   & 43.34      & 44.29 & 8.28       & 0     \\
SWIFT J1512.0-2119  &$  1.99^{+0.18}_{-0.09}$ & 44.16  &$  2203.3\pm  1804.3$&$  2313.5\pm    49.5$ &$    0.85\pm    0.10$   & 43.81      & 44.72 & 8.01       & 0     \\
SWIFT J1513.8-8125  &$  1.65^{+0.37}_{-0.21}$ & 44.53  &$ 10786.0\pm   731.0$&$  5313.2\pm   311.3$ &$    0.00\pm    0.00$   & 44.10      & 44.98 & 9.20       & 0     \\
SWIFT J1521.8+0334  &$  1.74^{+0.28}_{-0.13}$ & 44.81  &$  7261.0\pm   189.7$&$  4594.7\pm   257.7$ &$    0.15\pm    0.05$   & 44.06      & 44.94 & 9.00       & 0     \\
SWIFT J1535.9+5751  &$  1.78^{+0.07}_{-0.12}$ & 43.71  &$  4131.3\pm   109.1$&$  2511.5\pm   126.2$ &$    0.33\pm    0.02$   & 43.17      & 44.15 & 7.47       & 1     \\
SWIFT J1542.0-1410  &$  1.74^{+0.31}_{-0.64}$ & 44.53  &$  8302.9\pm   401.7$&$  3711.4\pm   249.1$ &$    0.63\pm    0.05$   & 43.95      & 44.85 & 8.58       & 0     \\
SWIFT J1547.5+2050  &$  1.61^{+0.16}_{-0.17}$ & 45.23  &$  5531.3\pm   113.1$&$  4733.6\pm   173.1$ &$    0.16\pm    0.03$   & 45.16      & 45.97 & 9.55       & 0     \\
SWIFT J1607.2+4834  &$  1.73^{+0.11}_{-0.08}$ & 44.74  &$  6648.9\pm   608.4$&$  4373.6\pm   243.0$ &$    0.27\pm    0.15$   & 43.64      & 44.57 & 8.71       & 0     \\
SWIFT J1614.0+6544  &$  1.92^{+0.09}_{-0.08}$ & 44.73  &$  9597.5\pm   563.1$&$  5369.2\pm    99.3$ &$    0.40\pm    0.04$   & 44.77      & 45.61 & 8.61       & 1     \\
SWIFT J1618.7-5930  &$  1.60^{+0.09}_{-0.10}$ & 43.77  &$  4169.1\pm   305.0$&$  2090.5\pm   169.2$ &$    0.61\pm    0.24$   & 43.57      & 44.50 & 7.90       & 0     \\
SWIFT J1656.1-5200  &$  1.71^{+0.12}_{-0.05}$ & 44.50  &$  4692.1\pm  1796.4$&$  3682.0\pm   730.1$ &$    0.20\pm    0.04$   & 44.23      & 45.10 & 8.87       & 0     \\
SWIFT J1708.6+2155  &$  1.63^{+0.12}_{-0.15}$ & 44.12  &$  6399.4\pm  1893.4$&$  3304.3\pm   237.1$ &$    0.41\pm    0.04$   & 43.16      & 44.14 & 8.18       & 0     \\
SWIFT J1723.2+3418  &$  1.98^{+0.12}_{-0.12}$ & 45.38  &$  4271.4\pm   529.2$&$  5422.6\pm   177.6$ &$    0.30\pm    0.05$   & 44.89      & 45.72 & 9.48       & 0     \\
SWIFT J1723.5+3630  &$  1.74^{+0.11}_{-0.14}$ & 43.87  &$  2632.5\pm    29.7$&$  1813.5\pm    61.7$ &$    0.96\pm    0.08$   & 42.75      & 43.78 & 7.25       & 0     \\
SWIFT J1741.9-1211  &$  2.09^{+0.06}_{-0.05}$ & 44.15  &$  5327.6\pm   431.6$&$  2926.0\pm   154.7$ &$    0.71\pm    0.04$   & 42.88      & 43.89 & 7.83       & 0     \\
SWIFT J1745.4+2906  &$  1.73^{+0.09}_{-0.09}$ & 44.82  &$ 13408.9\pm   979.8$&$  6300.3\pm    85.0$ &$    0.02\pm    0.01$   & 43.49      & 44.43 & 9.05       & 0     \\
SWIFT J1747.7-2253  &$  2.12^{+0.28}_{-0.19}$ & 43.97  &$ 14067.8\pm   766.5$&$  5897.9\pm    35.6$ &$    0.36\pm    0.02$   & 43.19      & 44.17 & 8.71       & 0     \\
SWIFT J1747.8+6837B &$  2.29^{+0.14}_{-0.14}$ & 43.70  &$  2052.0\pm    68.1$&$  1290.5\pm    22.5$ &$    0.32\pm    0.00$   & 43.48      & 44.42 & 7.55       & 0     \\
SWIFT J1802.8-1455  &$  1.52^{+0.12}_{-0.07}$ & 44.07  &$  5339.6\pm   828.1$&$  2715.8\pm    29.8$ &$    0.19\pm    0.01$   & 43.31      & 44.27 & 8.16       & 0     \\
SWIFT J1807.9+1124  &$  1.70^{+0.11}_{-0.05}$ & 44.58  &$ 11768.4\pm   237.3$&$  4601.0\pm    99.5$ &$    0.10\pm    0.03$   & 43.65      & 44.58 & 8.82       & 0     \\
SWIFT J1822.0+6421  &$  2.22^{+0.03}_{-0.02}$ & 45.73  &$  5655.9\pm   408.8$&$  1979.1\pm   199.7$ &$    0.12\pm    0.06$   & 46.01      & 46.79 & 9.21       & 0     \\
SWIFT J1835.0+3240  &$  2.07^{+0.02}_{-0.01}$ & 44.84  &$  5863.6\pm   159.8$&$  5743.4\pm   270.4$ &$    0.26\pm    0.11$   & 43.86      & 44.77 & 9.05       & 0     \\
SWIFT J1842.0+7945  &$  1.74^{+0.06}_{-0.03}$ & 44.88  &$ 11984.2\pm  2149.9$&$  5023.9\pm   858.5$ &$    0.03\pm    0.04$   & 44.43      & 45.29 & 9.12       & 1     \\
SWIFT J1844.5-6221  &$  1.94^{+0.05}_{-0.05}$ & 43.25  &$  1900.0\pm    74.8$&$  1435.7\pm    26.0$ &$    0.00\pm    0.00$   & 42.89      & 43.90 & 7.48       & 0     \\
SWIFT J1921.1-5842  &$  2.05^{+0.11}_{-0.11}$ & 44.24  &$  5734.1\pm   133.5$&$  4359.8\pm   800.1$ &$    0.56\pm    0.04$   & 43.78      & 44.69 & 8.66       & 0     \\
SWIFT J1925.0+5041  &$  1.71^{+0.26}_{-0.22}$ & 44.06  &$ 10611.2\pm   359.4$&$  4448.3\pm   283.9$ &$    0.00\pm    0.00$   & 43.94      & 44.84 & 8.97       & 0     \\
SWIFT J1930.5+3414  &$  1.86^{+0.21}_{-0.95}$ & 44.43  &$  5421.4\pm   855.3$&$  2524.3\pm  1075.0$ &$    0.42\pm    0.06$   & 43.61      & 44.53 & 8.16       & 0     \\
SWIFT J1933.9+3258  &$  2.03^{+0.05}_{-0.05}$ & 44.23  &$  3712.9\pm   249.1$&$  1830.7\pm   603.0$ &$    0.21\pm    0.01$   & 43.99      & 44.88 & 8.14       & 0     \\
SWIFT J1937.5-0613  &$  2.44^{+0.07}_{-0.05}$ & 42.74  &$  1536.6\pm    53.5$&$   915.1\pm    55.0$ &$    1.15\pm    0.05$   & 43.03      & 44.02 & 6.72       & 0     \\
SWIFT J1938.1-5108  &$  1.90^{+0.06}_{-0.09}$ & 43.75  &$  3210.4\pm   205.3$&$  1942.8\pm    60.5$ &$    0.56\pm    0.02$   & 42.81      & 43.83 & 7.49       & 0     \\
SWIFT J1940.4-3015  &$  1.91^{+0.10}_{-0.08}$ & 44.15  &$  6330.1\pm   467.7$&$  2827.8\pm    66.5$ &$    0.43\pm    0.14$   & 43.41      & 44.36 & 8.15       & 0     \\
SWIFT J1942.6-1024  &$  1.91^{+0.14}_{-0.10}$ & 42.68  &$  4326.8\pm   792.3$&$  4240.6\pm  1139.7$ &$    0.73\pm    0.07$   & 42.12      & 43.24 & 7.42       & 1     \\
SWIFT J2007.0-3433  &$  1.70^{+0.08}_{-0.10}$ & 43.41  &$  2013.6\pm   684.7$&$  1534.6\pm   137.1$ &$    0.81\pm    0.10$   & 42.21      & 43.31 & 6.90       & 0     \\
SWIFT J2035.2+2604  &$  2.08^{+0.39}_{-0.47}$ & 43.79  &$  2401.9\pm   269.3$&$  2419.4\pm    70.2$ &$    0.86\pm    0.05$   & 43.13      & 44.11 & 7.72       & 0     \\
SWIFT J2042.3+7507  &$  2.04^{+0.16}_{-0.21}$ & 45.16  &$ 10314.0\pm   198.4$&$  5411.8\pm    38.5$ &$    0.12\pm    0.00$   & 45.45      & 46.25 & 9.82       & 0     \\
SWIFT J2044.0+2832  &$  2.00^{+0.25}_{-0.16}$ & 44.01  &$  2432.8\pm    66.2$&$  2130.5\pm    94.0$ &$    0.78\pm    0.06$   & 45.43      & 46.23 & 8.75       & 0     \\
SWIFT J2044.2-1045  &$  1.67^{+0.09}_{-0.04}$ & 44.42  &$  3208.0\pm    93.6$&$  2014.2\pm    31.4$ &$    0.00\pm    0.00$   & 44.19      & 45.07 & 8.31       & 1     \\
SWIFT J2109.1-0942  &$  2.22^{+0.45}_{-0.33}$ & 43.51  &$  2727.2\pm   159.6$&$  1626.1\pm   155.2$ &$    0.57\pm    0.01$   & 43.68      & 44.60 & 7.75       & 0     \\
SWIFT J2114.4+8206  &$  1.79^{+0.04}_{-0.04}$ & 44.79  &$  6060.7\pm   196.1$&$  2811.0\pm    32.5$ &$    0.49\pm    0.05$   & 45.08      & 45.90 & 8.93       & 0     \\
﻿SWIFT J2116.3+2512  &$  1.84^{+0.16}_{-0.14}$ & 44.77  &$ 11961.3\pm  1060.1$&$  5025.2\pm   280.5$ &$    0.00\pm    0.00$    & 43.92      & 44.82 & 9.06       & 0     \\
SWIFT J2118.9+3336  &$  1.69^{+0.23}_{-0.19}$ & 43.90  &$  6158.4\pm   747.6$&$  3028.7\pm   235.1$ &$    0.30\pm    0.18$   & 43.52      & 44.46 & 8.32       & 0     \\
SWIFT J2124.6+5057  &$  1.82^{+0.11}_{-0.03}$ & 44.01  &$  2076.6\pm   526.4$&$   888.5\pm   218.6$ &$    0.00\pm    0.00$   & 44.78      & 45.61 & 7.97       & 0     \\
SWIFT J2156.1+4728  &$  1.52^{+0.16}_{-0.08}$ & 43.51  &$  4013.8\pm   190.6$&$  1684.8\pm    77.6$ &$    2.09\pm    0.29$   & 42.51      & 43.57 & 6.64       & 0     \\
SWIFT J2135.5-6222  &$  1.79^{+0.13}_{-0.06}$ & 44.39  &$  5361.4\pm   570.6$&$  2355.5\pm    97.3$ &$    0.56\pm    0.03$   & 43.76      & 44.67 & 8.12       & 0     \\
SWIFT J2139.7+5951  &$  1.74^{+1.34}_{-0.29}$ & 44.55  &$  3088.6\pm   536.5$&$  2316.3\pm   129.4$ &$    0.28\pm    0.02$   & 44.20      & 45.08 & 8.42       & 0     \\
SWIFTJ2157.4-0615   &$  1.79^{+0.15}_{-0.11}$ & 45.05  &$  4231.6\pm   455.8$&$  2167.3\pm    45.6$ &$    0.00\pm    0.00$   & 45.55      & 46.35 & 9.12       & 0     \\
SWIFT J2217.0+1413  &$  1.95^{+0.40}_{-0.17}$ & 44.07  &$  5275.0\pm   269.7$&$  2136.8\pm   157.7$ &$    0.38\pm    0.09$   & 44.26      & 45.13 & 8.34       & 0     \\
SWIFT J2223.9-0207  &$  1.74^{+0.17}_{-0.18}$ & 44.56  &$  7051.5\pm   395.9$&$  2956.4\pm   177.6$ &$    0.15\pm    0.02$   & 43.54      & 44.47 & 8.36       & 0     \\
SWIFT J2229.9+6646  &$  1.67^{+0.08}_{-0.20}$ & 44.86  &$  5331.6\pm   292.5$&$  2682.9\pm    46.7$ &$    0.17\pm    0.01$   & 43.66      & 44.59 & 8.33       & 0     \\
SWIFT J2240.2+0801  &$  1.92^{+0.08}_{-0.08}$ & 43.37  &$  3235.8\pm   181.1$&$  1942.4\pm   254.1$ &$    0.71\pm    0.33$   & 43.62      & 44.55 & 7.83       & 0     \\
SWIFT J2250.7-0854  &$  1.49^{+0.51}_{-0.28}$ & 44.08  &$  5166.7\pm   153.2$&$  2659.8\pm    48.5$ &$    0.50\pm    0.08$   & 42.48      & 43.54 & 7.63       & 0     \\
SWIFT J2254.1-1734  &$  1.59^{+0.35}_{-0.05}$ & 45.01  &$  4490.8\pm  1009.8$&$  2717.4\pm   118.9$ &$    0.03\pm    0.02$   & 44.49      & 45.34 & 8.79       & 0     \\
SWIFT J2256.5+0526  &$  2.18^{+0.15}_{-0.13}$ & 44.05  &$  1826.9\pm    44.3$&$  1620.8\pm    35.0$ &$    0.37\pm    0.03$   & 43.63      & 44.56 & 7.80       & 0     \\
SWIFT J2259.7+2458  &$  1.87^{+0.06}_{-0.06}$ & 43.81  &$  1928.1\pm   120.7$&$  1154.0\pm    57.2$ &$    0.10\pm    0.21$   & 42.97      & 43.97 & 7.29       & 0     \\
SWIFT J2303.3+0852  &$  2.09^{+0.14}_{-0.09}$ & 43.58  &$  2015.4\pm  1281.5$&$  2386.7\pm    75.0$ &$    0.48\pm    0.07$   & 43.51      & 44.45 & 7.15       & 1     \\
SWIFT J2304.8-0843  &$  1.81^{+0.01}_{-0.01}$ & 44.77  &$  9151.3\pm   967.8$&$  3853.8\pm   424.2$ &$    0.06\pm    0.03$   & 43.70      & 44.62 & 8.71       & 0     \\
SWIFT J2307.1+0433  &$  1.57^{+0.13}_{-0.12}$ & 43.76  &$  8164.1\pm   625.1$&$  3496.6\pm  1368.9$ &$    0.00\pm    0.00$   & 43.34      & 44.29 & 8.47       & 0     \\
SWIFT J2308.1+4014  &$  1.56^{+0.36}_{-0.24}$ & 44.17  &$  8437.0\pm   696.5$&$  3536.9\pm   279.5$ &$    0.23\pm    0.14$   & 43.59      & 44.52 & 8.51       & 0     \\
SWIFT J2318.9+0013  &$  1.88^{+0.03}_{-0.03}$ & 43.99  &$  6596.2\pm   366.8$&$  2814.5\pm   184.9$ &$    0.74\pm    0.14$   & 44.31      & 45.17 & 8.47       & 0     \\
SWIFT J2325.6+2157  &$  2.10^{+0.06}_{-0.03}$ & 44.81  &$  5736.2\pm   260.0$&$  3002.0\pm   131.5$ &$    0.29\pm    0.01$   & 44.37      & 45.24 & 8.72       & 0     \\
SWIFT J2351.9-0109  &$  1.78^{+0.35}_{-0.20}$ & 45.19  &$  5656.5\pm   478.2$&$  2852.0\pm    35.9$ &$    0.43\pm    0.10$   & 44.72      & 45.56 & 8.79       & 0     \\

   \hline
\end{tabular}
\end{lrbox}
\scalebox{0.75}{\usebox{\tablebox}}
\\(This table is available in its entirety in machine-readable form.)

\end{table*}

\end{document}